\documentclass[
 floatfix, 
 reprint,
 amsmath,amssymb,
 aps,
]{revtex4-2}
\usepackage{algorithm}    
\usepackage{algpseudocode} 
\usepackage{graphicx}
\usepackage{dcolumn}
\usepackage{bm}
\usepackage{xr-hyper}
\usepackage{xr}
\usepackage{cancel}
\usepackage{tikz}
\newcommand*\circled[1]{\tikz[baseline=(char.base)]{
            \node[shape=circle,draw,inner sep=2pt] (char) {#1};}}
\usepackage[inkscapelatex = false]{svg}
\usepackage[font={footnotesize}]{caption}

\usepackage{lineno}
\usepackage{ragged2e} 
\def\s{\sigma}
\def\g{\gamma}    
\def\a{\alpha}   
\def\b{\beta} 
\def\s{\sigma}    
\def \k{\kappa}  
\def \e{\epsilon}   
\def \r{\rho} 
\def \ve{\varepsilon}
\def \th{\vec{\theta}}   
\def \d{\delta} 
\def \k{\kappa}    
\def \l{\lambda} 
\def \z{\zeta}
\def \x{\xi}      
\def \n{\eta}
 
\def \O{\Omega}   
\def \S{\Sigma} 
\def \G{\Gamma}   
\def \D{} 
\def \Lam{\Lambda} 
\def \h{\hbar}   

\def \E{\rm{E}} 
\def \Var{\rm{Var}} 
\def \Cov{\rm{Cov}} 
 
\def \f{\frac} 
\def \del{\partial}    
 
\def \hf{\tfrac{1}{2}} 
\def \HF{\dfrac{1}{2}}  
\def \HQ{\dfrac{1}{4}}
 
\def \ord{\mathcal{O}} 
\def \ra{\rightarrow} 
\def \wo{\setminus} 
\def \>{\rangle} 
\def \<{\langle} 
\def\dg{^\dagger} 
\def\lba{\left(}    
\def\rba{\right)} 
\def\lbc{\left[} 
\def\rbc{\right]} 
\def\lbb{\left\{} 
\def\rbb{\right\}} 
\def \bra{\langle} 
\def \ket{\rangle} 
\def\ord{\mathcal{O}}

\def\be{\begin{equation}} 
\def\ee{\end{equation}} 
\def\longrightharpoonup{\relbar\joinrel\rightharpoonup}
\def\longleftharpoondown{\leftharpoondown\joinrel\relbar}

\def\longrightleftharpoons{
  \mathop{
    \vcenter{
      \hbox{
      \ooalign{
        \raise1pt\hbox{$\longrightharpoonup\joinrel$}\crcr
    \lower1pt\hbox{$\longleftharpoondown\joinrel$}
    }
      }
    }
  }
}

\newcommand \bea {\begin{eqnarray}} 
\newcommand \eea {\end{eqnarray}} 
 
\newcommand{\nn} {\nonumber}

\def\s{\sigma}
\def \g{\gamma}    \def \a{\alpha}   

\def \b{\beta} 
\def \s{\sigma}    \def \k{\kappa}  
\def \e{\epsilon}   \def \r{\rho} 
\def \ve{\varepsilon}
\def \th{\vec{\theta}}   \def \d{\delta} 
\def \k{\kappa}    

\def \l{\lambda} 
\def \z{\zeta}
\def \x{\xi}      
\def \n{\eta}
 
\def \O{\Omega}   \def \S{\Sigma} 
\def \G{\Gamma}   \def \D{} 
\def \Lam{\Lambda} 

\usepackage{xcolor}
\usepackage{graphicx}

\usepackage[hidelinks]{hyperref}

\begin{document}

\title{Diverse communities promote the coexistence of closely-related strains \\through emergent equalization and stabilization}

\author{Naven Narayanan Venkatanarayanan$^1$}
\email{navenv@ncbs.res.in}
\author{Akshit Goyal$^2$}
\email{akshitg@icts.res.in}

\affiliation{$^1$National Centre for Biological Sciences, Tata Institute of Fundamental Research, Bengaluru 560065.}
\affiliation{$^2$International Centre for Theoretical Sciences, Tata Institute of Fundamental Research, Bengaluru 560089.}

\begin{abstract}
\noindent
Microbial communities harbor extensive fine-scale diversity: closely-related strains of the same species coexist alongside many distantly-related taxa. Yet strain coexistence remains poorly understood, largely because most studies neglect the diverse communities in which strains are embedded. Here we combine community ecology and statistical physics to study the dynamics of closely-related strains in a community context. We demonstrate that in a diverse community, indirect interactions between strains---mediated through the surrounding community members---can be as strong as direct ones. These community-mediated feedbacks cause conspecific strains to behave as if they have correlated growth rates and reduced competition. Using modern coexistence theory, we show that these effects correspond to equalizing and stabilizing mechanisms which together promote strain coexistence. The same equalizing and stabilizing mechanisms also qualitatively transform strain abundance correlations: strains that compete strongly and show negative correlations in isolation instead show positive correlations in a community, appearing mutualistic despite being competitors. Our results demonstrate that strain dynamics are emergent consequences of the surrounding community, and that capturing community feedbacks does not require the full interaction network; only a small number of emergent parameters.
\end{abstract}

\maketitle
\noindent
Microbial communities, ranging from the human gut to marine ecosystems, are characterized by immense biodiversity~\cite{fierer2007metagenomic,shu2022,rosen2015fine,morgan2013biodiversity,ding2016environmental}. While most studies have focused on diversity at the species level, recent high-resolution metagenomic studies have revealed an equally striking layer of ``fine-scale'' diversity: the long-term coexistence of multiple closely-related strains within the same species, both in natural and evolved communities~\cite{zhao2019,garud2019,roodgar2019,truong2017,good2017b,goyal_interactions_2022,poyet2019library}. This strain-level variation often determines the functional output of the microbiome, including pathogen susceptibility and metabolic health~\cite{leventhal2018,van2020diversity,zhang2016strain,yan2020strain,vatanen2019genomic,kalan2019strain,park2022strain,brito2016tracking}. Yet despite its importance, we lack a quantitative framework that can predict when and why closely-related strains coexist.

Current ecological theory suggests that it should be very unlikely for closely-related strains to stably coexist, since their high niche overlap means that even small fitness differences can lead to strains competitively excluding each other~\cite{hutchinson1961,macarthur1967limiting,chesson2000mechanisms,chase2009ecological,adler2007niche,spaak2021mapping,hillerislambers2012rethinking,mayfield2010opposing}. Yet empirical observations repeatedly show that strains routinely coexist for extended periods of time~\cite{goyal_interactions_2022,good2017b,poyet2019library,rosen2015fine,papula2025extensive}. To explain this puzzle, several alternate mechanisms have been proposed, e.g., frequency-dependent selection~\cite{ascensao2025frequency, walton2025community}, phage-mediated interactions~\cite{thingstad2014theoretical} and metabolic niche partitioning~\cite{park2022strain}. 

However, all these mechanisms operate at the level of isolated strain pairs and neglect the diverse community in which strains are embedded~\cite{goyal2025paradox}. In natural settings, strains do not interact in a vacuum---they are coupled to hundreds of co-occurring species which can mediate indirect interactions between strains~\cite{fussmann2007eco,hendry2017eco,govaert2019eco,goyal2021ecology}. Whether these community-mediated feedbacks qualitatively change strain coexistence---and whether they distort the strain abundance correlations widely used to infer ecological interactions~\cite{goyal_interactions_2022,faust2012microbial,tackmann2019rapid,lopez2025imbalance,vanrossum2020,ho2022competition,berry2014deciphering,tikhonov2015interpreting,carr2019use}---remains an open question. 

Recent theoretical work using tools from statistical physics has made significant progress in understanding such diverse communities~\cite{li2025population,BuninGLV,advani2018statistical,tikhonov2018innovation,mahadevan2023,patro2025emergent,feng2024emergent}. These approaches show how collective community-mediated feedbacks emerge from the many ecological interactions in a diverse community. However, they neglect strain structure and focus almost exclusively on the species level. Thus, we lack a framework that accounts for strains within species and reveals how community feedbacks shape the coexistence of closely-related strains.

\begin{figure*}[t]
    \centering
    \includegraphics[width=0.72\textwidth]{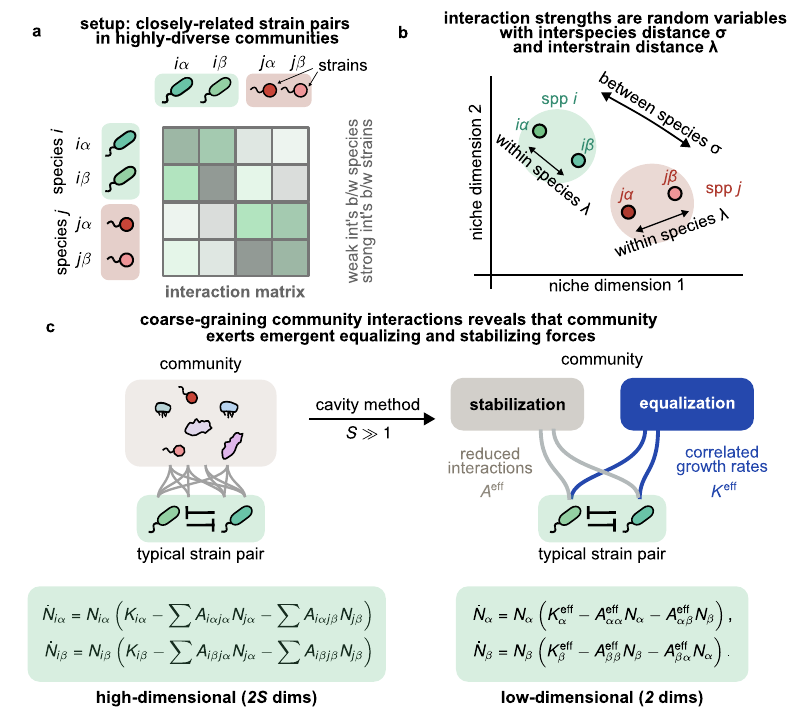}
    \caption{\justifying\textsf{\textbf{Dynamics of closely-related strains in highly-diverse ecological communities.} (a) Schematic of model setup with $S\gg1$ species (we show two: $i$ and $j$) in a community having two distinct strains $\alpha$ and $\beta$. All strains interact through a Generalized Lotka-Volterra (GLV) model with interaction matrix $A_{i\alpha j\beta}$ (Eqs.~\eqref{eq:Asetup1}--\eqref{eq:Asetup2}). (b) Schematic showing that interaction strengths are random variables with a characteristic species dissimilarity $\sigma$ and interstrain interaction dissimilarity $\lambda$. For conceptual illustration we show these strains in a high-dimensional niche space. (c) Using the cavity method, we coarse-grain the GLV model describing the full highly-diverse community (left) into an effective two-strain GLV model (right) for a typical strain pair of the same species. This model shows that strains behave as if they have different apparent growth rates $K^{\mathrm{eff}}$ and interaction strengths $A^{\mathrm{eff}}$.}}
    \label{fig:schematic}
\end{figure*}

Here, we bridge this gap by developing a unified theoretical framework to study the dynamics of closely-related strains embedded in diverse communities. We start with a minimal model of a diverse ecological community containing strains within species, where each species consists of a pair of strains~\cite{goyal_interactions_2022,good2017b,garud2019,roodgar2019}. Using the cavity method, we coarse-grain this full community into an effective description at the level of strains. Our central finding is that a diverse community collectively exerts two ecological forces on closely-related strains: (1) an emergent equalizing force, which correlates apparent strain growth rates, and (2) an emergent stabilizing force, which reduces their apparent competitive interaction strengths. Using modern coexistence theory, we show that these community-mediated equalizing and stabilizing forces significantly promote strain coexistence~\cite{chesson2000, spaak2021mapping, adler2007niche}. Strains can coexist in a community even when, in isolation, they would exclude each other. Further, these forces also transform strain abundance correlations: strains that compete strongly in isolation can appear mutualistic in a community. Our framework highlights that community context can enhance strain coexistence and reveals how to interpret strain abundance correlations in a community.

\section*{Results}

\textbf{\textsf{An ecological model of strain dynamics in highly-diverse communities.}} We aim to understand how diverse ecological communities shape the dynamics of closely-related strains. To do so, we consider a simple ecological model where $S \gg 1$ species interact with each other (Fig.~\ref{fig:schematic}a). Each species $i$ consists of two closely-related strains, $\alpha$ and $\beta$, with abundances $N_{i\alpha}$ and $N_{i\beta}$. We assume two strains per species not just because it is simple and tractable, but also because it is motivated by experimental observations~\cite{goyal_interactions_2022,good2017b,roodgar2019,zhao2019,poyet2019library}. Strain dynamics are described by the following set of $2S$ ordinary differential equations, which extend the canonical Generalized Lotka-Volterra (GLV) model to include strain structure (Appendix A):
\begin{equation}
\frac{dN_{i\alpha}}{dt} = N_{i\alpha} \left( K_{i\alpha} - \sum_{\substack{j=1}}^S A_{i\alpha j\alpha} N_{j\alpha} - \sum_{\substack{j=1}}^S A_{i\alpha j\beta} N_{j\beta} \right)
\label{eq:main_GLV_dynamics}
\end{equation}
where $K_{i\alpha}$ is the intrinsic growth rate of strain $\alpha$ of species $i$, and $A_{i\alpha j\beta}$ is the strength of interaction from strain $\beta$ of species $j$. 
Given the large number of species, we follow in the tradition of complex systems by modeling the underlying parameters as quenched random variables drawn from a Gaussian distribution  (Methods)~\cite{li2025population,BuninGLV,advani2018statistical,tikhonov2018innovation,mahadevan2023,patro2025emergent,feng2024emergent}. 

To model strains within species, we parameterize the interaction matrix $A$ at two levels (Fig.~\ref{fig:schematic}a--b). At the species level, the strength of interactions $A_{i\alpha j\alpha}$ and $A_{i\alpha j\beta}$  between strains from different species $i$ and $j$ is characterized by a standard deviation $\sigma$. These interactions take the form:
\begin{align}
A_{i\alpha j\alpha} &= \frac{\mu}{S} + \frac{\sigma}{\sqrt{S}} a_{ij} + \frac{\lambda}{\sqrt{S}} x_{ij\alpha}, \label{eq:Asetup1}\\
A_{i\alpha j\beta} &= \frac{\mu}{S} + \frac{\sigma}{\sqrt{S}} a_{ij} - \frac{\lambda}{\sqrt{S}} x_{ij\alpha},
\end{align}
where $\mu$ sets the mean interspecies competition, $a_{ij}$ is a random variable capturing species-level variation (identical for both strains of the same species $j$), and $x_{ij\alpha}$ captures strain-level variation. The scaling of interaction strengths with the number of species $S$ merely ensures a good thermodynamic limit $S\to\infty$ (Appendix A) and is consistent with interactions between phylogenetically distant species being weaker than between conspecific strains. Both $a_{ij}$ and $x_{i j\alpha}$ are uncorrelated with zero mean and unit variance. Note that in this setting, strains $\alpha$ and $\beta$ of species $j$ affect a distant strain $i\alpha$ with nearly identical strength---differing only by the $\pm \lambda$ terms. Importantly, since $x_{ij\a}$ has mean $0$, no strain is systematically better than the other. Thus as $\lambda$ increases, so do the phenotypic differences between strains, and consequently, their interactions with other taxa. Since the precise genotype-phenotype map connecting strain genomes with ecological interactions is still unknown, this is effectively equivalent to the common simplifying assumption of a smooth genotype-phenotype map~\cite{lemos2024phylogeny}. 

At the strain level, the strength of direct interactions between strains $\alpha$ and $\beta$ of the same species $A_{i\alpha i\beta}$ are much stronger and differ by a smaller amount characterized by $\lambda$. These interactions are given by:
\begin{equation}
A_{i\alpha i\beta} = \tilde{\mu} + \lambda y_{i}, \quad A_{i\beta i\alpha} = \tilde{\mu} - \lambda y_{i}, \quad A_{i\alpha i\alpha} = 1,
\label{eq:Asetup2}
\end{equation}
where $\tilde{\mu}\sim \mathcal{O}(1)$ reflects the average direct interaction strength between closely-related strains; $y_i$ is a standard normal random variable that helps capture differences in the consumption preferences between both strains of species $i$, given by $2\lambda y_i$. When $\lambda = 0$, strains of the same species have identical interactions with the rest of the community. As $\lambda$ increases, they become increasingly distinct in how they interact. Since strains are more similar than species, we assume strain interaction dissimilarity $\lambda$ is less than species interaction dissimilarity $\sigma$.

We also assume that the intrinsic growth rates, $K_{i\alpha}$ and $K_{i\beta}$ of two closely-related strains $i\alpha$ and $i\beta$ are independent for simplicity. However as we explain later, relaxing this assumption does not qualitatively affect our results (Appendix~F). Together, these assumptions capture the hierarchical organization of community diversity, i.e., strains within species. While one can use more complex ecological models, e.g., consumer-resource models~\cite{goyal2018diversity,marsland2019available,goldford2018emergent}, given that such models often map to GLV models~\cite{goyal2025universal,advani2018statistical,letten2017linking}, we suspect that our results might also apply more broadly.

\vskip 10pt
\textbf{\textsf{Community-mediated feedbacks correlate strain growth rates and reduce competition.}} To understand how closely-related strains behave in our model, we use the cavity method from statistical physics (Fig.~\ref{fig:schematic}c, Appendix B). The key idea is to ``coarse-grain'' the surrounding community and ask how it shapes the dynamics of strains $\alpha$ and $\beta$ of a focal species. Our central finding is that high-diversity community dynamics reduce to an effective two-strain model at steady-state:
\begin{align}
0=\frac{dN_{\alpha}}{dt} &= N_{\alpha} \left( K_{\alpha}^{\mathrm{eff}} - A_{\alpha\alpha}^{\mathrm{eff}} N_{\alpha} - A_{\alpha\beta}^{\mathrm{eff}} N_{\beta} \right), \label{eq:effModel1}\\
0=\frac{dN_{\beta}}{dt} &= N_{\beta} \left( K_{\beta}^{\mathrm{eff}} - A_{\beta\beta}^{\mathrm{eff}} N_{\beta} - A_{\beta\alpha}^{\mathrm{eff}} N_{\alpha} \right).
\label{eq:effModel2}
\end{align}

Note that here we have dropped the species index $i$ to highlight that we are focusing on a typical strain pair of a given species. Crucially, the effective parameters $K^{\mathrm{eff}}$ and $A^{\mathrm{eff}}$ are not the bare parameters of the original model in Eq.~\eqref{eq:main_GLV_dynamics}. Instead, they are significantly modified by community-mediated feedbacks (Fig.~\ref{fig:schematic}c; Appendix C). Note that the growth rates and direct interactions between strains themselves do not change. Instead, interactions with the rest of the community result in strains behaving as if they were directly interacting with a different apparent strength, and growing with a different apparent rate.

Community feedbacks affect strains via two distinct mechanisms. The first mechanism, which we term ``emergent equalization'', modifies the effective or apparent growth rates of closely-related strains $K_{\alpha}^{\mathrm{eff}}$ and $K_{\beta}^{\mathrm{eff}}$. Because closely-related strains interact similarly with other species in the community, their effective growth rates become correlated and more equalized $\mathrm{corr}(K_\alpha^{\mathrm{eff}},K_\beta^{\mathrm{eff}})=\rho_K$ (Appendix B). In the absence of a community we assume that the growth rates of both strains are uncorrelated and different, $\mathrm{corr}(K_\alpha,K_\beta)=0$. However, this assumption is not central to our results: even if there is a correlation in isolation, community feedbacks will further equalize growth rates by increasing this correlation. 

The second mechanism, which we term ``emergent stabilization'', modifies the effective or apparent interaction strengths $A^{\mathrm{eff}}$. Community feedbacks reduce inter-strain competition to $\tilde{\mu}^{\mathrm{eff}} \approx \tilde{\mu} - 2\sigma^2 \nu$, where $\nu$ is a susceptibility that measures the community's average response to perturbations (Appendix C). This reduction in competition can be understood by realizing that in a community, strains interact not just directly with each other, but also indirectly through other community members. The reason that indirect interactions typically buffer competition, rather than intensify it, is because of an ``enemy of my enemy is my friend'' effect: a strain that inhibits another distantly-related taxon is likely to indirectly benefit its closely-related strain partner due to lowered inhibition by this taxon. The reduction factor $2\s^2\nu$ thus reflects the net sum of such effects through all taxa in the community. This expression shows that as communities become more heterogeneous ($\s$ increases) and diverse ($\nu$ increases), they can buffer strain interactions more effectively, resulting in more stabilized competition.
Combined, these equalizing and stabilizing forces represent the two ways in which community-mediated feedbacks reshape strain dynamics.

\begin{figure}[t]
\centering
\hspace{-10pt}
\includegraphics[width =1.03\columnwidth]{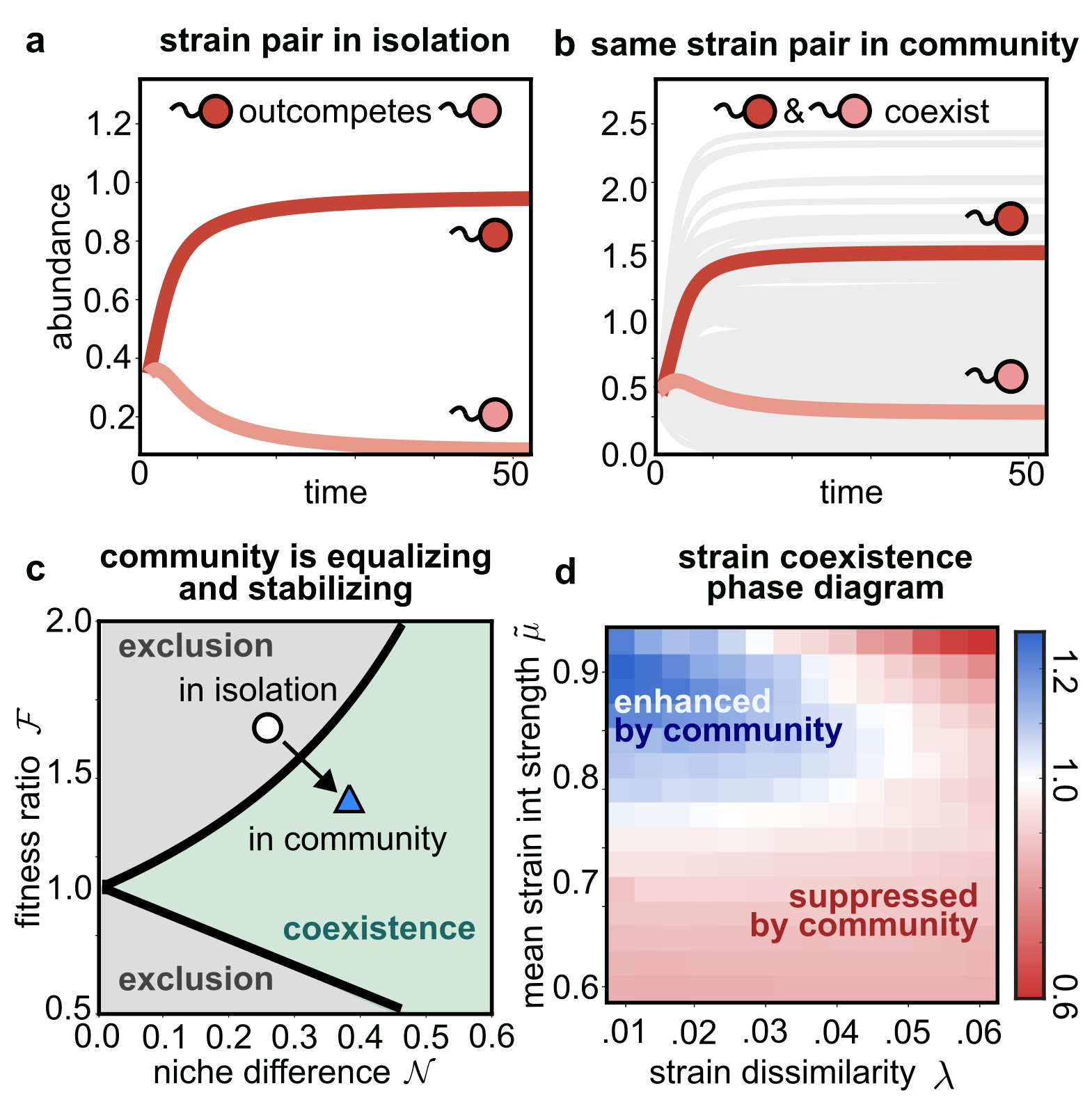} 
  \vspace{-10pt}
  \caption{\justifying\textsf{\textbf{Community feedbacks enhance strain coexistence.} (a) Two closely-related strains competing strongly in isolation: the strain with higher growth rate excludes the other. (b) The same two strains (red) embedded in a diverse community (gray) now stably coexist. (c) Explanation using modern coexistence theory due to niche differences $\mathcal{N}$ and fitness ratio $\mathcal{F}$. In isolation, strains in (a) sit in the exclusion region (circle). Community feedbacks serve as both an equalizing mechanism (vertical arrow) and stabilizing mechanism (horizontal arrow), together pushing the strains into the coexistence region (triangle). (d) Phase diagram of the ratio of coexistence probability in community versus isolation, with strain interaction dissimilarity $\lambda$ and direct interstrain interaction strength $\tilde{\mu}$. Communities enhance coexistence (blue) when strains are similar and compete strongly.}}
  \label{fig:coexistence}
\end{figure}

\vspace{10pt}
\textbf{\textsf{Strain coexistence can be enhanced by surrounding communities.}} We now explore the consequences of community-mediated feedbacks on strain coexistence. To illustrate, we show an example of two closely-related strains $\alpha$ and $\beta$ with intrinsic growth rates $K_\alpha$ and $K_\beta$. In isolation, these strains compete directly with interaction strength $\tilde{\mu}\approx 1$ (Methods). Because they compete strongly, their niches overlap significantly --- in isolation, the strain with higher growth rate will competitively exclude the other at steady state (Fig.~\ref{fig:coexistence}a). However, when we embed the same strain pair in a diverse community, we find that they now stably coexist (Fig.~\ref{fig:coexistence}b). 

The key to understanding this result lies in the community-mediated feedbacks we identified above. In the language of modern coexistence theory (MCT)~\cite{chesson2000, spaak2021mapping, adler2007niche, barabas2018chesson}, coexistence requires that stabilizing forces (niche differences) exceed equalizing forces (fitness ratio). For a two-strain GLV model, this criterion defines a coexistence region in the plane of fitness ratio $\mathcal{F}=(K_\alpha\sqrt{A_{\beta\alpha}A_{\beta\beta}})/(K_\beta\sqrt{A_{\alpha\beta}A_{\alpha\alpha}})$ versus niche difference $\mathcal{N}=1 - \sqrt{A_{\alpha\beta}A_{\beta\alpha}}/\sqrt{A_{\alpha\alpha}A_{\beta\beta}}$, measuring 1 minus the niche overlap due to strain competition (Fig.~\ref{fig:coexistence}c)~\cite{chesson2018updates}. Strains coexist when niche differences are large enough to overcome fitness ratio; otherwise, one strain excludes the other (Appendix D).

Without community context, closely-related strains typically sit in the exclusion region of this diagram (open circle in Fig.~\ref{fig:coexistence}c). Their low niche difference $\mathcal{N}$ places them outside the the coexistence boundary. 
Community feedbacks change this picture by moving strains along two axes simultaneously. Closely-related strains (small $\lambda$) have the same effective interactions; hence their effective fitness ratio is simply the ratio of their effective growth rates, $\mathcal{F}^{\mathrm{eff}} \approx K_{\alpha}^{\mathrm{eff}}/K_{\beta}^{\mathrm{eff}}$. Community feedbacks make these effective growth rates more correlated, pushing $\mathcal{F}^{\mathrm{eff}}$ toward 1---this serves as an equalizing mechanism that moves strains downward on the diagram (Fig.~\ref{fig:coexistence}c). 
Community feedbacks also reduce the effective strain interaction strengths, which increases their effective niche differences 
$\mathcal{N}^{\mathrm{eff}} = 1 - \sqrt{A_{\alpha\beta}^{\mathrm{eff}}A_{\beta\alpha}^{\mathrm{eff}}}/\sqrt{A_{\alpha\alpha}^{\mathrm{eff}}A_{\beta\beta}^{\mathrm{eff}}}$---this serves as a stabilizing mechanism that moves strains rightward (Appendix E). Together, the equalizing and stabilizing effects of the community work together to push strain pairs towards the coexistence region (filled triangle in Fig.~\ref{fig:coexistence}c).

Does community context always promote coexistence? Using extensive numerical simulations, we asked whether a random strain pair is more likely to coexist in a community than in isolation across a range of parameters (Methods). Fig.~\ref{fig:coexistence}d shows the relative ratio of coexistence probabilities (community versus isolation) as a function of strain interaction dissimilarity $\lambda$ and the mean direct interaction strength between strains $\tilde{\mu}$. 

Strikingly, when strains are very closely-related (small $\lambda$) and have strong direct interactions (large $\tilde{\mu}$), their coexistence is enhanced by a diverse community than in isolation (Fig.~\ref{fig:coexistence}d, blue). This can be understood by noting that the equalizing and stabilizing effects of a community are strongest \emph{only} when strains are very similar and interact strongly (Appendix E). 

As strains become more dissimilar (high $\lambda$), their effective fitness ratio $\mathcal{F}^{\mathrm{eff}}$ begins to increase (Appendix E). Moreover, the effective strain interactions become stronger and polarized. This significantly weakens both the equalizing and stabilizing effects of the community. A similar weakening of the community's equalizing effect occurs when strains have weak direct competition (low $\tilde{\mu}$). Indeed, as strains either get more dissimilar or have weaker interactions, community feedbacks begin to act in the opposite direction, even serving as ``anti-equalizing'' and ``anti-stabilizing'' mechanisms (Appendix E). As a result, dissimilar strains get pushed out of the coexistence region by the community, which now suppreses their coexistence (Fig.~\ref{fig:coexistence}d, red). Together, these results show that community-mediated feedbacks enhance coexistence the most for strains that compete strongest in isolation and are most closely-related.

\begin{figure}[t]
\centering
\hspace{-10pt}
\includegraphics[width=1.03\columnwidth]{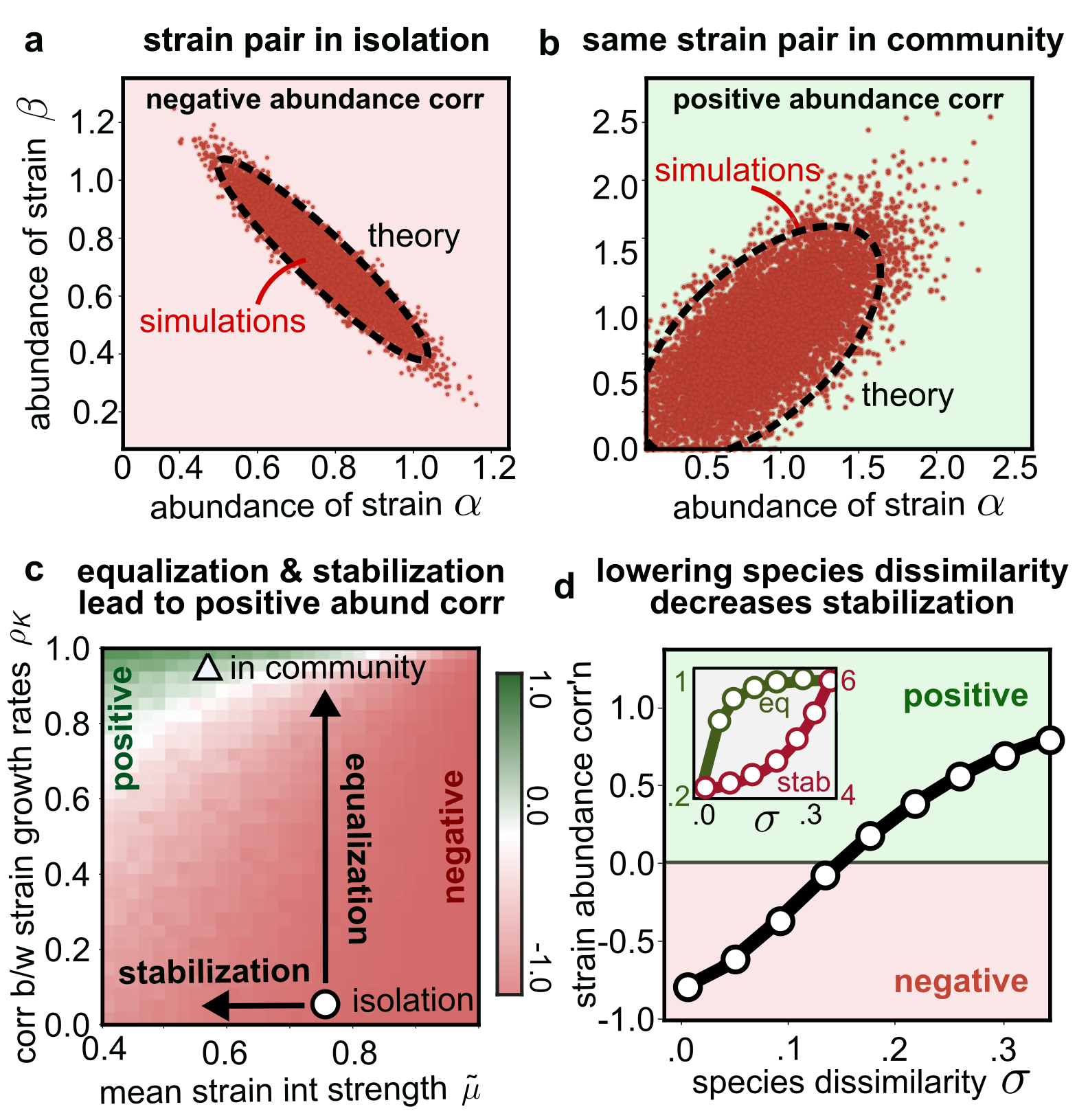} 
  \vspace{-10pt}
  \caption{\justifying\textsf{\textbf{Competing strains appear mutualistic due to community feedbacks.} (a) Two closely-related strains $\alpha$ and $\beta$ competing in isolation show a strong negative abundance correlation. Red ellipse shows 90\% confidence ellipse from theory. (b) The same strain pair embedded in a diverse community now shows a positive abundance correlation; ellipse shows cavity theory prediction. (c) Explanation using the effective two-strain model (Eqs.~\eqref{eq:effModel1}--\eqref{eq:effModel2}). Heatmap shows strain abundance correlation as a function of effective inter-strain interaction strength $\tilde{\mu}^{\text{eff}}$ and effective growth rate correlation $\rho_K$. Community feedbacks both stabilize (reduce $\tilde{\mu}^{\text{eff}}$) and equalize strains (increase $\rho_K$), together moving the strain pair from negative (circle) to positive (triangle) correlations. Positive correlations make strains appear mutualistic in a community despite having the strongest competition in isolation. (d) $\rho_{\alpha\beta}$ as a function of species dissimilarity $\sigma$, comparing simulations (dots) and cavity theory (curve). Inset: the strength of both equalization ($\rho_K$, green) and stabilization ($\langle 1/\mathrm{det}(A^{\mathrm{eff}})^2 \rangle$, magenta) increases with $\sigma$.}}
    \label{fig:dynamics}
\end{figure}

\vspace{10pt}

\textbf{\textsf{Competing strains appear mutualistic due to community feedbacks.}} Community-mediated feedbacks do not just transform strain coexistence; they can also fundamentally change the apparent nature of their interactions. In particular, they can change the sign of the correlation between abundances of closely-related strains. Such correlations have often been used to infer strain interactions from data~\cite{goyal_interactions_2022, faust2012microbial,faust2015metagenomics,tikhonov2015interpreting,freilich2011competitive,berry2014deciphering,barberan2012using,zhou2010functional}.

To illustrate this, we consider several pairs of closely-related strains $\alpha$ and $\beta$ which coexist in isolation. In this setting, their abundances are strongly negatively correlated across realizations: whichever strain has the higher intrinsic growth rate, $K_\a$ or $K_\beta$, tends to dominate over the other (Fig.~\ref{fig:dynamics}a). This is consistent with the interpretation that these strains compete strongly. However, when we embed the same strain pairs in a diverse community, this picture changes dramatically---the abundance correlations between closely-related strains reverses sign and becomes strongly positive (Fig.~\ref{fig:dynamics}b). Strains that competed strongly in isolation---and could even exclude each other---now appear mutualistic in a community context. Thus, community feedbacks can completely reverse the nature of the apparent interaction between closely-related strains.

The key to understanding this reversal lies in the community-mediated equalization and stabilization mechanisms we identified above. In the effective two-strain model (Eqs.~\eqref{eq:effModel1}--\eqref{eq:effModel2}), strain abundance correlations depend on both the effective inter-strain interaction strength $\tilde{\mu}^{\text{eff}}$ and the correlation between effective growth rates $\rho_K$. We can visualize this dependence by plotting the strain abundance correlation using simulations of a two-strain GLV model (Fig.~\ref{fig:dynamics}c). In isolation, the strain pair sits in the region of negative correlations (open circle in Fig.~\ref{fig:dynamics}c)---with strongly competitive direct interactions and no correlation between strain growth rates. Community feedbacks move the strain pair along two axes simultaneously: stabilization reduces the effective interaction strength $\tilde{\mu}^{\text{eff}}$ between strains, while equalization dramatically increases the correlation $\rho_K$ between their effective growth rates. Together, these forces push the strain pair into the region of positive correlations (filled triangle in Fig.~\ref{fig:dynamics}c).

How do these effects depend on the properties of the surrounding community? In Fig.~\ref{fig:dynamics}d, we show the strain abundance correlation as a function of species dissimilarity $\sigma$, which measures the heterogeneity of species interactions in the community. As $\sigma$ increases, the correlation increases and eventually becomes positive. Our cavity theory (solid curve) accurately captures this trend. The underlying reason for this increase in strain abundance correlations is that both community-mediated forces strengthen with increasing $\sigma$. Both the equalizing force $\rho_K$ and stabilizing force $\langle 1/\mathrm{det}(A^{\mathrm{eff}})^2 \rangle$ (Appendix~B) grow as communities become more heterogeneous (Fig.~\ref{fig:dynamics}d, inset). More heterogeneous communities thus exert stronger feedbacks on closely-related strains, making their effective growth rates more correlated and their effective competition weaker, which together drive the transition to positive abundance correlations. It is these positive abundance correlations that give the appearance that the strains are mutualistic, even though they compete strongly. Together, these results show that in a diverse community, the sign of strain abundance correlations rarely reflect the sign of direct interactions between strains. 

\vspace{10pt}
\textbf{\textsf{Strain dissimilarity governs strength of community feedbacks.}} 
Having established that community feedbacks can reverse the sign of strain abundance correlations, we now ask how the strength of these feedbacks depends on the phenotypic dissimilarity between strains. To build intuition, we return to the effective two-strain model in Eqs.~\eqref{eq:effModel1}--\eqref{eq:effModel2}. In this model, strain abundances are obtained by inverting the effective interaction matrix $A^{\text{eff}}$, so that $N = (A^{\text{eff}})^{-1} K^{\text{eff}}$. This means that two quantities govern strain abundance correlations. The first is the equalization strength, measured by $\rho_K = \text{corr}(K^{\text{eff}}_\alpha, K^{\text{eff}}_\beta)$, which captures how similarly the community shapes both strains' apparent growth rates. When equalization is strong, a community that promotes the growth of one strain also tends to promote its closely-related partner. The second is the stabilization strength, measured by $\langle 1/\mathrm{det}(A^{\mathrm{eff}})^2 \rangle$, which arises naturally from inverting the interaction matrix $A^{\text{eff}}$ and captures how sensitively strain abundances respond to growth rate differences. When community feedbacks reduce effective competition between strains, small differences in growth rates no longer translate into large differences in abundance---instead, correlated growth rates translate more directly into correlated abundances.

\begin{figure} [t]
    \centering
    \hspace{-10pt}
    \includegraphics[width=1.03\columnwidth]{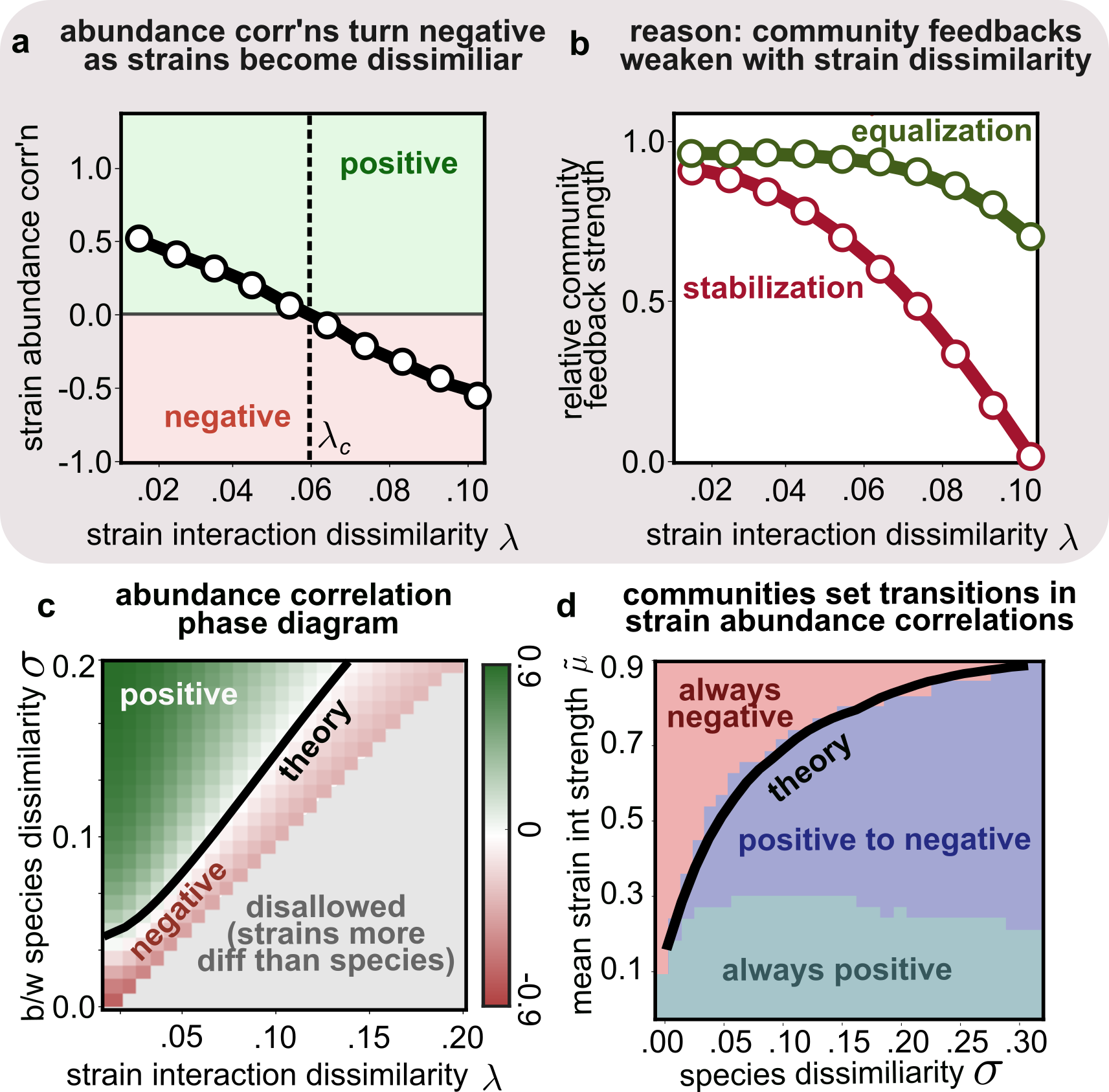}
    \caption{\justifying\textsf{\textbf{Strain dissimilarity governs strength of community feedbacks and strain abundance correlations.} (a) Strain abundance correlation versus strain interaction dissimilarity $\lambda$ for a community with fixed species dissimilarity $\sigma$. Correlations transition from positive to negative at a critical dissimilarity $\lambda_c$ (dotted line). (b) The reason: both equalization ($\rho_K$, green) and stabilization ($\langle 1/\mathrm{det}(A^{\mathrm{eff}})^2 \rangle$, magenta) weaken with increasing $\lambda$. Shown are values normalized to the maximum. When both these community feedbacks weaken, abundance correlations return to being negative, as they would in isolation. (c) Phase diagram of strain abundance correlations in $(\lambda, \sigma)$ space. The black line marks the theoretically predicted boundary from the cavity method separating positive and negative correlations. Gray region is disallowed since strains cannot be more dissimilar than species. (d) Regimes of strain abundance correlations in $(\sigma, \tilde{\mu})$ space: always negative (top), transitioning from positive to negative with increasing $\lambda$ (middle), and always positive (bottom). Black line shows the analytical boundary from the cavity method.}}
    \label{fig:mechanisms}
\end{figure}

Both equalization and stabilization rely on the fact that closely-related strains interact similarly with the rest of the community. As strains become more dissimilar (increasing $\lambda$), this breaks down: they begin to experience the community differently, weakening both forces (Fig.~\ref{fig:mechanisms}b). Consequently, strain abundance correlations decrease with increasing $\lambda$. Such abundance correlation patterns have been empirically observed in data from microbial communities~\cite{goyal_interactions_2022,garud2019,sireci2023environmental}. In this case, we observe that abundance correlations eventually become negative beyond a critical dissimilarity $\lambda_c$ (Fig.~\ref{fig:mechanisms}a). This occurs when community feedbacks become too weak to overcome direct competition, and once this happens, strains behave much as they would in isolation (Appendix~F). Thus, the critical dissimilarity $\lambda_c$ represents an emergent phenotypic scale below which strains appear mutualistic due to the community, and above which they behave as if they do not ``see'' the community at all.

What determines this scale? The answer is the species dissimilarity $\sigma$, which controls the overall strength of community feedbacks. Communities with more heterogeneous interactions (larger $\sigma$) exert stronger equalization and stabilization, pushing $\lambda_c$ to higher values and making positive correlations possible over a wider range of strain dissimilarities (Fig.~\ref{fig:mechanisms}c). The cavity method accurately predicts the boundary separating positive and negative correlations (black line). The mean direct interaction strength $\tilde{\mu}$ plays a complementary role: stronger direct competition between strains makes it harder for community feedbacks to overcome them. As explained before, these feedbacks are especially weak when species interactions in the community are very similar (low $\sigma$). Combined the effects of direct strain interactions and community-wide species dissimilarity can be summarized in the form of a final diagram, shown in Fig.~\ref{fig:mechanisms}d. This results in three distinct regimes in $(\sigma, \tilde{\mu})$ space---strain abundance correlations which remain always positive, always negative, and transitioning with increasing strain interaction dissimilarity $\lambda$ (Appendix~F). Thus, strain abundance correlations can exhibit a variety of behaviors depending on the balance between direct interactions and indirect community-mediated ones.

These results reinforce a central message of our study: community feedbacks are crucial for interpreting strain abundance correlations. Importantly, rather than attempting to measure the full interaction network, to understand strain coexistence and correlations in a highly-diverse community, it may be sufficient to characterize two simple community-mediated feedbacks: equalization and stabilization.

\section*{Discussion}

The coexistence of closely-related strains within microbial communities presents a major puzzle. Here, we attempt to understand the ecological drivers that shape strain-level diversity and dynamics. Our study reveals that strain dynamics cannot be understood in isolation---they are emergent consequences of the diverse communities in which strains are embedded. Combining community ecology and statistical physics, we show that community-mediated feedbacks reshape strain interactions through two distinct mechanisms: emergent equalization, which correlates the effective growth rates of conspecific strains, and emergent stabilization, which modifies their effective competitive interactions. Through these forces, community feedbacks qualitatively change the conditions for strain coexistence and the sign of their abundance correlations in ways that cannot be captured by studying strain pairs in isolation. 

The mechanisms of emergent equalization and stabilization connect naturally to modern coexistence theory, which attributes coexistence to stabilizing niche differences and equalizing fitness ratio~\cite{adler2007niche}. We find that community-mediated equalization and stabilization are strongest when strains are phenotypically most similar and compete most strongly in isolation. This counterintuitive result may explain why even strains which differ only slightly in their phenotype routinely coexist in diverse microbial communities. In this way, our work contributes to debates on the role of stabilizing and equalizing mechanisms in promoting the coexistence of phylogenetically similar (or different) species~\cite{hillerislambers2012rethinking,mayfield2010opposing}. Further, it also contributes a new strain-centered mechanism to a growing interest in ``emergent coexistence'', the ability of species pairs to coexist in larger communities, but not pairwise~\cite{aguade2025emergent, miller2025multispecies,chang2023emergent}.

Our findings also complement recent data analysis~\cite{sireci2023environmental,goyal_interactions_2022} which showed that species and strain abundance correlations decay with increasing phylogenetic distance, and attributed it to population-independent environmental equalization. Our framework reveals that, at least for strains, similar patterns can also arise in a totally population-dependent manner. Namely, internal community feedbacks can act like environmental equalization, without any changes in the external environment. This might have implications for other studies interpreting equalization and stabilization~\cite{bupu2025balance}. More broadly, our results show that abiotic and biotic factors affecting strains need not be independent; a community can simultaneously exert effects that appear like both. This highlights that one must infer ecological processes from strain abundance data with care.


Our model naturally involves some simplifying assumptions which can be relaxed in future work. First, we considered communities in which each species harbors exactly two strains. While this is somewhat consistent with experimental data~\cite{goyal_interactions_2022,good2017b,roodgar2019,zhao2019}, extending our work to allow more strains per species will allow us to generalize our results to more complex contexts. Second, our approach considers unstructured interactions and does not incorporate metabolic and spatial structure. Whether the presence of such structure would create deviations from our predictions~\cite{pearl2025structured} is important to understand in future work. Third, we focused on steady-state (equilibrium) communities, yet microbial communities often exhibit fluctuating abundance dynamics~\cite{grilli2020macroecological, sireci2023environmental,goyal_interactions_2022,roodgar2019}. Generalizing our results to these dynamical non-equilibrium scenarios also remains an avenue for future work. Finally, we consider community assembly from a pool already containing strains within species, but neglect the complex eco-evolutionary dynamics that result in such a pool. In the future, it would be valuable to understand the conditions in which such a strain structure can spontaneously emerge via evolution.

Our work also makes empirically verifiable predictions. The predicted transition from positive to negative correlations with increasing phenotypic dissimilarity could be tested using metagenomic data from communities where strains can be inferred from genomic variation alone. The dependence of this transition on merely the statistical properties of community interactions, i.e., species dissimilarity $\sigma$, enables interpreting strain interactions from abundance correlations once these interaction statistics can be measured. Our results offer quantitative guidance for correctly comparing strain dynamics across sites, individuals or environments in light of their surrounding communities. 

Finally, our work takes the first steps to systematically incorporate the hierarchical organization of biological diversity in ecological dynamics. While we focused at the level of strains within species, our work can be built upon to include a more complete hierarchy of diversity, from phyla to families to strains. Doing so will help concretely connect models of the same community at different resolutions, i.e., a coarse-grained model at the family level~\cite{lee2025functional,goldford2018emergent} with a fine-grained one at the strain level~\cite{goyal_interactions_2022}. We envision this leading to a unified theory of community dynamics at different levels of biological organization.


\vskip 10pt
\textsf{\textbf{Acknowledgments.}} We would like to thank P. Mehta, M. Sireci, S. Pollak, J. Grilli, M. Polz, M. Barbier, E. Blumenthal, G. Chure, J. McEnany, S.Y. Li, and Z. Feng for helpful discussions. This research was supported in part by the International Centre for Theoretical Sciences (ICTS) for participating in the program --- Winter School on Quantitative Systems Biology 2025 (Code: ICTS/Prog-qsb/2025/1). AG acknowledges support from the Ashok and Gita Vaish Junior Researcher Award, the ANRF Ramanujan Fellowship, as well the DAE, Govt. of India, under project no. RTI4001.

\bibliography{main}

\begin{thebibliography}{76}%
\makeatletter
\providecommand \@ifxundefined [1]{%
 \@ifx{#1\undefined}
}%
\providecommand \@ifnum [1]{%
 \ifnum #1\expandafter \@firstoftwo
 \else \expandafter \@secondoftwo
 \fi
}%
\providecommand \@ifx [1]{%
 \ifx #1\expandafter \@firstoftwo
 \else \expandafter \@secondoftwo
 \fi
}%
\providecommand \natexlab [1]{#1}%
\providecommand \enquote  [1]{``#1''}%
\providecommand \bibnamefont  [1]{#1}%
\providecommand \bibfnamefont [1]{#1}%
\providecommand \citenamefont [1]{#1}%
\providecommand \href@noop [0]{\@secondoftwo}%
\providecommand \href [0]{\begingroup \@sanitize@url \@href}%
\providecommand \@href[1]{\@@startlink{#1}\@@href}%
\providecommand \@@href[1]{\endgroup#1\@@endlink}%
\providecommand \@sanitize@url [0]{\catcode `\\12\catcode `\$12\catcode `\&12\catcode `\#12\catcode `\^12\catcode `\_12\catcode `\%12\relax}%
\providecommand \@@startlink[1]{}%
\providecommand \@@endlink[0]{}%
\providecommand \url  [0]{\begingroup\@sanitize@url \@url }%
\providecommand \@url [1]{\endgroup\@href {#1}{\urlprefix }}%
\providecommand \urlprefix  [0]{URL }%
\providecommand \Eprint [0]{\href }%
\providecommand \doibase [0]{https://doi.org/}%
\providecommand \selectlanguage [0]{\@gobble}%
\providecommand \bibinfo  [0]{\@secondoftwo}%
\providecommand \bibfield  [0]{\@secondoftwo}%
\providecommand \translation [1]{[#1]}%
\providecommand \BibitemOpen [0]{}%
\providecommand \bibitemStop [0]{}%
\providecommand \bibitemNoStop [0]{.\EOS\space}%
\providecommand \EOS [0]{\spacefactor3000\relax}%
\providecommand \BibitemShut  [1]{\csname bibitem#1\endcsname}%
\let\auto@bib@innerbib\@empty
\bibitem [{\citenamefont {Fierer}\ \emph {et~al.}(2007)\citenamefont {Fierer}, \citenamefont {Breitbart}, \citenamefont {Nulton}, \citenamefont {Salamon}, \citenamefont {Lozupone}, \citenamefont {Jones}, \citenamefont {Robeson}, \citenamefont {Edwards}, \citenamefont {Felts}, \citenamefont {Rayhawk} \emph {et~al.}}]{fierer2007metagenomic}%
  \BibitemOpen
  \bibfield  {author} {\bibinfo {author} {\bibfnamefont {N.}~\bibnamefont {Fierer}}, \bibinfo {author} {\bibfnamefont {M.}~\bibnamefont {Breitbart}}, \bibinfo {author} {\bibfnamefont {J.}~\bibnamefont {Nulton}}, \bibinfo {author} {\bibfnamefont {P.}~\bibnamefont {Salamon}}, \bibinfo {author} {\bibfnamefont {C.}~\bibnamefont {Lozupone}}, \bibinfo {author} {\bibfnamefont {R.}~\bibnamefont {Jones}}, \bibinfo {author} {\bibfnamefont {M.}~\bibnamefont {Robeson}}, \bibinfo {author} {\bibfnamefont {R.~A.}\ \bibnamefont {Edwards}}, \bibinfo {author} {\bibfnamefont {B.}~\bibnamefont {Felts}}, \bibinfo {author} {\bibfnamefont {S.}~\bibnamefont {Rayhawk}}, \emph {et~al.},\ }\bibfield  {title} {\bibinfo {title} {Metagenomic and small-subunit rrna analyses reveal the genetic diversity of bacteria, archaea, fungi, and viruses in soil},\ }\href@noop {} {\bibfield  {journal} {\bibinfo  {journal} {Applied and environmental microbiology}\ }\textbf {\bibinfo {volume} {73}},\ \bibinfo {pages} {7059} (\bibinfo {year}
  {2007})}\BibitemShut {NoStop}%
\bibitem [{\citenamefont {Shu}\ and\ \citenamefont {Huang}(2022)}]{shu2022}%
  \BibitemOpen
  \bibfield  {author} {\bibinfo {author} {\bibfnamefont {W.-S.}\ \bibnamefont {Shu}}\ and\ \bibinfo {author} {\bibfnamefont {L.-N.}\ \bibnamefont {Huang}},\ }\bibfield  {title} {\bibinfo {title} {Microbial diversity in extreme environments},\ }\href {https://doi.org/10.1038/s41579-021-00648-y} {\bibfield  {journal} {\bibinfo  {journal} {Nature Reviews Microbiology}\ }\textbf {\bibinfo {volume} {20}},\ \bibinfo {pages} {219} (\bibinfo {year} {2022})},\ \bibinfo {note} {publisher: Nature Publishing Group}\BibitemShut {NoStop}%
\bibitem [{\citenamefont {Rosen}\ \emph {et~al.}(2015)\citenamefont {Rosen}, \citenamefont {Davison}, \citenamefont {Bhaya},\ and\ \citenamefont {Fisher}}]{rosen2015fine}%
  \BibitemOpen
  \bibfield  {author} {\bibinfo {author} {\bibfnamefont {M.~J.}\ \bibnamefont {Rosen}}, \bibinfo {author} {\bibfnamefont {M.}~\bibnamefont {Davison}}, \bibinfo {author} {\bibfnamefont {D.}~\bibnamefont {Bhaya}},\ and\ \bibinfo {author} {\bibfnamefont {D.~S.}\ \bibnamefont {Fisher}},\ }\bibfield  {title} {\bibinfo {title} {Fine-scale diversity and extensive recombination in a quasisexual bacterial population occupying a broad niche},\ }\href@noop {} {\bibfield  {journal} {\bibinfo  {journal} {Science}\ }\textbf {\bibinfo {volume} {348}},\ \bibinfo {pages} {1019} (\bibinfo {year} {2015})}\BibitemShut {NoStop}%
\bibitem [{\citenamefont {Morgan}\ \emph {et~al.}(2013)\citenamefont {Morgan}, \citenamefont {Segata},\ and\ \citenamefont {Huttenhower}}]{morgan2013biodiversity}%
  \BibitemOpen
  \bibfield  {author} {\bibinfo {author} {\bibfnamefont {X.~C.}\ \bibnamefont {Morgan}}, \bibinfo {author} {\bibfnamefont {N.}~\bibnamefont {Segata}},\ and\ \bibinfo {author} {\bibfnamefont {C.}~\bibnamefont {Huttenhower}},\ }\bibfield  {title} {\bibinfo {title} {Biodiversity and functional genomics in the human microbiome},\ }\href@noop {} {\bibfield  {journal} {\bibinfo  {journal} {Trends in genetics}\ }\textbf {\bibinfo {volume} {29}},\ \bibinfo {pages} {51} (\bibinfo {year} {2013})}\BibitemShut {NoStop}%
\bibitem [{\citenamefont {Ding}\ \emph {et~al.}(2016)\citenamefont {Ding}, \citenamefont {Fu}, \citenamefont {Ding}, \citenamefont {Lu}, \citenamefont {Cheng},\ and\ \citenamefont {Zeng}}]{ding2016environmental}%
  \BibitemOpen
  \bibfield  {author} {\bibinfo {author} {\bibfnamefont {J.}~\bibnamefont {Ding}}, \bibinfo {author} {\bibfnamefont {L.}~\bibnamefont {Fu}}, \bibinfo {author} {\bibfnamefont {Z.-W.}\ \bibnamefont {Ding}}, \bibinfo {author} {\bibfnamefont {Y.-Z.}\ \bibnamefont {Lu}}, \bibinfo {author} {\bibfnamefont {S.~H.}\ \bibnamefont {Cheng}},\ and\ \bibinfo {author} {\bibfnamefont {R.~J.}\ \bibnamefont {Zeng}},\ }\bibfield  {title} {\bibinfo {title} {Environmental evaluation of coexistence of denitrifying anaerobic methane-oxidizing archaea and bacteria in a paddy field},\ }\href@noop {} {\bibfield  {journal} {\bibinfo  {journal} {Applied microbiology and biotechnology}\ }\textbf {\bibinfo {volume} {100}},\ \bibinfo {pages} {439} (\bibinfo {year} {2016})}\BibitemShut {NoStop}%
\bibitem [{\citenamefont {Zhao}\ \emph {et~al.}(2019)\citenamefont {Zhao}, \citenamefont {Lieberman}, \citenamefont {Poyet}, \citenamefont {Kauffman}, \citenamefont {Gibbons}, \citenamefont {Groussin}, \citenamefont {Xavier},\ and\ \citenamefont {Alm}}]{zhao2019}%
  \BibitemOpen
  \bibfield  {author} {\bibinfo {author} {\bibfnamefont {S.}~\bibnamefont {Zhao}}, \bibinfo {author} {\bibfnamefont {T.~D.}\ \bibnamefont {Lieberman}}, \bibinfo {author} {\bibfnamefont {M.}~\bibnamefont {Poyet}}, \bibinfo {author} {\bibfnamefont {K.~M.}\ \bibnamefont {Kauffman}}, \bibinfo {author} {\bibfnamefont {S.~M.}\ \bibnamefont {Gibbons}}, \bibinfo {author} {\bibfnamefont {M.}~\bibnamefont {Groussin}}, \bibinfo {author} {\bibfnamefont {R.~J.}\ \bibnamefont {Xavier}},\ and\ \bibinfo {author} {\bibfnamefont {E.~J.}\ \bibnamefont {Alm}},\ }\bibfield  {title} {\bibinfo {title} {Adaptive evolution within gut microbiomes of healthy people},\ }\href {https://doi.org/10.1016/j.chom.2019.03.007} {\bibfield  {journal} {\bibinfo  {journal} {Cell Host \& Microbe}\ }\textbf {\bibinfo {volume} {25}},\ \bibinfo {pages} {656} (\bibinfo {year} {2019})},\ \bibinfo {note} {publisher: Elsevier}\BibitemShut {NoStop}%
\bibitem [{\citenamefont {Garud}\ \emph {et~al.}(2019)\citenamefont {Garud}, \citenamefont {Good}, \citenamefont {Hallatschek},\ and\ \citenamefont {Pollard}}]{garud2019}%
  \BibitemOpen
  \bibfield  {author} {\bibinfo {author} {\bibfnamefont {N.~R.}\ \bibnamefont {Garud}}, \bibinfo {author} {\bibfnamefont {B.~H.}\ \bibnamefont {Good}}, \bibinfo {author} {\bibfnamefont {O.}~\bibnamefont {Hallatschek}},\ and\ \bibinfo {author} {\bibfnamefont {K.~S.}\ \bibnamefont {Pollard}},\ }\bibfield  {title} {\bibinfo {title} {Evolutionary dynamics of bacteria in the gut microbiome within and across hosts},\ }\href {https://doi.org/10.1371/journal.pbio.3000102} {\bibfield  {journal} {\bibinfo  {journal} {PLOS Biology}\ }\textbf {\bibinfo {volume} {17}},\ \bibinfo {pages} {e3000102} (\bibinfo {year} {2019})},\ \bibinfo {note} {publisher: Public Library of Science}\BibitemShut {NoStop}%
\bibitem [{\citenamefont {Roodgar}\ \emph {et~al.}(2019)\citenamefont {Roodgar}, \citenamefont {Good}, \citenamefont {Garud}, \citenamefont {Martis}, \citenamefont {Avula}, \citenamefont {Zhou}, \citenamefont {Lancaster}, \citenamefont {Lee}, \citenamefont {Babveyh}, \citenamefont {Nesamoney}, \citenamefont {Pollard},\ and\ \citenamefont {Snyder}}]{roodgar2019}%
  \BibitemOpen
  \bibfield  {author} {\bibinfo {author} {\bibfnamefont {M.}~\bibnamefont {Roodgar}}, \bibinfo {author} {\bibfnamefont {B.~H.}\ \bibnamefont {Good}}, \bibinfo {author} {\bibfnamefont {N.~R.}\ \bibnamefont {Garud}}, \bibinfo {author} {\bibfnamefont {S.}~\bibnamefont {Martis}}, \bibinfo {author} {\bibfnamefont {M.}~\bibnamefont {Avula}}, \bibinfo {author} {\bibfnamefont {W.}~\bibnamefont {Zhou}}, \bibinfo {author} {\bibfnamefont {S.}~\bibnamefont {Lancaster}}, \bibinfo {author} {\bibfnamefont {H.}~\bibnamefont {Lee}}, \bibinfo {author} {\bibfnamefont {A.}~\bibnamefont {Babveyh}}, \bibinfo {author} {\bibfnamefont {S.}~\bibnamefont {Nesamoney}}, \bibinfo {author} {\bibfnamefont {K.~S.}\ \bibnamefont {Pollard}},\ and\ \bibinfo {author} {\bibfnamefont {M.~P.}\ \bibnamefont {Snyder}},\ }\bibfield  {title} {\bibinfo {title} {Longitudinal linked read sequencing reveals ecological and evolutionary responses of a human gut microbiome during antibiotic treatment},\ }\href {https://doi.org/10.1101/2019.12.21.886093}
  {\bibfield  {journal} {\bibinfo  {journal} {bioRxiv}\ ,\ \bibinfo {pages} {2019.12.21.886093}} (\bibinfo {year} {2019})},\ \bibinfo {note} {publisher: Cold Spring Harbor Laboratory Section: New Results}\BibitemShut {NoStop}%
\bibitem [{\citenamefont {Truong}\ \emph {et~al.}(2017)\citenamefont {Truong}, \citenamefont {Tett}, \citenamefont {Pasolli}, \citenamefont {Huttenhower},\ and\ \citenamefont {Segata}}]{truong2017}%
  \BibitemOpen
  \bibfield  {author} {\bibinfo {author} {\bibfnamefont {D.~T.}\ \bibnamefont {Truong}}, \bibinfo {author} {\bibfnamefont {A.}~\bibnamefont {Tett}}, \bibinfo {author} {\bibfnamefont {E.}~\bibnamefont {Pasolli}}, \bibinfo {author} {\bibfnamefont {C.}~\bibnamefont {Huttenhower}},\ and\ \bibinfo {author} {\bibfnamefont {N.}~\bibnamefont {Segata}},\ }\bibfield  {title} {\bibinfo {title} {Microbial strain-level population structure and genetic diversity from metagenomes},\ }\href {https://doi.org/10.1101/gr.216242.116} {\bibfield  {journal} {\bibinfo  {journal} {Genome Research}\ }\textbf {\bibinfo {volume} {27}},\ \bibinfo {pages} {626} (\bibinfo {year} {2017})}\BibitemShut {NoStop}%
\bibitem [{\citenamefont {Good}\ \emph {et~al.}(2017)\citenamefont {Good}, \citenamefont {McDonald}, \citenamefont {Barrick}, \citenamefont {Lenski},\ and\ \citenamefont {Desai}}]{good2017b}%
  \BibitemOpen
  \bibfield  {author} {\bibinfo {author} {\bibfnamefont {B.~H.}\ \bibnamefont {Good}}, \bibinfo {author} {\bibfnamefont {M.~J.}\ \bibnamefont {McDonald}}, \bibinfo {author} {\bibfnamefont {J.~E.}\ \bibnamefont {Barrick}}, \bibinfo {author} {\bibfnamefont {R.~E.}\ \bibnamefont {Lenski}},\ and\ \bibinfo {author} {\bibfnamefont {M.~M.}\ \bibnamefont {Desai}},\ }\bibfield  {title} {\bibinfo {title} {The dynamics of molecular evolution over 60,000 generations},\ }\href {https://doi.org/10.1038/nature24287} {\bibfield  {journal} {\bibinfo  {journal} {Nature}\ }\textbf {\bibinfo {volume} {551}},\ \bibinfo {pages} {45} (\bibinfo {year} {2017})},\ \bibinfo {note} {number: 7678 Publisher: Nature Publishing Group}\BibitemShut {NoStop}%
\bibitem [{\citenamefont {Goyal}\ \emph {et~al.}(2022)\citenamefont {Goyal}, \citenamefont {Bittleston}, \citenamefont {Leventhal}, \citenamefont {Lu},\ and\ \citenamefont {Cordero}}]{goyal_interactions_2022}%
  \BibitemOpen
  \bibfield  {author} {\bibinfo {author} {\bibfnamefont {A.}~\bibnamefont {Goyal}}, \bibinfo {author} {\bibfnamefont {L.~S.}\ \bibnamefont {Bittleston}}, \bibinfo {author} {\bibfnamefont {G.~E.}\ \bibnamefont {Leventhal}}, \bibinfo {author} {\bibfnamefont {L.}~\bibnamefont {Lu}},\ and\ \bibinfo {author} {\bibfnamefont {O.~X.}\ \bibnamefont {Cordero}},\ }\bibfield  {title} {\bibinfo {title} {Interactions between strains govern the eco-evolutionary dynamics of microbial communities},\ }\href {https://doi.org/10.7554/eLife.74987} {\bibfield  {journal} {\bibinfo  {journal} {eLife}\ }\textbf {\bibinfo {volume} {11}},\ \bibinfo {pages} {e74987} (\bibinfo {year} {2022})},\ \bibinfo {note} {publisher: eLife Sciences Publications, Ltd}\BibitemShut {NoStop}%
\bibitem [{\citenamefont {Poyet}\ \emph {et~al.}(2019)\citenamefont {Poyet}, \citenamefont {Groussin}, \citenamefont {Gibbons}, \citenamefont {Avila-Pacheco}, \citenamefont {Jiang}, \citenamefont {Kearney}, \citenamefont {Perrotta}, \citenamefont {Berdy}, \citenamefont {Zhao}, \citenamefont {Lieberman} \emph {et~al.}}]{poyet2019library}%
  \BibitemOpen
  \bibfield  {author} {\bibinfo {author} {\bibfnamefont {M.}~\bibnamefont {Poyet}}, \bibinfo {author} {\bibfnamefont {M.}~\bibnamefont {Groussin}}, \bibinfo {author} {\bibfnamefont {S.~M.}\ \bibnamefont {Gibbons}}, \bibinfo {author} {\bibfnamefont {J.}~\bibnamefont {Avila-Pacheco}}, \bibinfo {author} {\bibfnamefont {X.}~\bibnamefont {Jiang}}, \bibinfo {author} {\bibfnamefont {S.~M.}\ \bibnamefont {Kearney}}, \bibinfo {author} {\bibfnamefont {A.~R.}\ \bibnamefont {Perrotta}}, \bibinfo {author} {\bibfnamefont {B.}~\bibnamefont {Berdy}}, \bibinfo {author} {\bibfnamefont {S.}~\bibnamefont {Zhao}}, \bibinfo {author} {\bibfnamefont {T.}~\bibnamefont {Lieberman}}, \emph {et~al.},\ }\bibfield  {title} {\bibinfo {title} {A library of human gut bacterial isolates paired with longitudinal multiomics data enables mechanistic microbiome research},\ }\href@noop {} {\bibfield  {journal} {\bibinfo  {journal} {Nature medicine}\ }\textbf {\bibinfo {volume} {25}},\ \bibinfo {pages} {1442} (\bibinfo {year} {2019})}\BibitemShut
  {NoStop}%
\bibitem [{\citenamefont {Leventhal}\ \emph {et~al.}(2018)\citenamefont {Leventhal}, \citenamefont {Boix}, \citenamefont {Kuechler}, \citenamefont {Enke}, \citenamefont {Sliwerska}, \citenamefont {Holliger},\ and\ \citenamefont {Cordero}}]{leventhal2018}%
  \BibitemOpen
  \bibfield  {author} {\bibinfo {author} {\bibfnamefont {G.~E.}\ \bibnamefont {Leventhal}}, \bibinfo {author} {\bibfnamefont {C.}~\bibnamefont {Boix}}, \bibinfo {author} {\bibfnamefont {U.}~\bibnamefont {Kuechler}}, \bibinfo {author} {\bibfnamefont {T.~N.}\ \bibnamefont {Enke}}, \bibinfo {author} {\bibfnamefont {E.}~\bibnamefont {Sliwerska}}, \bibinfo {author} {\bibfnamefont {C.}~\bibnamefont {Holliger}},\ and\ \bibinfo {author} {\bibfnamefont {O.~X.}\ \bibnamefont {Cordero}},\ }\bibfield  {title} {\bibinfo {title} {Strain-level diversity drives alternative community types in millimetre-scale granular biofilms},\ }\href {https://doi.org/10.1038/s41564-018-0242-3} {\bibfield  {journal} {\bibinfo  {journal} {Nature Microbiology}\ }\textbf {\bibinfo {volume} {3}},\ \bibinfo {pages} {1295} (\bibinfo {year} {2018})},\ \bibinfo {note} {number: 11 Publisher: Nature Publishing Group}\BibitemShut {NoStop}%
\bibitem [{\citenamefont {Van~Rossum}\ \emph {et~al.}(2020{\natexlab{a}})\citenamefont {Van~Rossum}, \citenamefont {Ferretti}, \citenamefont {Maistrenko},\ and\ \citenamefont {Bork}}]{van2020diversity}%
  \BibitemOpen
  \bibfield  {author} {\bibinfo {author} {\bibfnamefont {T.}~\bibnamefont {Van~Rossum}}, \bibinfo {author} {\bibfnamefont {P.}~\bibnamefont {Ferretti}}, \bibinfo {author} {\bibfnamefont {O.~M.}\ \bibnamefont {Maistrenko}},\ and\ \bibinfo {author} {\bibfnamefont {P.}~\bibnamefont {Bork}},\ }\bibfield  {title} {\bibinfo {title} {Diversity within species: interpreting strains in microbiomes},\ }\href@noop {} {\bibfield  {journal} {\bibinfo  {journal} {Nature Reviews Microbiology}\ }\textbf {\bibinfo {volume} {18}},\ \bibinfo {pages} {491} (\bibinfo {year} {2020}{\natexlab{a}})}\BibitemShut {NoStop}%
\bibitem [{\citenamefont {Zhang}\ and\ \citenamefont {Zhao}(2016)}]{zhang2016strain}%
  \BibitemOpen
  \bibfield  {author} {\bibinfo {author} {\bibfnamefont {C.}~\bibnamefont {Zhang}}\ and\ \bibinfo {author} {\bibfnamefont {L.}~\bibnamefont {Zhao}},\ }\bibfield  {title} {\bibinfo {title} {Strain-level dissection of the contribution of the gut microbiome to human metabolic disease},\ }\href@noop {} {\bibfield  {journal} {\bibinfo  {journal} {Genome medicine}\ }\textbf {\bibinfo {volume} {8}},\ \bibinfo {pages} {41} (\bibinfo {year} {2016})}\BibitemShut {NoStop}%
\bibitem [{\citenamefont {Yan}\ \emph {et~al.}(2020)\citenamefont {Yan}, \citenamefont {Nguyen}, \citenamefont {Franzosa},\ and\ \citenamefont {Huttenhower}}]{yan2020strain}%
  \BibitemOpen
  \bibfield  {author} {\bibinfo {author} {\bibfnamefont {Y.}~\bibnamefont {Yan}}, \bibinfo {author} {\bibfnamefont {L.~H.}\ \bibnamefont {Nguyen}}, \bibinfo {author} {\bibfnamefont {E.~A.}\ \bibnamefont {Franzosa}},\ and\ \bibinfo {author} {\bibfnamefont {C.}~\bibnamefont {Huttenhower}},\ }\bibfield  {title} {\bibinfo {title} {Strain-level epidemiology of microbial communities and the human microbiome},\ }\href@noop {} {\bibfield  {journal} {\bibinfo  {journal} {Genome medicine}\ }\textbf {\bibinfo {volume} {12}},\ \bibinfo {pages} {71} (\bibinfo {year} {2020})}\BibitemShut {NoStop}%
\bibitem [{\citenamefont {Vatanen}\ \emph {et~al.}(2019)\citenamefont {Vatanen}, \citenamefont {Plichta}, \citenamefont {Somani}, \citenamefont {M{\"u}nch}, \citenamefont {Arthur}, \citenamefont {Hall}, \citenamefont {Rudolf}, \citenamefont {Oakeley}, \citenamefont {Ke}, \citenamefont {Young} \emph {et~al.}}]{vatanen2019genomic}%
  \BibitemOpen
  \bibfield  {author} {\bibinfo {author} {\bibfnamefont {T.}~\bibnamefont {Vatanen}}, \bibinfo {author} {\bibfnamefont {D.~R.}\ \bibnamefont {Plichta}}, \bibinfo {author} {\bibfnamefont {J.}~\bibnamefont {Somani}}, \bibinfo {author} {\bibfnamefont {P.~C.}\ \bibnamefont {M{\"u}nch}}, \bibinfo {author} {\bibfnamefont {T.~D.}\ \bibnamefont {Arthur}}, \bibinfo {author} {\bibfnamefont {A.~B.}\ \bibnamefont {Hall}}, \bibinfo {author} {\bibfnamefont {S.}~\bibnamefont {Rudolf}}, \bibinfo {author} {\bibfnamefont {E.~J.}\ \bibnamefont {Oakeley}}, \bibinfo {author} {\bibfnamefont {X.}~\bibnamefont {Ke}}, \bibinfo {author} {\bibfnamefont {R.~A.}\ \bibnamefont {Young}}, \emph {et~al.},\ }\bibfield  {title} {\bibinfo {title} {Genomic variation and strain-specific functional adaptation in the human gut microbiome during early life},\ }\href@noop {} {\bibfield  {journal} {\bibinfo  {journal} {Nature microbiology}\ }\textbf {\bibinfo {volume} {4}},\ \bibinfo {pages} {470} (\bibinfo {year} {2019})}\BibitemShut {NoStop}%
\bibitem [{\citenamefont {Kalan}\ \emph {et~al.}(2019)\citenamefont {Kalan}, \citenamefont {Meisel}, \citenamefont {Loesche}, \citenamefont {Horwinski}, \citenamefont {Soaita}, \citenamefont {Chen}, \citenamefont {Uberoi}, \citenamefont {Gardner},\ and\ \citenamefont {Grice}}]{kalan2019strain}%
  \BibitemOpen
  \bibfield  {author} {\bibinfo {author} {\bibfnamefont {L.~R.}\ \bibnamefont {Kalan}}, \bibinfo {author} {\bibfnamefont {J.~S.}\ \bibnamefont {Meisel}}, \bibinfo {author} {\bibfnamefont {M.~A.}\ \bibnamefont {Loesche}}, \bibinfo {author} {\bibfnamefont {J.}~\bibnamefont {Horwinski}}, \bibinfo {author} {\bibfnamefont {I.}~\bibnamefont {Soaita}}, \bibinfo {author} {\bibfnamefont {X.}~\bibnamefont {Chen}}, \bibinfo {author} {\bibfnamefont {A.}~\bibnamefont {Uberoi}}, \bibinfo {author} {\bibfnamefont {S.~E.}\ \bibnamefont {Gardner}},\ and\ \bibinfo {author} {\bibfnamefont {E.~A.}\ \bibnamefont {Grice}},\ }\bibfield  {title} {\bibinfo {title} {Strain-and species-level variation in the microbiome of diabetic wounds is associated with clinical outcomes and therapeutic efficacy},\ }\href@noop {} {\bibfield  {journal} {\bibinfo  {journal} {Cell host \& microbe}\ }\textbf {\bibinfo {volume} {25}},\ \bibinfo {pages} {641} (\bibinfo {year} {2019})}\BibitemShut {NoStop}%
\bibitem [{\citenamefont {Park}\ \emph {et~al.}(2022)\citenamefont {Park}, \citenamefont {Rao}, \citenamefont {Coyte}, \citenamefont {Kuziel}, \citenamefont {Zhang}, \citenamefont {Huang}, \citenamefont {Franzosa}, \citenamefont {Weng}, \citenamefont {Huttenhower},\ and\ \citenamefont {Rakoff-Nahoum}}]{park2022strain}%
  \BibitemOpen
  \bibfield  {author} {\bibinfo {author} {\bibfnamefont {S.-Y.}\ \bibnamefont {Park}}, \bibinfo {author} {\bibfnamefont {C.}~\bibnamefont {Rao}}, \bibinfo {author} {\bibfnamefont {K.~Z.}\ \bibnamefont {Coyte}}, \bibinfo {author} {\bibfnamefont {G.~A.}\ \bibnamefont {Kuziel}}, \bibinfo {author} {\bibfnamefont {Y.}~\bibnamefont {Zhang}}, \bibinfo {author} {\bibfnamefont {W.}~\bibnamefont {Huang}}, \bibinfo {author} {\bibfnamefont {E.~A.}\ \bibnamefont {Franzosa}}, \bibinfo {author} {\bibfnamefont {J.-K.}\ \bibnamefont {Weng}}, \bibinfo {author} {\bibfnamefont {C.}~\bibnamefont {Huttenhower}},\ and\ \bibinfo {author} {\bibfnamefont {S.}~\bibnamefont {Rakoff-Nahoum}},\ }\bibfield  {title} {\bibinfo {title} {Strain-level fitness in the gut microbiome is an emergent property of glycans and a single metabolite},\ }\href@noop {} {\bibfield  {journal} {\bibinfo  {journal} {Cell}\ }\textbf {\bibinfo {volume} {185}},\ \bibinfo {pages} {513} (\bibinfo {year} {2022})}\BibitemShut {NoStop}%
\bibitem [{\citenamefont {Brito}\ and\ \citenamefont {Alm}(2016)}]{brito2016tracking}%
  \BibitemOpen
  \bibfield  {author} {\bibinfo {author} {\bibfnamefont {I.~L.}\ \bibnamefont {Brito}}\ and\ \bibinfo {author} {\bibfnamefont {E.~J.}\ \bibnamefont {Alm}},\ }\bibfield  {title} {\bibinfo {title} {Tracking strains in the microbiome: insights from metagenomics and models},\ }\href@noop {} {\bibfield  {journal} {\bibinfo  {journal} {Frontiers in Microbiology}\ }\textbf {\bibinfo {volume} {7}},\ \bibinfo {pages} {712} (\bibinfo {year} {2016})}\BibitemShut {NoStop}%
\bibitem [{\citenamefont {Hutchinson}(1961)}]{hutchinson1961}%
  \BibitemOpen
  \bibfield  {author} {\bibinfo {author} {\bibfnamefont {G.~E.}\ \bibnamefont {Hutchinson}},\ }\bibfield  {title} {\bibinfo {title} {The {Paradox} of the {Plankton}},\ }\href {https://doi.org/10.1086/282171} {\bibfield  {journal} {\bibinfo  {journal} {The American Naturalist}\ }\textbf {\bibinfo {volume} {95}},\ \bibinfo {pages} {137} (\bibinfo {year} {1961})},\ \bibinfo {note} {publisher: The University of Chicago Press}\BibitemShut {NoStop}%
\bibitem [{\citenamefont {MacArthur}\ and\ \citenamefont {Levins}(1967)}]{macarthur1967limiting}%
  \BibitemOpen
  \bibfield  {author} {\bibinfo {author} {\bibfnamefont {R.}~\bibnamefont {MacArthur}}\ and\ \bibinfo {author} {\bibfnamefont {R.}~\bibnamefont {Levins}},\ }\bibfield  {title} {\bibinfo {title} {The limiting similarity, convergence, and divergence of coexisting species},\ }\href@noop {} {\bibfield  {journal} {\bibinfo  {journal} {The american naturalist}\ }\textbf {\bibinfo {volume} {101}},\ \bibinfo {pages} {377} (\bibinfo {year} {1967})}\BibitemShut {NoStop}%
\bibitem [{\citenamefont {Chesson}(2000{\natexlab{a}})}]{chesson2000mechanisms}%
  \BibitemOpen
  \bibfield  {author} {\bibinfo {author} {\bibfnamefont {P.}~\bibnamefont {Chesson}},\ }\bibfield  {title} {\bibinfo {title} {Mechanisms of maintenance of species diversity},\ }\href@noop {} {\bibfield  {journal} {\bibinfo  {journal} {Annual review of Ecology and Systematics}\ }\textbf {\bibinfo {volume} {31}},\ \bibinfo {pages} {343} (\bibinfo {year} {2000}{\natexlab{a}})}\BibitemShut {NoStop}%
\bibitem [{\citenamefont {Chase}\ and\ \citenamefont {Leibold}(2009)}]{chase2009ecological}%
  \BibitemOpen
  \bibfield  {author} {\bibinfo {author} {\bibfnamefont {J.~M.}\ \bibnamefont {Chase}}\ and\ \bibinfo {author} {\bibfnamefont {M.~A.}\ \bibnamefont {Leibold}},\ }\href@noop {} {\emph {\bibinfo {title} {Ecological niches: linking classical and contemporary approaches}}}\ (\bibinfo  {publisher} {University of Chicago Press},\ \bibinfo {year} {2009})\BibitemShut {NoStop}%
\bibitem [{\citenamefont {Adler}\ \emph {et~al.}(2007)\citenamefont {Adler}, \citenamefont {HilleRisLambers},\ and\ \citenamefont {Levine}}]{adler2007niche}%
  \BibitemOpen
  \bibfield  {author} {\bibinfo {author} {\bibfnamefont {P.~B.}\ \bibnamefont {Adler}}, \bibinfo {author} {\bibfnamefont {J.}~\bibnamefont {HilleRisLambers}},\ and\ \bibinfo {author} {\bibfnamefont {J.~M.}\ \bibnamefont {Levine}},\ }\bibfield  {title} {\bibinfo {title} {A niche for neutrality},\ }\href@noop {} {\bibfield  {journal} {\bibinfo  {journal} {Ecology letters}\ }\textbf {\bibinfo {volume} {10}},\ \bibinfo {pages} {95} (\bibinfo {year} {2007})}\BibitemShut {NoStop}%
\bibitem [{\citenamefont {Spaak}\ \emph {et~al.}(2021)\citenamefont {Spaak}, \citenamefont {Godoy},\ and\ \citenamefont {De~Laender}}]{spaak2021mapping}%
  \BibitemOpen
  \bibfield  {author} {\bibinfo {author} {\bibfnamefont {J.~W.}\ \bibnamefont {Spaak}}, \bibinfo {author} {\bibfnamefont {O.}~\bibnamefont {Godoy}},\ and\ \bibinfo {author} {\bibfnamefont {F.}~\bibnamefont {De~Laender}},\ }\bibfield  {title} {\bibinfo {title} {Mapping species niche and fitness differences for communities with multiple interaction types},\ }\href@noop {} {\bibfield  {journal} {\bibinfo  {journal} {Oikos}\ }\textbf {\bibinfo {volume} {130}},\ \bibinfo {pages} {2065} (\bibinfo {year} {2021})}\BibitemShut {NoStop}%
\bibitem [{\citenamefont {HilleRisLambers}\ \emph {et~al.}(2012)\citenamefont {HilleRisLambers}, \citenamefont {Adler}, \citenamefont {Harpole}, \citenamefont {Levine},\ and\ \citenamefont {Mayfield}}]{hillerislambers2012rethinking}%
  \BibitemOpen
  \bibfield  {author} {\bibinfo {author} {\bibfnamefont {J.}~\bibnamefont {HilleRisLambers}}, \bibinfo {author} {\bibfnamefont {P.~B.}\ \bibnamefont {Adler}}, \bibinfo {author} {\bibfnamefont {W.~S.}\ \bibnamefont {Harpole}}, \bibinfo {author} {\bibfnamefont {J.~M.}\ \bibnamefont {Levine}},\ and\ \bibinfo {author} {\bibfnamefont {M.~M.}\ \bibnamefont {Mayfield}},\ }\bibfield  {title} {\bibinfo {title} {Rethinking community assembly through the lens of coexistence theory},\ }\href@noop {} {\bibfield  {journal} {\bibinfo  {journal} {Annual review of ecology, evolution, and systematics}\ }\textbf {\bibinfo {volume} {43}},\ \bibinfo {pages} {227} (\bibinfo {year} {2012})}\BibitemShut {NoStop}%
\bibitem [{\citenamefont {Mayfield}\ and\ \citenamefont {Levine}(2010)}]{mayfield2010opposing}%
  \BibitemOpen
  \bibfield  {author} {\bibinfo {author} {\bibfnamefont {M.~M.}\ \bibnamefont {Mayfield}}\ and\ \bibinfo {author} {\bibfnamefont {J.~M.}\ \bibnamefont {Levine}},\ }\bibfield  {title} {\bibinfo {title} {Opposing effects of competitive exclusion on the phylogenetic structure of communities},\ }\href@noop {} {\bibfield  {journal} {\bibinfo  {journal} {Ecology letters}\ }\textbf {\bibinfo {volume} {13}},\ \bibinfo {pages} {1085} (\bibinfo {year} {2010})}\BibitemShut {NoStop}%
\bibitem [{\citenamefont {Papula}\ \emph {et~al.}(2025)\citenamefont {Papula}, \citenamefont {Birzu},\ and\ \citenamefont {Fisher}}]{papula2025extensive}%
  \BibitemOpen
  \bibfield  {author} {\bibinfo {author} {\bibfnamefont {A.}~\bibnamefont {Papula}}, \bibinfo {author} {\bibfnamefont {G.}~\bibnamefont {Birzu}},\ and\ \bibinfo {author} {\bibfnamefont {D.~S.}\ \bibnamefont {Fisher}},\ }\bibfield  {title} {\bibinfo {title} {Extensive recombination, selection, and asexual blooms shape the diversity of the dominant clade of prochlorococcus},\ }\href@noop {} {\bibfield  {journal} {\bibinfo  {journal} {bioRxiv}\ ,\ \bibinfo {pages} {2025}} (\bibinfo {year} {2025})}\BibitemShut {NoStop}%
\bibitem [{\citenamefont {Ascensao}\ \emph {et~al.}(2025)\citenamefont {Ascensao}, \citenamefont {Abedi}, \citenamefont {Prasad},\ and\ \citenamefont {Hallatschek}}]{ascensao2025frequency}%
  \BibitemOpen
  \bibfield  {author} {\bibinfo {author} {\bibfnamefont {J.~A.}\ \bibnamefont {Ascensao}}, \bibinfo {author} {\bibfnamefont {K.~D.}\ \bibnamefont {Abedi}}, \bibinfo {author} {\bibfnamefont {A.~N.}\ \bibnamefont {Prasad}},\ and\ \bibinfo {author} {\bibfnamefont {O.}~\bibnamefont {Hallatschek}},\ }\bibfield  {title} {\bibinfo {title} {Frequency-dependent fitness effects are ubiquitous},\ }\href@noop {} {\bibfield  {journal} {\bibinfo  {journal} {bioRxiv}\ } (\bibinfo {year} {2025})}\BibitemShut {NoStop}%
\bibitem [{\citenamefont {Walton}\ \emph {et~al.}(2025)\citenamefont {Walton}, \citenamefont {Xu}, \citenamefont {Sharma}, \citenamefont {Gellert}, \citenamefont {Yeh}, \citenamefont {Cremer}, \citenamefont {Xue}, \citenamefont {Petrov},\ and\ \citenamefont {Good}}]{walton2025community}%
  \BibitemOpen
  \bibfield  {author} {\bibinfo {author} {\bibfnamefont {S.~J.}\ \bibnamefont {Walton}}, \bibinfo {author} {\bibfnamefont {Q.}~\bibnamefont {Xu}}, \bibinfo {author} {\bibfnamefont {R.}~\bibnamefont {Sharma}}, \bibinfo {author} {\bibfnamefont {H.~R.}\ \bibnamefont {Gellert}}, \bibinfo {author} {\bibfnamefont {C.-F.}\ \bibnamefont {Yeh}}, \bibinfo {author} {\bibfnamefont {J.}~\bibnamefont {Cremer}}, \bibinfo {author} {\bibfnamefont {K.~S.}\ \bibnamefont {Xue}}, \bibinfo {author} {\bibfnamefont {D.~A.}\ \bibnamefont {Petrov}},\ and\ \bibinfo {author} {\bibfnamefont {B.~H.}\ \bibnamefont {Good}},\ }\bibfield  {title} {\bibinfo {title} {Community coalescence reveals strong selection and coexistence within species in complex microbial communities},\ }\href@noop {} {\bibfield  {journal} {\bibinfo  {journal} {bioRxiv}\ ,\ \bibinfo {pages} {2025}} (\bibinfo {year} {2025})}\BibitemShut {NoStop}%
\bibitem [{\citenamefont {Thingstad}\ \emph {et~al.}(2014)\citenamefont {Thingstad}, \citenamefont {V{\aa}ge}, \citenamefont {Storesund}, \citenamefont {Sandaa},\ and\ \citenamefont {Giske}}]{thingstad2014theoretical}%
  \BibitemOpen
  \bibfield  {author} {\bibinfo {author} {\bibfnamefont {T.~F.}\ \bibnamefont {Thingstad}}, \bibinfo {author} {\bibfnamefont {S.}~\bibnamefont {V{\aa}ge}}, \bibinfo {author} {\bibfnamefont {J.~E.}\ \bibnamefont {Storesund}}, \bibinfo {author} {\bibfnamefont {R.-A.}\ \bibnamefont {Sandaa}},\ and\ \bibinfo {author} {\bibfnamefont {J.}~\bibnamefont {Giske}},\ }\bibfield  {title} {\bibinfo {title} {A theoretical analysis of how strain-specific viruses can control microbial species diversity},\ }\href@noop {} {\bibfield  {journal} {\bibinfo  {journal} {Proceedings of the National Academy of Sciences}\ }\textbf {\bibinfo {volume} {111}},\ \bibinfo {pages} {7813} (\bibinfo {year} {2014})}\BibitemShut {NoStop}%
\bibitem [{\citenamefont {Goyal}\ and\ \citenamefont {Chure}(2025)}]{goyal2025paradox}%
  \BibitemOpen
  \bibfield  {author} {\bibinfo {author} {\bibfnamefont {A.}~\bibnamefont {Goyal}}\ and\ \bibinfo {author} {\bibfnamefont {G.}~\bibnamefont {Chure}},\ }\bibfield  {title} {\bibinfo {title} {Paradox of the sub-plankton: Plausible mechanisms and open problems underlying strain-level diversity in microbial communities},\ }\href@noop {} {\bibfield  {journal} {\bibinfo  {journal} {Environmental Microbiology}\ }\textbf {\bibinfo {volume} {27}},\ \bibinfo {pages} {e70094} (\bibinfo {year} {2025})}\BibitemShut {NoStop}%
\bibitem [{\citenamefont {Fussmann}\ \emph {et~al.}(2007)\citenamefont {Fussmann}, \citenamefont {Loreau},\ and\ \citenamefont {Abrams}}]{fussmann2007eco}%
  \BibitemOpen
  \bibfield  {author} {\bibinfo {author} {\bibfnamefont {G.~F.}\ \bibnamefont {Fussmann}}, \bibinfo {author} {\bibfnamefont {M.}~\bibnamefont {Loreau}},\ and\ \bibinfo {author} {\bibfnamefont {P.~A.}\ \bibnamefont {Abrams}},\ }\bibfield  {title} {\bibinfo {title} {Eco-evolutionary dynamics of communities and ecosystems},\ }\href@noop {} {\bibfield  {journal} {\bibinfo  {journal} {Functional ecology}\ ,\ \bibinfo {pages} {465}} (\bibinfo {year} {2007})}\BibitemShut {NoStop}%
\bibitem [{\citenamefont {Hendry}(2017)}]{hendry2017eco}%
  \BibitemOpen
  \bibfield  {author} {\bibinfo {author} {\bibfnamefont {A.~P.}\ \bibnamefont {Hendry}},\ }\href@noop {} {\emph {\bibinfo {title} {Eco-evolutionary dynamics}}}\ (\bibinfo  {publisher} {Princeton university press},\ \bibinfo {year} {2017})\BibitemShut {NoStop}%
\bibitem [{\citenamefont {Govaert}\ \emph {et~al.}(2019)\citenamefont {Govaert}, \citenamefont {Fronhofer}, \citenamefont {Lion}, \citenamefont {Eizaguirre}, \citenamefont {Bonte}, \citenamefont {Egas}, \citenamefont {Hendry}, \citenamefont {De~Brito~Martins}, \citenamefont {Meli{\'a}n}, \citenamefont {Raeymaekers} \emph {et~al.}}]{govaert2019eco}%
  \BibitemOpen
  \bibfield  {author} {\bibinfo {author} {\bibfnamefont {L.}~\bibnamefont {Govaert}}, \bibinfo {author} {\bibfnamefont {E.~A.}\ \bibnamefont {Fronhofer}}, \bibinfo {author} {\bibfnamefont {S.}~\bibnamefont {Lion}}, \bibinfo {author} {\bibfnamefont {C.}~\bibnamefont {Eizaguirre}}, \bibinfo {author} {\bibfnamefont {D.}~\bibnamefont {Bonte}}, \bibinfo {author} {\bibfnamefont {M.}~\bibnamefont {Egas}}, \bibinfo {author} {\bibfnamefont {A.~P.}\ \bibnamefont {Hendry}}, \bibinfo {author} {\bibfnamefont {A.}~\bibnamefont {De~Brito~Martins}}, \bibinfo {author} {\bibfnamefont {C.~J.}\ \bibnamefont {Meli{\'a}n}}, \bibinfo {author} {\bibfnamefont {J.~A.}\ \bibnamefont {Raeymaekers}}, \emph {et~al.},\ }\bibfield  {title} {\bibinfo {title} {Eco-evolutionary feedbacks—theoretical models and perspectives},\ }\href@noop {} {\bibfield  {journal} {\bibinfo  {journal} {Functional Ecology}\ }\textbf {\bibinfo {volume} {33}},\ \bibinfo {pages} {13} (\bibinfo {year} {2019})}\BibitemShut {NoStop}%
\bibitem [{\citenamefont {Goyal}\ \emph {et~al.}(2021)\citenamefont {Goyal}, \citenamefont {Wang}, \citenamefont {Dubinkina},\ and\ \citenamefont {Maslov}}]{goyal2021ecology}%
  \BibitemOpen
  \bibfield  {author} {\bibinfo {author} {\bibfnamefont {A.}~\bibnamefont {Goyal}}, \bibinfo {author} {\bibfnamefont {T.}~\bibnamefont {Wang}}, \bibinfo {author} {\bibfnamefont {V.}~\bibnamefont {Dubinkina}},\ and\ \bibinfo {author} {\bibfnamefont {S.}~\bibnamefont {Maslov}},\ }\bibfield  {title} {\bibinfo {title} {Ecology-guided prediction of cross-feeding interactions in the human gut microbiome},\ }\href@noop {} {\bibfield  {journal} {\bibinfo  {journal} {Nature communications}\ }\textbf {\bibinfo {volume} {12}},\ \bibinfo {pages} {1335} (\bibinfo {year} {2021})}\BibitemShut {NoStop}%
\bibitem [{\citenamefont {Faust}\ and\ \citenamefont {Raes}(2012)}]{faust2012microbial}%
  \BibitemOpen
  \bibfield  {author} {\bibinfo {author} {\bibfnamefont {K.}~\bibnamefont {Faust}}\ and\ \bibinfo {author} {\bibfnamefont {J.}~\bibnamefont {Raes}},\ }\bibfield  {title} {\bibinfo {title} {Microbial interactions: from networks to models},\ }\href@noop {} {\bibfield  {journal} {\bibinfo  {journal} {Nature Reviews Microbiology}\ }\textbf {\bibinfo {volume} {10}},\ \bibinfo {pages} {538} (\bibinfo {year} {2012})}\BibitemShut {NoStop}%
\bibitem [{\citenamefont {Tackmann}\ \emph {et~al.}(2019)\citenamefont {Tackmann}, \citenamefont {Rodrigues},\ and\ \citenamefont {von Mering}}]{tackmann2019rapid}%
  \BibitemOpen
  \bibfield  {author} {\bibinfo {author} {\bibfnamefont {J.}~\bibnamefont {Tackmann}}, \bibinfo {author} {\bibfnamefont {J.~F.~M.}\ \bibnamefont {Rodrigues}},\ and\ \bibinfo {author} {\bibfnamefont {C.}~\bibnamefont {von Mering}},\ }\bibfield  {title} {\bibinfo {title} {Rapid inference of direct interactions in large-scale ecological networks from heterogeneous microbial sequencing data},\ }\href@noop {} {\bibfield  {journal} {\bibinfo  {journal} {Cell systems}\ }\textbf {\bibinfo {volume} {9}},\ \bibinfo {pages} {286} (\bibinfo {year} {2019})}\BibitemShut {NoStop}%
\bibitem [{\citenamefont {L{\'o}pez}\ \emph {et~al.}(2025)\citenamefont {L{\'o}pez}, \citenamefont {Bonachela}, \citenamefont {Dominguez-Bello}, \citenamefont {Manhart}, \citenamefont {Levin}, \citenamefont {Blaser},\ and\ \citenamefont {Mu{\~n}oz}}]{lopez2025imbalance}%
  \BibitemOpen
  \bibfield  {author} {\bibinfo {author} {\bibfnamefont {R.~C.}\ \bibnamefont {L{\'o}pez}}, \bibinfo {author} {\bibfnamefont {J.~A.}\ \bibnamefont {Bonachela}}, \bibinfo {author} {\bibfnamefont {M.~G.}\ \bibnamefont {Dominguez-Bello}}, \bibinfo {author} {\bibfnamefont {M.}~\bibnamefont {Manhart}}, \bibinfo {author} {\bibfnamefont {S.~A.}\ \bibnamefont {Levin}}, \bibinfo {author} {\bibfnamefont {M.~J.}\ \bibnamefont {Blaser}},\ and\ \bibinfo {author} {\bibfnamefont {M.~A.}\ \bibnamefont {Mu{\~n}oz}},\ }\bibfield  {title} {\bibinfo {title} {Imbalance in gut microbial interactions as a marker of health and disease},\ }\href@noop {} {\bibfield  {journal} {\bibinfo  {journal} {bioRxiv}\ ,\ \bibinfo {pages} {2025}} (\bibinfo {year} {2025})}\BibitemShut {NoStop}%
\bibitem [{\citenamefont {Van~Rossum}\ \emph {et~al.}(2020{\natexlab{b}})\citenamefont {Van~Rossum}, \citenamefont {Ferretti}, \citenamefont {Maistrenko},\ and\ \citenamefont {Bork}}]{vanrossum2020}%
  \BibitemOpen
  \bibfield  {author} {\bibinfo {author} {\bibfnamefont {T.}~\bibnamefont {Van~Rossum}}, \bibinfo {author} {\bibfnamefont {P.}~\bibnamefont {Ferretti}}, \bibinfo {author} {\bibfnamefont {O.~M.}\ \bibnamefont {Maistrenko}},\ and\ \bibinfo {author} {\bibfnamefont {P.}~\bibnamefont {Bork}},\ }\bibfield  {title} {\bibinfo {title} {Diversity within species: interpreting strains in microbiomes},\ }\href {https://doi.org/10.1038/s41579-020-0368-1} {\bibfield  {journal} {\bibinfo  {journal} {Nature Reviews Microbiology}\ }\textbf {\bibinfo {volume} {18}},\ \bibinfo {pages} {491} (\bibinfo {year} {2020}{\natexlab{b}})},\ \bibinfo {note} {number: 9 Publisher: Nature Publishing Group}\BibitemShut {NoStop}%
\bibitem [{\citenamefont {Ho}\ \emph {et~al.}(2022)\citenamefont {Ho}, \citenamefont {Good},\ and\ \citenamefont {Huang}}]{ho2022competition}%
  \BibitemOpen
  \bibfield  {author} {\bibinfo {author} {\bibfnamefont {P.-Y.}\ \bibnamefont {Ho}}, \bibinfo {author} {\bibfnamefont {B.~H.}\ \bibnamefont {Good}},\ and\ \bibinfo {author} {\bibfnamefont {K.~C.}\ \bibnamefont {Huang}},\ }\bibfield  {title} {\bibinfo {title} {Competition for fluctuating resources reproduces statistics of species abundance over time across wide-ranging microbiotas},\ }\href@noop {} {\bibfield  {journal} {\bibinfo  {journal} {Elife}\ }\textbf {\bibinfo {volume} {11}},\ \bibinfo {pages} {e75168} (\bibinfo {year} {2022})}\BibitemShut {NoStop}%
\bibitem [{\citenamefont {Berry}\ and\ \citenamefont {Widder}(2014)}]{berry2014deciphering}%
  \BibitemOpen
  \bibfield  {author} {\bibinfo {author} {\bibfnamefont {D.}~\bibnamefont {Berry}}\ and\ \bibinfo {author} {\bibfnamefont {S.}~\bibnamefont {Widder}},\ }\bibfield  {title} {\bibinfo {title} {Deciphering microbial interactions and detecting keystone species with co-occurrence networks},\ }\href@noop {} {\bibfield  {journal} {\bibinfo  {journal} {Frontiers in microbiology}\ }\textbf {\bibinfo {volume} {5}},\ \bibinfo {pages} {219} (\bibinfo {year} {2014})}\BibitemShut {NoStop}%
\bibitem [{\citenamefont {Tikhonov}\ \emph {et~al.}(2015)\citenamefont {Tikhonov}, \citenamefont {Leach},\ and\ \citenamefont {Wingreen}}]{tikhonov2015interpreting}%
  \BibitemOpen
  \bibfield  {author} {\bibinfo {author} {\bibfnamefont {M.}~\bibnamefont {Tikhonov}}, \bibinfo {author} {\bibfnamefont {R.~W.}\ \bibnamefont {Leach}},\ and\ \bibinfo {author} {\bibfnamefont {N.~S.}\ \bibnamefont {Wingreen}},\ }\bibfield  {title} {\bibinfo {title} {Interpreting 16s metagenomic data without clustering to achieve sub-otu resolution},\ }\href@noop {} {\bibfield  {journal} {\bibinfo  {journal} {The ISME journal}\ }\textbf {\bibinfo {volume} {9}},\ \bibinfo {pages} {68} (\bibinfo {year} {2015})}\BibitemShut {NoStop}%
\bibitem [{\citenamefont {Carr}\ \emph {et~al.}(2019)\citenamefont {Carr}, \citenamefont {Diener}, \citenamefont {Baliga},\ and\ \citenamefont {Gibbons}}]{carr2019use}%
  \BibitemOpen
  \bibfield  {author} {\bibinfo {author} {\bibfnamefont {A.}~\bibnamefont {Carr}}, \bibinfo {author} {\bibfnamefont {C.}~\bibnamefont {Diener}}, \bibinfo {author} {\bibfnamefont {N.~S.}\ \bibnamefont {Baliga}},\ and\ \bibinfo {author} {\bibfnamefont {S.~M.}\ \bibnamefont {Gibbons}},\ }\bibfield  {title} {\bibinfo {title} {Use and abuse of correlation analyses in microbial ecology},\ }\href@noop {} {\bibfield  {journal} {\bibinfo  {journal} {The ISME journal}\ }\textbf {\bibinfo {volume} {13}},\ \bibinfo {pages} {2647} (\bibinfo {year} {2019})}\BibitemShut {NoStop}%
\bibitem [{\citenamefont {Li}\ \emph {et~al.}(2025)\citenamefont {Li}, \citenamefont {Feng}, \citenamefont {Goyal},\ and\ \citenamefont {Mehta}}]{li2025population}%
  \BibitemOpen
  \bibfield  {author} {\bibinfo {author} {\bibfnamefont {S.~Y.}\ \bibnamefont {Li}}, \bibinfo {author} {\bibfnamefont {Z.}~\bibnamefont {Feng}}, \bibinfo {author} {\bibfnamefont {A.}~\bibnamefont {Goyal}},\ and\ \bibinfo {author} {\bibfnamefont {P.}~\bibnamefont {Mehta}},\ }\bibfield  {title} {\bibinfo {title} {Population genetics in complex ecological communities},\ }\href@noop {} {\bibfield  {journal} {\bibinfo  {journal} {arXiv preprint arXiv:2509.23977}\ } (\bibinfo {year} {2025})}\BibitemShut {NoStop}%
\bibitem [{\citenamefont {Bunin}(2017)}]{BuninGLV}%
  \BibitemOpen
  \bibfield  {author} {\bibinfo {author} {\bibfnamefont {G.}~\bibnamefont {Bunin}},\ }\bibfield  {title} {\bibinfo {title} {Ecological communities with lotka-volterra dynamics},\ }\href@noop {} {\bibfield  {journal} {\bibinfo  {journal} {Physical Review E}\ }\textbf {\bibinfo {volume} {95}},\ \bibinfo {pages} {042414} (\bibinfo {year} {2017})}\BibitemShut {NoStop}%
\bibitem [{\citenamefont {Advani}\ \emph {et~al.}(2018)\citenamefont {Advani}, \citenamefont {Bunin},\ and\ \citenamefont {Mehta}}]{advani2018statistical}%
  \BibitemOpen
  \bibfield  {author} {\bibinfo {author} {\bibfnamefont {M.}~\bibnamefont {Advani}}, \bibinfo {author} {\bibfnamefont {G.}~\bibnamefont {Bunin}},\ and\ \bibinfo {author} {\bibfnamefont {P.}~\bibnamefont {Mehta}},\ }\bibfield  {title} {\bibinfo {title} {Statistical physics of community ecology: a cavity solution to macarthur’s consumer resource model},\ }\href@noop {} {\bibfield  {journal} {\bibinfo  {journal} {Journal of Statistical Mechanics: Theory and Experiment}\ }\textbf {\bibinfo {volume} {2018}},\ \bibinfo {pages} {033406} (\bibinfo {year} {2018})}\BibitemShut {NoStop}%
\bibitem [{\citenamefont {Tikhonov}\ and\ \citenamefont {Monasson}(2018)}]{tikhonov2018innovation}%
  \BibitemOpen
  \bibfield  {author} {\bibinfo {author} {\bibfnamefont {M.}~\bibnamefont {Tikhonov}}\ and\ \bibinfo {author} {\bibfnamefont {R.}~\bibnamefont {Monasson}},\ }\bibfield  {title} {\bibinfo {title} {Innovation rather than improvement: A solvable high-dimensional model highlights the limitations of scalar fitness},\ }\href@noop {} {\bibfield  {journal} {\bibinfo  {journal} {Journal of Statistical Physics}\ }\textbf {\bibinfo {volume} {172}},\ \bibinfo {pages} {74} (\bibinfo {year} {2018})}\BibitemShut {NoStop}%
\bibitem [{\citenamefont {Mahadevan}\ \emph {et~al.}(2023)\citenamefont {Mahadevan}, \citenamefont {Pearce},\ and\ \citenamefont {Fisher}}]{mahadevan2023}%
  \BibitemOpen
  \bibfield  {author} {\bibinfo {author} {\bibfnamefont {A.}~\bibnamefont {Mahadevan}}, \bibinfo {author} {\bibfnamefont {M.~T.}\ \bibnamefont {Pearce}},\ and\ \bibinfo {author} {\bibfnamefont {D.~S.}\ \bibnamefont {Fisher}},\ }\bibfield  {title} {\bibinfo {title} {Spatiotemporal ecological chaos enables gradual evolutionary diversification without niches or tradeoffs},\ }\href {https://doi.org/10.7554/eLife.82734} {\bibfield  {journal} {\bibinfo  {journal} {eLife}\ }\textbf {\bibinfo {volume} {12}},\ \bibinfo {pages} {e82734} (\bibinfo {year} {2023})},\ \bibinfo {note} {publisher: eLife Sciences Publications, Ltd}\BibitemShut {NoStop}%
\bibitem [{\citenamefont {Patro}\ \emph {et~al.}(2025)\citenamefont {Patro}, \citenamefont {Taylor},\ and\ \citenamefont {Goyal}}]{patro2025emergent}%
  \BibitemOpen
  \bibfield  {author} {\bibinfo {author} {\bibfnamefont {N.~K.}\ \bibnamefont {Patro}}, \bibinfo {author} {\bibfnamefont {W.}~\bibnamefont {Taylor}},\ and\ \bibinfo {author} {\bibfnamefont {A.}~\bibnamefont {Goyal}},\ }\bibfield  {title} {\bibinfo {title} {Emergent self-inhibition governs the landscape of stable states in complex ecosystems},\ }\href@noop {} {\bibfield  {journal} {\bibinfo  {journal} {arXiv preprint arXiv:2511.06697}\ } (\bibinfo {year} {2025})}\BibitemShut {NoStop}%
\bibitem [{\citenamefont {Feng}\ \emph {et~al.}(2024)\citenamefont {Feng}, \citenamefont {Marsland~III}, \citenamefont {Rocks},\ and\ \citenamefont {Mehta}}]{feng2024emergent}%
  \BibitemOpen
  \bibfield  {author} {\bibinfo {author} {\bibfnamefont {Z.}~\bibnamefont {Feng}}, \bibinfo {author} {\bibfnamefont {R.}~\bibnamefont {Marsland~III}}, \bibinfo {author} {\bibfnamefont {J.~W.}\ \bibnamefont {Rocks}},\ and\ \bibinfo {author} {\bibfnamefont {P.}~\bibnamefont {Mehta}},\ }\bibfield  {title} {\bibinfo {title} {Emergent competition shapes top-down versus bottom-up control in multi-trophic ecosystems},\ }\href@noop {} {\bibfield  {journal} {\bibinfo  {journal} {PLOS Computational Biology}\ }\textbf {\bibinfo {volume} {20}},\ \bibinfo {pages} {e1011675} (\bibinfo {year} {2024})}\BibitemShut {NoStop}%
\bibitem [{\citenamefont {Chesson}(2000{\natexlab{b}})}]{chesson2000}%
  \BibitemOpen
  \bibfield  {author} {\bibinfo {author} {\bibfnamefont {P.}~\bibnamefont {Chesson}},\ }\bibfield  {title} {\bibinfo {title} {Mechanisms of maintenance of species diversity},\ }\href {https://doi.org/10.1146/annurev.ecolsys.31.1.343} {\bibfield  {journal} {\bibinfo  {journal} {Annual Review of Ecology and Systematics}\ }\textbf {\bibinfo {volume} {31}},\ \bibinfo {pages} {343} (\bibinfo {year} {2000}{\natexlab{b}})},\ \bibinfo {note} {\_eprint: https://doi.org/10.1146/annurev.ecolsys.31.1.343}\BibitemShut {NoStop}%
\bibitem [{\citenamefont {Lemos-Costa}\ \emph {et~al.}(2024)\citenamefont {Lemos-Costa}, \citenamefont {Miller},\ and\ \citenamefont {Allesina}}]{lemos2024phylogeny}%
  \BibitemOpen
  \bibfield  {author} {\bibinfo {author} {\bibfnamefont {P.}~\bibnamefont {Lemos-Costa}}, \bibinfo {author} {\bibfnamefont {Z.~R.}\ \bibnamefont {Miller}},\ and\ \bibinfo {author} {\bibfnamefont {S.}~\bibnamefont {Allesina}},\ }\bibfield  {title} {\bibinfo {title} {Phylogeny structures species' interactions in experimental ecological communities},\ }\href@noop {} {\bibfield  {journal} {\bibinfo  {journal} {Ecology Letters}\ }\textbf {\bibinfo {volume} {27}},\ \bibinfo {pages} {e14490} (\bibinfo {year} {2024})}\BibitemShut {NoStop}%
\bibitem [{\citenamefont {Goyal}\ and\ \citenamefont {Maslov}(2018)}]{goyal2018diversity}%
  \BibitemOpen
  \bibfield  {author} {\bibinfo {author} {\bibfnamefont {A.}~\bibnamefont {Goyal}}\ and\ \bibinfo {author} {\bibfnamefont {S.}~\bibnamefont {Maslov}},\ }\bibfield  {title} {\bibinfo {title} {Diversity, stability, and reproducibility in stochastically assembled microbial ecosystems},\ }\href@noop {} {\bibfield  {journal} {\bibinfo  {journal} {Physical Review Letters}\ }\textbf {\bibinfo {volume} {120}},\ \bibinfo {pages} {158102} (\bibinfo {year} {2018})}\BibitemShut {NoStop}%
\bibitem [{\citenamefont {Marsland~III}\ \emph {et~al.}(2019)\citenamefont {Marsland~III}, \citenamefont {Cui}, \citenamefont {Goldford}, \citenamefont {Sanchez}, \citenamefont {Korolev},\ and\ \citenamefont {Mehta}}]{marsland2019available}%
  \BibitemOpen
  \bibfield  {author} {\bibinfo {author} {\bibfnamefont {R.}~\bibnamefont {Marsland~III}}, \bibinfo {author} {\bibfnamefont {W.}~\bibnamefont {Cui}}, \bibinfo {author} {\bibfnamefont {J.}~\bibnamefont {Goldford}}, \bibinfo {author} {\bibfnamefont {A.}~\bibnamefont {Sanchez}}, \bibinfo {author} {\bibfnamefont {K.}~\bibnamefont {Korolev}},\ and\ \bibinfo {author} {\bibfnamefont {P.}~\bibnamefont {Mehta}},\ }\bibfield  {title} {\bibinfo {title} {Available energy fluxes drive a transition in the diversity, stability, and functional structure of microbial communities},\ }\href@noop {} {\bibfield  {journal} {\bibinfo  {journal} {PLoS computational biology}\ }\textbf {\bibinfo {volume} {15}},\ \bibinfo {pages} {e1006793} (\bibinfo {year} {2019})}\BibitemShut {NoStop}%
\bibitem [{\citenamefont {Goldford}\ \emph {et~al.}(2018)\citenamefont {Goldford}, \citenamefont {Lu}, \citenamefont {Baji{\'c}}, \citenamefont {Estrela}, \citenamefont {Tikhonov}, \citenamefont {Sanchez-Gorostiaga}, \citenamefont {Segr{\`e}}, \citenamefont {Mehta},\ and\ \citenamefont {Sanchez}}]{goldford2018emergent}%
  \BibitemOpen
  \bibfield  {author} {\bibinfo {author} {\bibfnamefont {J.~E.}\ \bibnamefont {Goldford}}, \bibinfo {author} {\bibfnamefont {N.}~\bibnamefont {Lu}}, \bibinfo {author} {\bibfnamefont {D.}~\bibnamefont {Baji{\'c}}}, \bibinfo {author} {\bibfnamefont {S.}~\bibnamefont {Estrela}}, \bibinfo {author} {\bibfnamefont {M.}~\bibnamefont {Tikhonov}}, \bibinfo {author} {\bibfnamefont {A.}~\bibnamefont {Sanchez-Gorostiaga}}, \bibinfo {author} {\bibfnamefont {D.}~\bibnamefont {Segr{\`e}}}, \bibinfo {author} {\bibfnamefont {P.}~\bibnamefont {Mehta}},\ and\ \bibinfo {author} {\bibfnamefont {A.}~\bibnamefont {Sanchez}},\ }\bibfield  {title} {\bibinfo {title} {Emergent simplicity in microbial community assembly},\ }\href@noop {} {\bibfield  {journal} {\bibinfo  {journal} {Science}\ }\textbf {\bibinfo {volume} {361}},\ \bibinfo {pages} {469} (\bibinfo {year} {2018})}\BibitemShut {NoStop}%
\bibitem [{\citenamefont {Goyal}\ \emph {et~al.}(2025)\citenamefont {Goyal}, \citenamefont {Rocks},\ and\ \citenamefont {Mehta}}]{goyal2025universal}%
  \BibitemOpen
  \bibfield  {author} {\bibinfo {author} {\bibfnamefont {A.}~\bibnamefont {Goyal}}, \bibinfo {author} {\bibfnamefont {J.~W.}\ \bibnamefont {Rocks}},\ and\ \bibinfo {author} {\bibfnamefont {P.}~\bibnamefont {Mehta}},\ }\bibfield  {title} {\bibinfo {title} {Universal niche geometry governs the response of ecosystems to environmental perturbations},\ }\href@noop {} {\bibfield  {journal} {\bibinfo  {journal} {PRX Life}\ }\textbf {\bibinfo {volume} {3}},\ \bibinfo {pages} {013010} (\bibinfo {year} {2025})}\BibitemShut {NoStop}%
\bibitem [{\citenamefont {Letten}\ \emph {et~al.}(2017)\citenamefont {Letten}, \citenamefont {Ke},\ and\ \citenamefont {Fukami}}]{letten2017linking}%
  \BibitemOpen
  \bibfield  {author} {\bibinfo {author} {\bibfnamefont {A.~D.}\ \bibnamefont {Letten}}, \bibinfo {author} {\bibfnamefont {P.-J.}\ \bibnamefont {Ke}},\ and\ \bibinfo {author} {\bibfnamefont {T.}~\bibnamefont {Fukami}},\ }\bibfield  {title} {\bibinfo {title} {Linking modern coexistence theory and contemporary niche theory},\ }\href@noop {} {\bibfield  {journal} {\bibinfo  {journal} {Ecological Monographs}\ }\textbf {\bibinfo {volume} {87}},\ \bibinfo {pages} {161} (\bibinfo {year} {2017})}\BibitemShut {NoStop}%
\bibitem [{\citenamefont {Barab{\'a}s}\ \emph {et~al.}(2018)\citenamefont {Barab{\'a}s}, \citenamefont {D'Andrea},\ and\ \citenamefont {Stump}}]{barabas2018chesson}%
  \BibitemOpen
  \bibfield  {author} {\bibinfo {author} {\bibfnamefont {G.}~\bibnamefont {Barab{\'a}s}}, \bibinfo {author} {\bibfnamefont {R.}~\bibnamefont {D'Andrea}},\ and\ \bibinfo {author} {\bibfnamefont {S.~M.}\ \bibnamefont {Stump}},\ }\bibfield  {title} {\bibinfo {title} {Chesson's coexistence theory},\ }\href@noop {} {\bibfield  {journal} {\bibinfo  {journal} {Ecological monographs}\ }\textbf {\bibinfo {volume} {88}},\ \bibinfo {pages} {277} (\bibinfo {year} {2018})}\BibitemShut {NoStop}%
\bibitem [{\citenamefont {Chesson}(2018)}]{chesson2018updates}%
  \BibitemOpen
  \bibfield  {author} {\bibinfo {author} {\bibfnamefont {P.}~\bibnamefont {Chesson}},\ }\bibfield  {title} {\bibinfo {title} {Updates on mechanisms of maintenance of species diversity},\ }\href@noop {} {\bibfield  {journal} {\bibinfo  {journal} {Journal of ecology}\ }\textbf {\bibinfo {volume} {106}},\ \bibinfo {pages} {1773} (\bibinfo {year} {2018})}\BibitemShut {NoStop}%
\bibitem [{\citenamefont {Faust}\ \emph {et~al.}(2015)\citenamefont {Faust}, \citenamefont {Lahti}, \citenamefont {Gonze}, \citenamefont {De~Vos},\ and\ \citenamefont {Raes}}]{faust2015metagenomics}%
  \BibitemOpen
  \bibfield  {author} {\bibinfo {author} {\bibfnamefont {K.}~\bibnamefont {Faust}}, \bibinfo {author} {\bibfnamefont {L.}~\bibnamefont {Lahti}}, \bibinfo {author} {\bibfnamefont {D.}~\bibnamefont {Gonze}}, \bibinfo {author} {\bibfnamefont {W.~M.}\ \bibnamefont {De~Vos}},\ and\ \bibinfo {author} {\bibfnamefont {J.}~\bibnamefont {Raes}},\ }\bibfield  {title} {\bibinfo {title} {Metagenomics meets time series analysis: unraveling microbial community dynamics},\ }\href@noop {} {\bibfield  {journal} {\bibinfo  {journal} {Current opinion in microbiology}\ }\textbf {\bibinfo {volume} {25}},\ \bibinfo {pages} {56} (\bibinfo {year} {2015})}\BibitemShut {NoStop}%
\bibitem [{\citenamefont {Freilich}\ \emph {et~al.}(2011)\citenamefont {Freilich}, \citenamefont {Zarecki}, \citenamefont {Eilam}, \citenamefont {Segal}, \citenamefont {Henry}, \citenamefont {Kupiec}, \citenamefont {Gophna}, \citenamefont {Sharan},\ and\ \citenamefont {Ruppin}}]{freilich2011competitive}%
  \BibitemOpen
  \bibfield  {author} {\bibinfo {author} {\bibfnamefont {S.}~\bibnamefont {Freilich}}, \bibinfo {author} {\bibfnamefont {R.}~\bibnamefont {Zarecki}}, \bibinfo {author} {\bibfnamefont {O.}~\bibnamefont {Eilam}}, \bibinfo {author} {\bibfnamefont {E.~S.}\ \bibnamefont {Segal}}, \bibinfo {author} {\bibfnamefont {C.~S.}\ \bibnamefont {Henry}}, \bibinfo {author} {\bibfnamefont {M.}~\bibnamefont {Kupiec}}, \bibinfo {author} {\bibfnamefont {U.}~\bibnamefont {Gophna}}, \bibinfo {author} {\bibfnamefont {R.}~\bibnamefont {Sharan}},\ and\ \bibinfo {author} {\bibfnamefont {E.}~\bibnamefont {Ruppin}},\ }\bibfield  {title} {\bibinfo {title} {Competitive and cooperative metabolic interactions in bacterial communities},\ }\href@noop {} {\bibfield  {journal} {\bibinfo  {journal} {Nature communications}\ }\textbf {\bibinfo {volume} {2}},\ \bibinfo {pages} {589} (\bibinfo {year} {2011})}\BibitemShut {NoStop}%
\bibitem [{\citenamefont {Barber{\'a}n}\ \emph {et~al.}(2012)\citenamefont {Barber{\'a}n}, \citenamefont {Bates}, \citenamefont {Casamayor},\ and\ \citenamefont {Fierer}}]{barberan2012using}%
  \BibitemOpen
  \bibfield  {author} {\bibinfo {author} {\bibfnamefont {A.}~\bibnamefont {Barber{\'a}n}}, \bibinfo {author} {\bibfnamefont {S.~T.}\ \bibnamefont {Bates}}, \bibinfo {author} {\bibfnamefont {E.~O.}\ \bibnamefont {Casamayor}},\ and\ \bibinfo {author} {\bibfnamefont {N.}~\bibnamefont {Fierer}},\ }\bibfield  {title} {\bibinfo {title} {Using network analysis to explore co-occurrence patterns in soil microbial communities},\ }\href@noop {} {\bibfield  {journal} {\bibinfo  {journal} {The ISME journal}\ }\textbf {\bibinfo {volume} {6}},\ \bibinfo {pages} {343} (\bibinfo {year} {2012})}\BibitemShut {NoStop}%
\bibitem [{\citenamefont {Zhou}\ \emph {et~al.}(2010)\citenamefont {Zhou}, \citenamefont {Deng}, \citenamefont {Luo}, \citenamefont {He}, \citenamefont {Tu}, \citenamefont {Zhi}, \citenamefont {Zhou}, \citenamefont {Deng}, \citenamefont {Luo}, \citenamefont {He} \emph {et~al.}}]{zhou2010functional}%
  \BibitemOpen
  \bibfield  {author} {\bibinfo {author} {\bibfnamefont {J.}~\bibnamefont {Zhou}}, \bibinfo {author} {\bibfnamefont {Y.}~\bibnamefont {Deng}}, \bibinfo {author} {\bibfnamefont {F.}~\bibnamefont {Luo}}, \bibinfo {author} {\bibfnamefont {Z.}~\bibnamefont {He}}, \bibinfo {author} {\bibfnamefont {Q.}~\bibnamefont {Tu}}, \bibinfo {author} {\bibfnamefont {X.}~\bibnamefont {Zhi}}, \bibinfo {author} {\bibfnamefont {J.}~\bibnamefont {Zhou}}, \bibinfo {author} {\bibfnamefont {Y.}~\bibnamefont {Deng}}, \bibinfo {author} {\bibfnamefont {F.}~\bibnamefont {Luo}}, \bibinfo {author} {\bibfnamefont {Z.}~\bibnamefont {He}}, \emph {et~al.},\ }\bibfield  {title} {\bibinfo {title} {Functional molecular ecological networks. mbio 1: e00169-10},\ }\href@noop {} {\bibfield  {journal} {\bibinfo  {journal} {View in Article}\ } (\bibinfo {year} {2010})}\BibitemShut {NoStop}%
\bibitem [{\citenamefont {Sireci}\ \emph {et~al.}(2023)\citenamefont {Sireci}, \citenamefont {Mu{\~n}oz},\ and\ \citenamefont {Grilli}}]{sireci2023environmental}%
  \BibitemOpen
  \bibfield  {author} {\bibinfo {author} {\bibfnamefont {M.}~\bibnamefont {Sireci}}, \bibinfo {author} {\bibfnamefont {M.~A.}\ \bibnamefont {Mu{\~n}oz}},\ and\ \bibinfo {author} {\bibfnamefont {J.}~\bibnamefont {Grilli}},\ }\bibfield  {title} {\bibinfo {title} {Environmental fluctuations explain the universal decay of species-abundance correlations with phylogenetic distance},\ }\href@noop {} {\bibfield  {journal} {\bibinfo  {journal} {Proceedings of the National Academy of Sciences}\ }\textbf {\bibinfo {volume} {120}},\ \bibinfo {pages} {e2217144120} (\bibinfo {year} {2023})}\BibitemShut {NoStop}%
\bibitem [{\citenamefont {Aguad{\'e}-Gorgori{\'o}}\ and\ \citenamefont {K{\'e}fi}(2025)}]{aguade2025emergent}%
  \BibitemOpen
  \bibfield  {author} {\bibinfo {author} {\bibfnamefont {G.}~\bibnamefont {Aguad{\'e}-Gorgori{\'o}}}\ and\ \bibinfo {author} {\bibfnamefont {S.}~\bibnamefont {K{\'e}fi}},\ }\bibfield  {title} {\bibinfo {title} {Emergent coexistence and the limits of reductionism in ecological communities},\ }\href@noop {} {\bibfield  {journal} {\bibinfo  {journal} {bioRxiv}\ ,\ \bibinfo {pages} {2025}} (\bibinfo {year} {2025})}\BibitemShut {NoStop}%
\bibitem [{\citenamefont {Miller}\ and\ \citenamefont {Max}(2025)}]{miller2025multispecies}%
  \BibitemOpen
  \bibfield  {author} {\bibinfo {author} {\bibfnamefont {Z.~R.}\ \bibnamefont {Miller}}\ and\ \bibinfo {author} {\bibfnamefont {D.}~\bibnamefont {Max}},\ }\bibfield  {title} {\bibinfo {title} {Multispecies coexistence emerges from pairwise exclusions in communities with competitive hierarchy},\ }\href@noop {} {\bibfield  {journal} {\bibinfo  {journal} {Ecology Letters}\ }\textbf {\bibinfo {volume} {28}},\ \bibinfo {pages} {e70206} (\bibinfo {year} {2025})}\BibitemShut {NoStop}%
\bibitem [{\citenamefont {Chang}\ \emph {et~al.}(2023)\citenamefont {Chang}, \citenamefont {Baji{\'c}}, \citenamefont {Vila}, \citenamefont {Estrela},\ and\ \citenamefont {Sanchez}}]{chang2023emergent}%
  \BibitemOpen
  \bibfield  {author} {\bibinfo {author} {\bibfnamefont {C.-Y.}\ \bibnamefont {Chang}}, \bibinfo {author} {\bibfnamefont {D.}~\bibnamefont {Baji{\'c}}}, \bibinfo {author} {\bibfnamefont {J.~C.}\ \bibnamefont {Vila}}, \bibinfo {author} {\bibfnamefont {S.}~\bibnamefont {Estrela}},\ and\ \bibinfo {author} {\bibfnamefont {A.}~\bibnamefont {Sanchez}},\ }\bibfield  {title} {\bibinfo {title} {Emergent coexistence in multispecies microbial communities},\ }\href@noop {} {\bibfield  {journal} {\bibinfo  {journal} {Science}\ }\textbf {\bibinfo {volume} {381}},\ \bibinfo {pages} {343} (\bibinfo {year} {2023})}\BibitemShut {NoStop}%
\bibitem [{\citenamefont {Bupu}\ \emph {et~al.}(2025)\citenamefont {Bupu}, \citenamefont {Shoemaker}, \citenamefont {Mazzarisi},\ and\ \citenamefont {Grilli}}]{bupu2025balance}%
  \BibitemOpen
  \bibfield  {author} {\bibinfo {author} {\bibfnamefont {A.}~\bibnamefont {Bupu}}, \bibinfo {author} {\bibfnamefont {W.~R.}\ \bibnamefont {Shoemaker}}, \bibinfo {author} {\bibfnamefont {O.}~\bibnamefont {Mazzarisi}},\ and\ \bibinfo {author} {\bibfnamefont {J.}~\bibnamefont {Grilli}},\ }\bibfield  {title} {\bibinfo {title} {A balance between environmental filtering and competitive exclusion modulates the macroecology of alternative stable states in microbial communities},\ }\href@noop {} {\bibfield  {journal} {\bibinfo  {journal} {bioRxiv}\ ,\ \bibinfo {pages} {2025}} (\bibinfo {year} {2025})}\BibitemShut {NoStop}%
\bibitem [{\citenamefont {Pearl~Mizrahi}\ \emph {et~al.}(2025)\citenamefont {Pearl~Mizrahi}, \citenamefont {Lee}, \citenamefont {Goyal}, \citenamefont {Owen},\ and\ \citenamefont {Gore}}]{pearl2025structured}%
  \BibitemOpen
  \bibfield  {author} {\bibinfo {author} {\bibfnamefont {S.}~\bibnamefont {Pearl~Mizrahi}}, \bibinfo {author} {\bibfnamefont {H.}~\bibnamefont {Lee}}, \bibinfo {author} {\bibfnamefont {A.}~\bibnamefont {Goyal}}, \bibinfo {author} {\bibfnamefont {E.}~\bibnamefont {Owen}},\ and\ \bibinfo {author} {\bibfnamefont {J.}~\bibnamefont {Gore}},\ }\bibfield  {title} {\bibinfo {title} {Structured interactions explain the absence of keystone species in synthetic microcosms},\ }\href@noop {} {\bibfield  {journal} {\bibinfo  {journal} {The ISME Journal}\ }\textbf {\bibinfo {volume} {19}},\ \bibinfo {pages} {wraf211} (\bibinfo {year} {2025})}\BibitemShut {NoStop}%
\bibitem [{\citenamefont {Grilli}(2020)}]{grilli2020macroecological}%
  \BibitemOpen
  \bibfield  {author} {\bibinfo {author} {\bibfnamefont {J.}~\bibnamefont {Grilli}},\ }\bibfield  {title} {\bibinfo {title} {Macroecological laws describe variation and diversity in microbial communities},\ }\href@noop {} {\bibfield  {journal} {\bibinfo  {journal} {Nature communications}\ }\textbf {\bibinfo {volume} {11}},\ \bibinfo {pages} {4743} (\bibinfo {year} {2020})}\BibitemShut {NoStop}%
\bibitem [{\citenamefont {Lee}\ \emph {et~al.}(2025)\citenamefont {Lee}, \citenamefont {Liu}, \citenamefont {Crocker}, \citenamefont {Wang}, \citenamefont {Huggins}, \citenamefont {Tikhonov}, \citenamefont {Mani},\ and\ \citenamefont {Kuehn}}]{lee2025functional}%
  \BibitemOpen
  \bibfield  {author} {\bibinfo {author} {\bibfnamefont {K.~K.}\ \bibnamefont {Lee}}, \bibinfo {author} {\bibfnamefont {S.}~\bibnamefont {Liu}}, \bibinfo {author} {\bibfnamefont {K.}~\bibnamefont {Crocker}}, \bibinfo {author} {\bibfnamefont {J.}~\bibnamefont {Wang}}, \bibinfo {author} {\bibfnamefont {D.~R.}\ \bibnamefont {Huggins}}, \bibinfo {author} {\bibfnamefont {M.}~\bibnamefont {Tikhonov}}, \bibinfo {author} {\bibfnamefont {M.}~\bibnamefont {Mani}},\ and\ \bibinfo {author} {\bibfnamefont {S.}~\bibnamefont {Kuehn}},\ }\bibfield  {title} {\bibinfo {title} {Functional regimes define soil microbiome response to environmental change},\ }\href@noop {} {\bibfield  {journal} {\bibinfo  {journal} {Nature}\ }\textbf {\bibinfo {volume} {644}},\ \bibinfo {pages} {1028} (\bibinfo {year} {2025})}\BibitemShut {NoStop}%
\bibitem [{\citenamefont {Begier}\ and\ \citenamefont {Hamdan}(1971)}]{begier1971correlation}%
  \BibitemOpen
  \bibfield  {author} {\bibinfo {author} {\bibfnamefont {M.~H.}\ \bibnamefont {Begier}}\ and\ \bibinfo {author} {\bibfnamefont {M.}~\bibnamefont {Hamdan}},\ }\bibfield  {title} {\bibinfo {title} {Correlation in a bivariate normal distribution with truncation in both variables},\ }\href@noop {} {\bibfield  {journal} {\bibinfo  {journal} {Australian Journal of Statistics}\ }\textbf {\bibinfo {volume} {13}},\ \bibinfo {pages} {77} (\bibinfo {year} {1971})}\BibitemShut {NoStop}%
\bibitem [{\citenamefont {Rosenbaum}(1961)}]{rosenbaum1961moments}%
  \BibitemOpen
  \bibfield  {author} {\bibinfo {author} {\bibfnamefont {S.}~\bibnamefont {Rosenbaum}},\ }\bibfield  {title} {\bibinfo {title} {Moments of a truncated bivariate normal distribution},\ }\href@noop {} {\bibfield  {journal} {\bibinfo  {journal} {Journal of the Royal Statistical Society Series B: Statistical Methodology}\ }\textbf {\bibinfo {volume} {23}},\ \bibinfo {pages} {405} (\bibinfo {year} {1961})}\BibitemShut {NoStop}%
\bibitem [{\citenamefont {Song}\ \emph {et~al.}(2019)\citenamefont {Song}, \citenamefont {Barab{\'a}s},\ and\ \citenamefont {Saavedra}}]{song2019consequences}%
  \BibitemOpen
  \bibfield  {author} {\bibinfo {author} {\bibfnamefont {C.}~\bibnamefont {Song}}, \bibinfo {author} {\bibfnamefont {G.}~\bibnamefont {Barab{\'a}s}},\ and\ \bibinfo {author} {\bibfnamefont {S.}~\bibnamefont {Saavedra}},\ }\bibfield  {title} {\bibinfo {title} {On the consequences of the interdependence of stabilizing and equalizing mechanisms},\ }\href@noop {} {\bibfield  {journal} {\bibinfo  {journal} {The American Naturalist}\ }\textbf {\bibinfo {volume} {194}},\ \bibinfo {pages} {627} (\bibinfo {year} {2019})}\BibitemShut {NoStop}%
\end{thebibliography}%

\clearpage
\begin{widetext}

\def\s{\sigma}
\def\g{\gamma}    
\def\a{\alpha}   
\def\b{\beta} 
\def\s{\sigma}    
\def \k{\kappa}  
\def \e{\epsilon}   
\def \r{\rho} 
\def \ve{\varepsilon}
\def \th{\vec{\theta}}   
\def \d{\delta} 
\def \k{\kappa}    
\def \l{\lambda} 
\def \z{\zeta}
\def \x{\xi}      
\def \n{{\tmu^{\mathrm{eff}}}}
 
\def \O{\Omega}   
\def \S{\Sigma} 
\def \G{\Gamma}   
\def \D{} 
\def \Lam{\Lambda} 
\def \h{\hbar}   

\def \E{\rm{E}} 
\def \Var{\rm{Var}} 
\def \Cov{\rm{Cov}} 
 
\def \f{\frac} 
\def \del{\partial}    
 
\def \hf{\tfrac{1}{2}} 
\def \HF{\dfrac{1}{2}}  
\def \HQ{\dfrac{1}{4}}
 
\def \ord{\mathcal{O}} 
\def \ra{\rightarrow} 
\def \wo{\setminus} 
\def \>{\rangle} 
\def \<{\langle} 
\def\dg{^\dagger} 
\def\lba{\left(}    
\def\rba{\right)} 
\def\lbc{\left[} 
\def\rbc{\right]} 
\def\lbb{\left\{} 
\def\rbb{\right\}} 
\def \bra{\langle} 
\def \ket{\rangle} 
\def\ord{\mathcal{O}} 
\newcommand{\tmu}{\tilde{\mu}}
\newcommand{\tmueff}{\tilde{\mu}^{\mathrm{eff}}}
\begin{widetext}

\appendix

\captionsetup{font=normalsize}
\renewcommand{\theequation}{S\arabic{equation}}
\renewcommand{\thefigure}{S\arabic{figure}}
\setcounter{figure}{0}
\setcounter{equation}{0}
\setcounter{section}{0}
\renewcommand{\thesection}{\Alph{section}}

\makeatother
{\hskip 150pt\textbf{SUPPLEMENTARY INFORMATION}}

\section*{Appendix A: Setup of highly-diverse GLV model with strain structure} 
\label{app:glvSetup}
We wish to understand the dynamics of highly-diverse, densely interacting communities (where the number of species $S \rightarrow \infty$) with fine-scale strain structure, i.e., where species are comprised of multiple strains. For simplicity, we first consider the case where each species has two strains $\alpha$ and $\beta$. We will derive the effective dynamics for a typical closely-related strain pair embedded in a highly-diverse community using the cavity method.  We will see that the dynamics describing a closely-related strain pair can be coarse-grained to a two-dimensional GLV model, where the impact of the community manifests as changes to effective model parameters such as growth rates, self-inhibition strengths and interstrain interactions. We derive these dynamics for communities governed by these GLV dynamics, but our results and conclusions are general and should apply to other niche models such as CRMs as well (albeit with greater mathematical tedium).

We work with a community following GLV dynamics described by the following set of ordinary differential equations:

\be
\frac{d N_{i\a}}{dt} = N_{i\a} \left(K_{i\a} - N_{i\a} - A_{i\a i\b} N_{i\b} - \sum_{\substack{j=1\\j \neq i}}^{S} A_{i\a j\a} N_{j\a} - \sum_{\substack{j=1\\j \neq i}}^{S} A_{i\a j\b} N_{j\b} \right) ,
\ee
where the intrinsic growth rates $K_{i\a}$ for strain $\a$ of species $i$, and interaction strengths $A_{i\a j\b}$ --- representing the inhibition from strain $\b$ of species $j$ on strain $\a$ of species $i$ ---  are random variables defined according to the following ensemble: 

\begin{align}
A_{i\a j \a} &= \frac{\mu}{S} + \s a_{ij} + \l x_{ij\a}\\
A_{i\a j \b} &= \frac{\mu}{S} + \s a_{ij} - \l x_{ij\a}\\
A_{i\b j \a} &= \frac{\mu}{S} + \s a_{ij} + \l x_{ij\b}\\
A_{i\b j \b} &= \frac{\mu}{S} + \s a_{ij} - \l x_{ij\b}\\
A_{i\b i \a} &= \tilde{\mu} + \l y_{i\a\b}\\
A_{i\a i \b} &= \tilde{\mu} - \l y_{i\a\b}\\
K_{i\a} &= K + \s_K z_{i\a}
\end{align}
where $a_{ij}$, $x_{ij\a}$, $y_{i\a\b}$ and $z_{i\a}$ are zero-mean random variables with the following statistics:

\begin{align}
\bra a_{ij} a_{kl} \ket &= {\d_{ik}\d_{jl} \over S} + {\r d_{jk}\d_{il}\over S} \label{eq:a_corr}\\
\bra x_{ij\a} x_{kl\b} \ket &= {\d_{ik}\d_{jl}\d_{\a\b} \over S} + {\r \d_{jk}\d_{il}\over S}\label{eq:xcorr}\\
\bra y_{i\a\b} y_{j\g\d} \ket &= {\d_{ij}\d_{\a\g}\d_{\b\d}}\\
\bra z_{i\a} z_{j\b} \ket &= \d_{ij} \d_{\a\b}.
\end{align}

This ensemble assumes that two closely-related strains $j\a$ and $j\b$ of species $j$ impact a distant strain (say $i\a$) with nearly the same randomly-chosen interaction strength. However, the impacts of $j\a$ and $j\b$ on $i\a$ are not identical. Instead they differ by a random modifier of strength proportional to $\l$. The parameter $\l$ thus quantifies the interaction dissimilarity between closely-related strains. $\l =0$ implies strains of the same species are identical; as $\l$ increases they become increasingly different. The parameter $\s$ similarly indicates the interaction dissimilarity between two distinct species $i$ and $j$.

We are interested in the collective dynamics of all strains in the community at steady-state.

\be
{d N_{i\a\wo 0} \over dt} = N_{i\a\wo 0} \left(K_{i\a} - N_{i\a\wo 0} - \sum_{j=1}^{S} A_{i\a j\a} N_{j\a\wo 0} - \sum_{j=1}^{S} A_{i\a j\b} N_{j\b\wo 0} \right) = 0.
\ee

\section*{Appendix B: Cavity derivation of effective two-strain model}
\label{app:cavityDerivation}
To derive effective equations for a typical pair of closely-related strains $i\a$ and $i\b$, we will use the cavity method. We start by assuming that a community with $S$ species --- with two strains per species --- is at steady state. To this community, we add a new cavity species $0$. The invading cavity species also consists of a pair of closely-related strains $0\a$ and $0\b$. After the two strains $0\a$ and $0\b$ of the cavity species invade, the dynamics of all other strains will change as:

\be
{d N_{i\a} \over dt} = N_{i\a} \left(K_{i\a} - \sum_{j=1}^{S} A_{i\a j\a} N_{j\a} - \sum_{j=1}^{S} A_{i\a j\b} N_{j\b} - A_{i\a 0\a} N_{0\a} - A_{i\a 0\b} N_{0\b} \right).
\ee

From the definition of our ensemble, we see that both $0\a$ and $0\b$ act as small perturbations to the growth rate $\d K_{j\a}$ of all other strains $j\a$. Thus, linear response is justified and we can write the perturbed abundance trajectories of strains $j\a$ and $j\b$ in the community as

\begin{align}
N_{j\a} &\approx N_{j\a\wo 0} - \sum_k \nu_{j\a k\a} \left(A_{k\a0\a} N_{0\a} + A_{k\a 0\b} N_{0\b}\right) - \sum_k \nu_{j\a k\b} \left(A_{k\b0\a} N_{0\a} + A_{k\b 0\b} N_{0\b}\right),\\
N_{j\b} &\approx N_{j\b\wo 0} - \sum_k \nu_{j\b k\a} \left(A_{k\a0\a} N_{0\a} + A_{k\a 0\b} N_{0\b}\right) - \sum_k \nu_{j\b k\b} \left(A_{k\b0\a} N_{0\a} + A_{k\b 0\b} N_{0\b}\right),
\end{align}
where $\nu_{j\a k\b} = \f{\d N_{j\a}}{\d K_{k\b}}$ is a susceptibility. Substituting these into the dynamics for the cavity strains $0\a$ and $0\b$, we get

\begin{align}
{dN_{0\a} \over dt} &= N_{0\a} \Big[ K_{0\a} - N_{0\a} - \sum_{j\neq 0} A_{0\a j\a} N_{j\a} - \sum_{j\neq 0} A_{0\a j\b} N_{j\b} - A_{0\a 0\b} N_{0\b}\Big],
\label{eq:lv_unsimplified_0alpha}\\
{dN_{0\b} \over dt} &= N_{0\b} \Big[ K_{0\b} - N_{0\b} - \sum_{j\neq 0} A_{0\b j\a} N_{j\a} - \sum_{j\neq 0} A_{0\b j\b} N_{j\b} - A_{0\b 0\a} N_{0\a}\Big].
\label{eq:lv_unsimplified_0beta}
\end{align}

The terms in the sum over $S$ species involve the sum of a large number of random variables. We will now analyze each of these terms in high-dimensional limit $S\rightarrow \infty$ separately. The set of terms which multiply ${\mu \over S}$, $\rho \sigma a_{pj}$ and $\s \sqrt{1-\rho^2} a_{mj}$ are as follows

\begin{align}
0 = N_{0\a} \Big[ & K_{0\a} - N_{0\a} - A_{0\a 0\b} N_{0\b} - \sum_{j\neq 0} A_{0\a j\a} N_{j\a\wo 0} - \sum_{j\neq 0} A_{0\a j\b} N_{j\b\wo 0} \nn\\
  & + \underbrace{\sum_{jk} A_{0\a j\a} \nu_{j\a k\a} A_{k\a 0\a}}_{\circled{\tiny{1}}} N_{0\a} + \underbrace{\sum_{jk} A_{0\a j\a} \nu_{j\a k\a} A_{k\a 0\b}}_{\circled{\tiny{2}}} N_{0\b}\nn\\
  & + \underbrace{\sum_{jk} A_{0\a j\a} \nu_{j\a k\b} A_{k\b 0\a}}_{\circled{\tiny{3}}} N_{0\a} + \underbrace{\sum_{jk} A_{0\a j\a} \nu_{j\a k\b} A_{k\b 0\b}}_{\circled{\tiny{4}}} N_{0\b}\nn\\
  & + \underbrace{\sum_{jk} A_{0\a j\b} \nu_{j\b k\a} A_{k\a 0\a}}_{\circled{\tiny{5}}} N_{0\a} + \underbrace{\sum_{jk} A_{0\a j\b} \nu_{j\b k\a} A_{k\a 0\b}}_{\circled{\tiny{6}}} N_{0\b}\nn\\
  & + \underbrace{\sum_{jk} A_{0\a j\b} \nu_{j\b k\b} A_{k\b 0\a}}_{\circled{\tiny{7}}} N_{0\a} + \underbrace{\sum_{jk} A_{0\a j\b} \nu_{j\b k\b} A_{k\b 0\b}}_{\circled{\tiny{8}}} N_{0\b}\Big].
  \label{eq:n0selffeedback}
\end{align}

\vskip 15pt

\textbf{Term \circled{\tiny{1}}:}
\be
\sum_{jk} A_{0\a j\a} \nu_{j\a k\a} A_{k\a 0\a} = \sum_{jk} \left({\mu\over S} + \s a_{0j} + \l x_{0j\a} \right) \nu_{j\a k\a} \left( {\mu\over S} + \s a_{k0} + \l x_{k0\a} \right)
\ee

Considering the expectation of this random variable, we get:

\be
\sum_{jk} \left\bra {\cancelto{}{\mu^2 \over S^2}} + \cancelto{}{{\mu\over S} \s a_{k0}} + \cancelto{}{{\mu\over S} \l x_{k0\a}} + \cancelto{}{{\mu\over S} \s a_{0j}} + \cancelto{}{{\mu\over S} \l x_{0j\a}} + \s^2 a_{0j} a_{k0} + \cancelto{}{\s \l a_{0j} x_{k0\a}} + \cancelto{}{\s\l a_{k0}x_{0j\a}} + \l^2 x_{0j\a} x_{k0\a}  \right\ket
\ee
where each of the cancelled terms are subleading in $\ord(S^{-1/2})$ and vanish as $S\rightarrow \infty$.

\begin{align}
\implies \circled{\tiny{1}} &=  \s^2 \left\bra \sum_{jk} a_{0j} a_{k0} \nu_{j\a k\a} \right\ket + \l^2 \left\bra \sum_{jk} x_{0j\a}x_{k0\a} \nu_{j\a k\a}\right\ket,\\
&= \s^2 S \left\bra a_{0j} a_{k0} \right\ket \left\bra \nu_{j\a k\a} \right\ket + \l^2 S \left\bra x_{0j\a}x_{k0\a}\right\ket \left\bra \nu_{j\a k\a}\right\ket,\\
&= \s^2 \r \left\bra \nu_{\a \a} \right\ket + \l^2 \r \left\bra \nu_{\a \a} \right\ket,
\label{eq:circt1}
\end{align}
where $\bra\nu_{\a\a}\ket$ is the mean of the $\a\a$ terms of the community-wide susceptibility matrix, defined as

\be
\left\bra \nu_{\a\a} \right\ket = \sum_{ij} {\del N_{i\a} \over \del K_{j\a}}. 
\label{eq:self_susceptibility}
\ee

Similarly, we can define the cross-susceptibility as 

\be
\left\bra \nu_{\a\b} \right\ket = \sum_{ij} {\del N_{i\a} \over \del K_{j\b}}. 
\label{eq:cross_susceptibility}
\ee

Following similar steps, we can compute all the other circled terms in eq (\ref{eq:n0selffeedback}).

\textbf{Term \circled{\tiny{2}}:}

\begin{align}
\sum_{jk} A_{0\b j\a} \nu_{j\a k\a} A_{k\a 0\b} &=  \s^2 \left\bra \sum_{jk} a_{0j} a_{k0} \nu_{j\a k\a} \right\ket - \l^2 \left\bra \sum_{jk} x_{0j\a}x_{k0\a} \nu_{j\a k\a}\right\ket,\\
&= \s^2 S \left\bra a_{0j} a_{k0} \right\ket \left\bra \nu_{j\a k\a} \right\ket - \l^2 S \left\bra x_{0j\a}x_{k0\a}\right\ket \left\bra \nu_{j\a k\a}\right\ket,\\
&= \s^2 \r \left\bra \nu_{\a \a} \right\ket - \l^2 \r \left\bra \nu_{\a \a} \right\ket,
\label{eq:circt2}
\end{align}

\textbf{Term \circled{\tiny{3}}:}

\begin{align}
\sum_{jk} A_{0\b j\a} \nu_{j\a k\b} A_{k\b 0\a} &=  \s^2 \left\bra \sum_{jk} a_{0j} a_{k0} \nu_{j\a k\b} \right\ket + \l^2 \left\bra \sum_{jk} x_{0j\a}x_{k0\b} \nu_{j\a k\b}\right\ket,\\
&= \s^2 S \left\bra a_{0j} a_{k0} \right\ket \left\bra \nu_{j\a k\b} \right\ket + \l^2 S \left\bra x_{0j\a}x_{k0\b}\right\ket \left\bra \nu_{j\a k\b}\right\ket,\\
&= \s^2 \r \left\bra \nu_{\a \b} \right\ket + \l^2 \r \left\bra \nu_{\a \b} \right\ket,
\label{eq:circt3}
\end{align}

\textbf{Term \circled{\tiny{4}}:}

\begin{align}
\sum_{jk} A_{0\b j\a} \nu_{j\a k\b} A_{k\b 0\b} &=  \s^2 \left\bra \sum_{jk} a_{0j} a_{k0} \nu_{j\a k\b} \right\ket - \l^2 \left\bra \sum_{jk} x_{0j\a}x_{k0\b} \nu_{j\a k\b}\right\ket,\\
&= \s^2 S \left\bra a_{0j} a_{k0} \right\ket \left\bra \nu_{j\a k\b} \right\ket - \l^2 S \left\bra x_{0j\a}x_{k0\b}\right\ket \left\bra \nu_{j\a k\b}\right\ket,\\
&= \s^2 \r \left\bra \nu_{\a \b} \right\ket - \l^2 \r \left\bra \nu_{\a \b} \right\ket,
\label{eq:circt4}
\end{align}

\textbf{Term \circled{\tiny{5}}:}

\begin{align}
\sum_{jk} A_{0\b j\b} \nu_{j\b k\a} A_{k\a 0\a} &=  \s^2 \left\bra \sum_{jk} a_{0j} a_{k0} \nu_{j\b k\a} \right\ket - \l^2 \left\bra \sum_{jk} x_{0j\a}x_{k0\a} \nu_{j\b k\a}\right\ket,\\
&= \s^2 S \left\bra a_{0j} a_{k0} \right\ket \left\bra \nu_{j\b k\a} \right\ket - \l^2 S \left\bra x_{0j\a}x_{k0\a}\right\ket \left\bra \nu_{j\b k\a}\right\ket,\\
&= \s^2 \r \left\bra \nu_{\b \a} \right\ket - \l^2 \r \left\bra \nu_{\b \a} \right\ket,
\label{eq:circt5}
\end{align}

\textbf{Term \circled{\tiny{6}}:}

\begin{align}
\sum_{jk} A_{0\b j\b} \nu_{j\b k\a} A_{k\a 0\b} &=  \s^2 \left\bra \sum_{jk} a_{0j} a_{k0} \nu_{j\b k\a} \right\ket + \l^2 \left\bra \sum_{jk} x_{0j\a}x_{k0\a} \nu_{j\b k\a}\right\ket,\\
&= \s^2 S \left\bra a_{0j} a_{k0} \right\ket \left\bra \nu_{j\b k\a} \right\ket + \l^2 S \left\bra x_{0j\a}x_{k0\a}\right\ket \left\bra \nu_{j\b k\a}\right\ket,\\
&= \s^2 \r \left\bra \nu_{\b \a} \right\ket + \l^2 \r \left\bra \nu_{\b \a} \right\ket,
\label{eq:circt6}
\end{align}

\textbf{Term \circled{\tiny{7}}:}

\begin{align}
\sum_{jk} A_{0\b j\b} \nu_{j\b k\b} A_{k\b 0\a} &=  \s^2 \left\bra \sum_{jk} a_{0j} a_{k0} \nu_{j\b k\b} \right\ket - \l^2 \left\bra \sum_{jk} x_{0j\a}x_{k0\b} \nu_{j\b k\b}\right\ket,\\
&= \s^2 S \left\bra a_{0j} a_{k0} \right\ket \left\bra \nu_{j\b k\b} \right\ket - \l^2 S \left\bra x_{0j\a}x_{k0\b}\right\ket \left\bra \nu_{j\b k\b}\right\ket,\\
&= \s^2 \r \left\bra \nu_{\b \b} \right\ket - \l^2 \r \left\bra \nu_{\b \b} \right\ket,
\label{eq:circt7}
\end{align}

\textbf{Term \circled{\tiny{8}}:}

\begin{align}
\sum_{jk} A_{0\b j\b} \nu_{j\b k\b} A_{k\b 0\b} &=  \s^2 \left\bra \sum_{jk} a_{0j} a_{k0} \nu_{j\b k\b} \right\ket + \l^2 \left\bra \sum_{jk} x_{0j\a}x_{k0\b} \nu_{j\b k\b}\right\ket,\\
&= \s^2 S \left\bra a_{0j} a_{k0} \right\ket \left\bra \nu_{j\b k\b} \right\ket + \l^2 S \left\bra x_{0j\a}x_{k0\b}\right\ket \left\bra \nu_{j\b k\b}\right\ket,\\
&= \s^2 \r \left\bra \nu_{\b \b} \right\ket + \l^2 \r \left\bra \nu_{\b \b} \right\ket,
\label{eq:circt8}
\end{align}

Similarly, the terms involving $N_{j\a\wo 0}$ and $N_{j\b\wo 0}$ can be simplified by recognizing that they represent the sum of many weakly correlated random variables and in the large $S$ limit become Gaussian fields. Thus we only need to compute their first and second moments, which we will calculate below. Similarly, the closely-related cavity strain $0\b$ has dynamics as follows

\begin{align}
0 = N_{0\b} \Big[ & K_{0\b} - N_{0\b} - A_{0\b 0\a} N_{0\a} - \sum_{j\neq 0} A_{0\b j\a} N_{j\a\wo 0} - \sum_{j\neq 0} A_{0\b j\b} N_{j\b\wo 0} \nn\\
  & + \underbrace{\sum_{jk} A_{0\b j\a} \nu_{j\a k\a} A_{k\a 0\a}}_{\circled{\tiny{1}}}N_{0\a} + \underbrace{\sum_{jk} A_{0\b j\a} \nu_{j\a k\a} A_{k\a 0\b}}_{\circled{\tiny{2}}}N_{0\b}\nn\\
  & + \underbrace{\sum_{jk} A_{0\b j\a} \nu_{j\a k\b} A_{k\b 0\a}}_{\circled{\tiny{3}}}N_{0\a} + \underbrace{\sum_{jk} A_{0\b j\a} \nu_{j\a k\b} A_{k\b 0\b}}_{\circled{\tiny{4}}}N_{0\b}\nn\\
  & + \underbrace{\sum_{jk} A_{0\b j\b} \nu_{j\b k\a} A_{k\a 0\a}}_{\circled{\tiny{5}}}N_{0\a} + \underbrace{\sum_{jk} A_{0\b j\b} \nu_{j\b k\a} A_{k\a 0\b}}_{\circled{\tiny{6}}}N_{0\b}\nn\\
  & + \underbrace{\sum_{jk} A_{0\b j\b} \nu_{j\b k\b} A_{k\b 0\a}}_{\circled{\tiny{7}}}N_{0\a} + \underbrace{\sum_{jk} A_{0\b j\b} \nu_{j\b k\b} A_{k\b 0\b}}_{\circled{\tiny{8}}}N_{0\b}\Big].
\end{align}

Both $K_{0\a}$ and $K_{0\b}$ are values centred around a mean $K$ and deviation $\d K_i$ with statistics $\langle\d K_i\rangle = 0$ and $\langle\d K_i^2\rangle = \s_K^2$.
The circled terms are almost the same as for the $0\a$ strain, except they all start with $A_{0\b}$ instead of $A_{0\a}$. This doesn't change what the terms evaluate to, because the terms involving $\r$ in Eq (\ref{eq:xcorr}) are independent of $\a$ and $\b$. Thus we can substitute the circled terms from previous calculations in Eqns (\ref{eq:circt1})--(\ref{eq:circt8}). One crucial behaviour that emerges due to our model setup is that although there are strain level differences within a species (modelled using $\l x_{0j\a}$ or $\l x_{0j\b}$), at the species level they encounter the same inter-species interactions i.e., $\s_{ij}$ remains strain independent. This leads to the emergence of correlational structures between the fluctuating parts of the effective growth rates of the two strains in addition to their means and variances. Calculating this abundance correlation of conspecific strains is crucial to understand their joint distributions. We can calculate the first, second, and cross-moments for the strains. First, we define for each of the strains

\begin{align}
S_\alpha &= \sum_j A_{0\alpha\, j\alpha} N_{j\alpha \wo 0} + \sum_j A_{0\alpha\, j\beta} N_{j\beta \wo 0}, \\
S_\beta  &= \sum_j A_{0\beta\, j\alpha} N_{j\alpha \wo 0} + \sum_j A_{0\beta\, j\beta} N_{j\beta \wo 0}.
\end{align}

Using the interaction coefficients defined for the ensemble (S2-S7), we can write

\begin{equation}
    \langle S_\a \rangle = \left \langle \sum_j\left(\frac{\mu}{S} + \s a_{0j} + \l x_{0j\a}\right) N_{j \a \wo0} + \sum_j\left(\frac{\mu}{S} + \s a_{0j} - \l x_{0j\a}\right) N_{j \b \wo 0} \right \rangle.
\end{equation}
Expanding out we get,

\begin{equation}
    \left \langle \sum_j \frac{\mu}{S} N_{j \a \wo 0} + \cancelto{}{\sum_j \s a_{0j} N_{j \a \wo 0}} + \cancelto {}{\sum_j \l x_{0j\a} N_{j \a \wo 0}} + \sum_j \frac{\mu}{S} N_{j \b \wo 0} + \cancelto{}{\sum_j \s a_{0j} N_{j \b \wo 0}} - \cancelto {}{\sum_j \l x_{0j\a} N_{j \b \wo 0}} \right \rangle
\end{equation}
  which can be reduced. Note that the exact same reduction can be performed for $S_\b$. From above, we can find that

\begin{equation}
    \langle S_\a \rangle = \langle S_\b \rangle = \mu (\langle N_\a + N_\b\rangle) 
\end{equation}

where $\langle N_\a \rangle$ and $\langle N_\b \rangle$ are the means of all the abundances of the $\a$ and $\b$ strains of the non-cavity species. Similarly we can calculate the second moments of $S_\a$ 

\begin{align}
    \langle S_\a^2 \rangle &=  \left \langle \sum_{jk}\left(\frac{\mu}{S} + \s a_{0j} + \l x_{0j\a}\right)\left(\frac{\mu}{S} + \s a_{0k} + \l x_{0k\a}\right) N_{j \a \wo 0} N_{k \a \wo 0} \right \rangle \\ \nn
    &+ \left \langle \sum_{jk}\left(\frac{\mu}{S} + \s a_{0j} - \l x_{0j\a}\right)\left(\frac{\mu}{S} + \s a_{0k} - \l x_{0k\a}\right) N_{j \b \wo 0} N_{k \b \wo 0} \right \rangle \\ \nn
    &+ 2 \left \langle \sum_{jk}\left(\frac{\mu}{S} + \s a_{0j} + \l x_{0j\a}\right)\left(\frac{\mu}{S} + \s a_{0k} - \l x_{0k\b}\right) N_{j \a \wo 0} N_{k \b \wo 0} \right \rangle
\end{align}

We can expand and remove all the subleading terms that go to 0. Following expansion, we get

\begin{align}
    \langle S_\a ^2\rangle & = \sum_{jk} \left \langle \frac{\mu^2}{S^2} N_{j \a \wo 0} N_{k \a \wo 0} \right \rangle + \sum_{jk} \left \langle \frac{\s^2}{S} a_{0j}a_{0k} N_{j \a \wo 0} N_{ k\a  \wo 0}\right \rangle 
    + \sum_{jk} \left \langle \frac{\l^2}{S} x_{0j\a} x_{0k\a} N_{j \a \wo 0} N_{ k\a  \wo 0}\right \rangle \\
    &+ \sum_{jk} \left \langle \frac{\mu^2}{S^2} N_{j \b \wo 0} N_{k \b \wo 0} \right \rangle + \sum_{jk} \left \langle \frac{\s^2}{S} a_{0j}a_{0k} N_{j \b \wo 0} N_{ k\b  \wo 0}\right \rangle 
    + \sum_{jk} \left \langle \frac{\l^2}{S} x_{0j\b} x_{0k\b} N_{j \b \wo 0} N_{ k\b  \wo 0}\right \rangle \\ 
    &+ 2 \left[ \sum_{jk} \left \langle \frac{\mu^2}{S^2} N_{j \a \wo 0} N_{k \b \wo 0} \right \rangle + \sum_{jk} \left \langle \frac{\s^2}{S} a_{0j}a_{0k} N_{j \a \wo 0} N_{ k \b  \wo 0}\right \rangle 
    + \sum_{jk} \left \langle \frac{\l^2}{S} x_{0j\a} x_{0k\b} N_{j \a \wo 0} N_{ k\b  \wo 0}\right \rangle \right].
\end{align}

Summing all the quantities, we are left with,

\begin{align}
    \langle S_\a ^2\rangle & = \mu ^2\left(\langle N_\a^2 \rangle + \langle N_\b^2 \rangle \right) + \s^2 \left(  \langle N_\a^2 \rangle + \langle N_\b^2 \rangle \right) +\l^2 \left( \langle N_\a^2 \rangle + \langle N_\b^2 \rangle \right) + 2 \left(\mu^2 \langle N_\a N_\b \rangle + \s^2 \langle N_\a N_\b \rangle - \l^2 \langle N_\a N_\b \rangle  \right)
\end{align}

The variance of the fluctuating component in strain $\b$ can be obtained using the same method and we find that by rearranging the terms and grouping them, we can rewrite it as 

\begin{align}
    \langle S_\a^2 \rangle &= \langle S_\b^2 \rangle = \mu^2 \langle (N_\a + N_\b)^2  \rangle +\s^2 \langle (N_\a + N_\b)^2  \rangle + \l^2 \langle (N_\a - N_\b)^2  \rangle
\end{align}

We can also obtain the variances of the two quantities by subtracting the first moments of the expressions

\begin{align}
    Var(S_\a) = Var(S_\b) = \s^2 \langle (N_\a + N_\b)^2  \rangle + \l^2 \langle (N_\a - N_\b)^2  \rangle
\end{align}

\vspace{1em}
\noindent

We can introduce for simplicity
\begin{equation}
M_j = N_{j\alpha \wo 0} + N_{j\beta \wo 0}, \qquad 
D_j = N_{j\alpha \wo 0} - N_{j\beta \wo 0}.
\end{equation}






where
\begin{align}
\langle M^2 \rangle &= \langle N_\alpha^2 \rangle + \langle N_\beta^2 \rangle + 2\langle N_\alpha N_\beta \rangle \\
\langle D^2 \rangle &= \langle N_\alpha^2 \rangle + \langle N_\beta^2 \rangle - 2\langle N_\alpha N_\beta \rangle
\end{align}

Using this new notation, we can calculate the cross-moment of the fluctuating parts of both strains i.e. $\langle S_\a, S_\b\rangle$

\begin{align}
    \langle S_\alpha S_\beta \rangle  = \left\langle \left(\sum_j \frac{\mu}{S} M_j + \frac{\s}{\sqrt{S}}\sum_j a_{0j}M_j+\frac{\l}{\sqrt{S}}\sum_j x_{0j\a}D_j\right)\left(\sum_k \frac{\mu}{S} M_k + \frac{\s}{\sqrt{S}}\sum_k a_{0k}M_k+\frac{\l}{\sqrt{S}}\sum_k x_{0k\a}D_k\right) \right\rangle
\end{align}

which, following taking the product is equal to,

\begin{align}
\langle S_\alpha S_\beta \rangle  = \mu^2 \sum_{jk} \langle M_j M_k \rangle + \s^2 \sum_{jk} \langle a_{0j} a_{0k} \rangle \langle M_j M_k \rangle
 - \lambda^2 \sum_{jk} \langle x_{0j\alpha} x_{0k\beta} \rangle \langle D_j D_k \rangle.
\end{align}


Since $x_{0j\alpha}$ and $x_{0j\beta}$ are independent, the third term vanishes. Further we can calculate the covariance of the fluctuating components of the two strains using the formula $\mathrm{Cov}(S_\a,S_\b) = \langle S_\alpha S_\beta \rangle - \langle S_\a \rangle \langle S_\b \rangle$. Using the formula, we can obtain
\begin{equation}
\mathrm{Cov}(S_\alpha,S_\beta)
= \frac{\s^2}{S} \sum_j \langle M_j^2 \rangle
= \s^2\Big( \langle N_\alpha^2\rangle + \langle N_\beta^2\rangle + 2\langle N_\alpha N_\beta\rangle \Big),
\label{eq:cov_fluct}
\end{equation}
which is strictly positive.


The variances of the two fluctuating parts are
\begin{equation}
\mathrm{Var}(S_\alpha)=\mathrm{Var}(S_\beta)
=\s^2\langle M^2\rangle+\lambda^2\langle D^2\rangle.
\label{eq: var_fluct}
\end{equation}
Thus, the correlation between $S_\alpha$ and $S_\beta$ (from \ref{eq:cov_fluct} and \ref{eq: var_fluct}) is
\begin{equation}
\rho_{\mathrm{int}} = \mathrm{Corr}(S_\alpha,S_\beta)
= \frac{\mathrm{Cov}(S_\a,S_\b)}{\sqrt{\mathrm{Var} (S_\a) \mathrm{Var} (S_\b)}}  =  \frac{\s^2\left \langle \left(N_\alpha + N_\beta \right)^2 \right \rangle}
{\s^2\left \langle \left(N_\alpha + N_\beta \right)^2 \right \rangle + \l^2 \left \langle \left(N_\alpha - N_\beta \right)^2 \right \rangle},
\label{eq:rho_K}
\end{equation}
where
\begin{align}
\langle M^2\rangle &= \langle N_\alpha^2\rangle+\langle N_\beta^2\rangle+2\langle N_\alpha N_\beta\rangle,\\
\langle D^2\rangle &= \langle N_\alpha^2\rangle+\langle N_\beta^2\rangle-2\langle N_\alpha N_\beta\rangle.
\end{align}

As $\lambda$ increases, the antisymmetric (difference) effects dominates, and the correlation between the two strains decreases from near $+1$ toward 0. This means that nearly-identical strains ($\l \approx0$) feel very similar effective environments and thus have correlated growth rates (i.e., effects from other species in the community) This effect is further compounded by the effects of the environment on the growth rates of the two strains. On one hand, the shared variance in interaction effects at the species level ($\s^2$) keeps the abundances of the two strain coupled ($\rho_{\mathrm{int}} \approx 1$, resulting in positive correlations between strains) while on the other hand strain interaction dissimilarity ($\l$) and environmental noise serves to decouple the abundance fluctuations. Note that there is an additional $\s_K^2$ terms that emerges from independent fluctuations around the mean growth rate $K$ which are drawn from a distribution $\mathcal{N}$$(0,\s_K^2)$.

As a result, in the large $S$ limit, the mean-field dynamics for the $0\a$ and $0\b$ strains become

\begin{align} 
0 = N_{0 \a} \Big[ & \underbrace{K - \mu (\bra N_\a \ket + \bra N_\b \ket) + \sqrt{\s^2\left \langle \left(N_\alpha + N_\beta \right)^2 \right \rangle + \l^2 \left \langle \left(N_\alpha - N_\beta \right)^2 \right \rangle} z^{(\mathrm{int})}_{0\a} + \sqrt{\s^2_K} z_{0\a}}_{\text{effective growth rate, } K^{\text{eff}}_{0\a}} \nn \\ 
  -&\underbrace{\Big( 1 - (\s^2 + \l^2) \r \left( \bra \nu_{\a\a}\ket + \bra \nu_{\a\b} \ket \right) - (\s^2 -\l^2)\r (\bra \nu_{\b\b} \ket + \bra \nu_{\b\a} \ket ) \Big)}_{\text{effective self-inhibition, } A^{\text{eff}}_{0\a0\a}}  N_{0\a} \nn \\
  -&\underbrace{\Big( \tilde{\mu} - \l y_{0\a\b} - (\s^2 + \l^2) \r ( \bra \nu_{\b\b} \ket + \bra \nu_{\b\a} \ket) - (\s^2 -\l^2) \r (\bra \nu_{\a\a} \ket + \bra \nu_{\a\b} \ket ) \Big)}_{\text{effective interaction with closely-related strain, } A^{\text{eff}}_{0\a0\b}} N_{0\b}  \Big],
\label{eq:eff0aGLV}
\end{align}
where $z^{(\mathrm{int})}_{0\a}$ and $z_{0\a}$ are independent standard normal variable withs zero mean and unit variance.

\begin{align} 
0 &= N_{0 \b} \Big[\underbrace{K - \mu (\bra N_\a \ket + \bra N_\b \ket)} \nn \\ 
&+ \underbrace{\sqrt{\s^2\left \langle \left(N_\alpha + N_\beta \right)^2 \right \rangle + \l^2 \left \langle \left(N_\alpha - N_\beta \right)^2 \right \rangle }\big(\rho_{\mathrm{int}} z^{(\mathrm{int})}_{0\a} + \sqrt{(1-\rho_{\mathrm{int}}^2)} z^{(\mathrm{int})}_{0\b}\big) + \sqrt{\s^2_K}z_{0\b}}_{\text{effective growth rate, } K^{\text{eff}}_{0\b}} \nn \\ 
  -&\underbrace{\Big( 1 - (\s^2 + \l^2) \r ( \bra \nu_{\b\b} \ket + \bra \nu_{\b\a} \ket) - (\s^2 -\l^2) \r (\bra \nu_{\a\a} \ket + \bra \nu_{\a\b} \ket ) \Big)}_{\text{effective self-inhibition, } A^{\text{eff}}_{0\b0\b}}  N_{0\b} \nn \\
  -&\underbrace{\Big( \tilde{\mu} + \l y_{0\a\b} - (\s^2 + \l^2) \r \left( \bra \nu_{\a\a}\ket + \bra \nu_{\a\b} \ket \right) - (\s^2 -\l^2)\r (\bra \nu_{\b\b} \ket + \bra \nu_{\b\a} \ket ) \Big)}_{\text{effective interaction with closely-related strain, } A^{\text{eff}}_{0\b0\a}} N_{0\a}  \Big].
\label{eq:eff0bGLV}
\end{align}

Eqns (\ref{eq:eff0aGLV}) and (\ref{eq:eff0bGLV}) describe a two-strain GLV model with effective parameters that contain the effect of the community on a prototypical closely-related strain pair $0\a$ and $0\b$. Note that $z^{(\mathrm{int})}_{0\b}, z_{0\b}$ are also independent random variables drawn from $\mathcal{N} (0,1)$. More simply, we can rewrite the above two equations as

\begin{align}
    0 &= N_{0\a}\left[(g + \s_g ~z^{(\mathrm{int})}_{0\a} + \s_K~z_{0\a}\right) - A_{0\a0\a}^\mathrm{eff}N_{0\a} - A_{0\a0\b}^\mathrm{eff}N_{0\b}] \label{eq: 2-spp-glv-sigG_alpha}  \\
    0 &= N_{0\b}\left[\left(g + \s_g ~\left(\rho_{\mathrm{int}}z^{(\mathrm{int})}_{0\a} + \sqrt{(1-\rho_{\mathrm{int}}^2)} z^{(\mathrm{int})}_{0\b} \right) + \s_K~z_{0\b} \right) - A_{0\b0\a}^\mathrm{eff}N_{0\a} - A_{0\b0\b}^\mathrm{eff}N_{0\b} \right]
    \label{eq: 2-spp-glv-sigG_beta}
\end{align}

where g refers to the mean of the effective carrying capacity equal to $K - \mu\left(\langle N_\a + N_\b \rangle \right) $ and $\s_g$ refers to the variance arising from the joint effect of all strains from the other species equal to $ \sqrt{\s^2\left \langle \left(N_\alpha + N_\beta \right)^2 \right \rangle + \l^2 \left \langle \left(N_\alpha - N_\beta \right)^2 \right \rangle}$ .

Looking at this form of the two-strain model at equilibrium, we can see the presence of an emergent correlation between the growth rates of the two strains indicate a coupling of growth rates (correlated by some strength $\rho_{\mathrm{int}}$ --- a phenomenon that is not observed in a simple 2 strain model in the absence of the community (see Appendix F below for a detailed calculation). From eqns. \ref{eq:rho_K}, \ref{eq: 2-spp-glv-sigG_alpha}, \ref{eq: 2-spp-glv-sigG_beta}, we can simply calculate the correlation between $K_\a$ and $K_\b$ as

\begin{equation}
    \mathrm{Corr}(K_\a,K_\b) = \r_K = \frac{\rho_{\mathrm{int}} \s_g^2}{\s_g^2 + \s_K^2} = \frac{\s^2\left \langle \left(N_\alpha + N_\beta \right)^2 \right \rangle}{\s^2\left \langle \left(N_\alpha + N_\beta \right)^2 \right \rangle + \l^2 \left \langle \left(N_\alpha - N_\beta \right)^2 \right \rangle + \s_K^2}
\end{equation}

More succinctly, we can write down the equations above in the form of the effective two species model as in equations \ref{eq:effModel1} and \ref{eq:effModel2} in the main text i.e.

\begin{align}
0 = \frac{dN_{\alpha}}{dt} &= N_{\alpha} \left( K_{\alpha}^{\mathrm{eff}} - A_{\alpha\alpha}^{\mathrm{eff}} N_{\alpha} - A_{\alpha\beta}^{\mathrm{eff}} N_{\beta} \right), \label{eq:supp_effModel1_eq}\\
0 = \frac{dN_{\beta}}{dt} &= N_{\beta} \left( K_{\beta}^{\mathrm{eff}} - A_{\beta\beta}^{\mathrm{eff}} N_{\beta} - A_{\beta\alpha}^{\mathrm{eff}} N_{\alpha} \right).
\end{align}

We further assume for simplicity that $\langle \nu_{\a\a} \rangle$ = $\langle \nu_{\b\b} \rangle$ and $\langle \nu_{\a\b} \rangle$ = $\langle \nu_{\b\a} \rangle$, and thus we can define $\nu = \langle \nu_{\a\a} \rangle + \langle \nu_{\a\b} \rangle$ and equal to $\langle \nu_{\b\a} \rangle + \langle \nu_{\b\b} \rangle$. Numerical simulations support the validity of this assumption. The effective parameters of the effective two-species model is then as follows

\begin{align}
    K_\a^{\text{eff}} &=  K - \mu (\bra N_\a \ket + \bra N_\b \ket) + \sqrt{\s^2\left \langle \left(N_\alpha + N_\beta \right)^2 \right \rangle + \l^2 \left \langle \left(N_\alpha - N_\beta \right)^2 \right \rangle} z^{(\mathrm{int})}_{0\a} + \sqrt{\s^2_K} z_{0\a} \\  
    K_\b^{\text{eff}} &= K - \mu (\bra N_\a \ket + \bra N_\b \ket) +\sqrt{\s^2\left \langle \left(N_\alpha + N_\beta \right)^2 \right \rangle + \l^2 \left \langle \left(N_\alpha - N_\beta \right)^2 \right \rangle }\big(\rho_{\mathrm{int}} z^{(\mathrm{int})}_{0\a} + \sqrt{(1-\rho_{\mathrm{int}}^2)} z^{(\mathrm{int})}_{0\b}\big) + \sqrt{\s^2_K}z_{0\b} \\
    A_{0\a0\a}^{\text{eff}} &= A_{0\b0\b}^{\text{eff}} = 1-2\s^2\rho \nu\\
    A_{0\a0\b}^{\text{eff}} &= (\tilde{\mu} - \l y - 2\s^2\rho \nu) \\
    A_{0\b0\a}^{\text{eff}} &= (\tilde{\mu} + \l y - 2\s^2\rho \nu)  
\end{align}

where $K_\a^{\text{eff}}, K_\b^{\text{eff}}$ are the effective growth rates of the two strains. The community serves to reduce the average growth rate by a term $\mu (\bra N_\a \ket + \bra N_\b \ket)$ and adds extra variance as well. $A_{0\a0\a}^{\text{eff}}, A_{0\b0\b}^{\text{eff}}$ are the effecting self-regulating terms for the two strains which is reduced due to the effects of the community relative the strains in isolation. Lastly, $A_{0\a0\b}^{\text{eff}}, A_{0\b0\a}^{\text{eff}}$ are the effective competition (or inter-strain interaction) coefficients. Note that the community serves to reduce the magnitude of competition within and between strains. This reduced competition shapes the coexistence dynamics of both strains \textit{sensu} ``enemy of the enemy is a friend" effect. Thus, $\r_K$ provides us information on the joint effects of independent variation of growth rates $\s_K^2$ and correlated fluctuations arising from interactions with the rest of the community. We can simplify these equations in the low $\l$ limit, which describes the dynamics of two closely related strains by setting $\l = 0$ which gives us

\begin{align}
    \lim_{\l\to 0^+} \, K_\a^{\text{eff}} &=  K - \mu (\bra N_\a \ket + \bra N_\b \ket) + \sqrt{\s^2\left \langle \left(N_\alpha + N_\beta \right)^2 \right \rangle} z^{(\mathrm{int})}_{0\a} + \sqrt{\s^2_K} z_{0\a} \\  
    \lim_{\l\to 0^+} \, K_\b^{\text{eff}} &= K - \mu (\bra N_\a \ket + \bra N_\b \ket) +\sqrt{\s^2\left \langle \left(N_\alpha + N_\beta \right)^2 \right \rangle}\big(\rho_{\mathrm{int}} z^{(\mathrm{int})}_{0\a} + \sqrt{(1-\rho_{\mathrm{int}}^2)} z^{(\mathrm{int})}_{0\b}\big) + \sqrt{\s^2_K}z_{0\b} \\
    \lim_{\l\to 0^+}A_{0\a0\a}^{\text{eff}} &= A_{0\b0\b}^{\text{eff}} = 1-2\s^2\rho \nu\\
    \lim_{\l\to 0^+} \, A_{0\a0\b}^{\text{eff}} &= (\tilde{\mu}- 2\s^2\rho \nu) \\
    \lim_{\l\to 0^+} \, A_{0\b0\a}^{\text{eff}} &= (\tilde{\mu} - 2\s^2\rho \nu)  
\end{align}

\section*{Appendix C: Solving the effective two-strain model using self-consistency relations} 
\label{app: mean field eqns}

When both strains survive at steady state (\ref{eq:eff0aGLV}) and (\ref{eq:eff0bGLV})), both per capita growth rates are zero. This yields a pair of equations in $N_{0\a}$ and $N_{0\b}$ which are solved simultaneously. When they cannot coexist, solutions to these equations are infeasible and correspond to at least one of the strains having negative abundance. The form of the per capita growth rates in eqns (\ref{eq:eff0aGLV}) and (\ref{eq:eff0bGLV}) represents two correlated Gaussians. Thus the joint abundance distribution of strains $0\a$ and $0\b$ is a bivariate Gaussian truncated below 0 for both variables $N_{0\a}$ and $N_{0\b}$. This truncated binormal is fully described by 5 parameters: the means and variances of the two variables and the correlation between them $\r_{\a\b}$. To compute these 5 parameters, we first set both per capita growth rates to 0 and simultaneously solve the resulting equations for $N_{0\a}^\pm$ and $N_{0\b}^\pm$:

\begin{align}
N_{0\a}^\pm &= \f{A_{0\b 0\b}^{\text{eff}} K_{0\a}^{\text{eff}} - A_{0\a 0\b}^{\text{eff}} K_{0\b}^{\text{eff}}}{A_{0\a 0\a}^{\text{eff}} A_{0\b 0\b}^{\text{eff}} - A_{0\b 0\a}^{\text{eff}} A_{0\a 0\b}^{\text{eff}}}, 
\label{eq:N0alphauntrunc}\\
N_{0\b}^\pm &= \f{A_{0\a 0\a}^{\text{eff}} K_{0\b}^{\text{eff}} - A_{0\b 0\a}^{\text{eff}} K_{0\a}^{\text{eff}}}{A_{0\a 0\a}^{\text{eff}} A_{0\b 0\b}^{\text{eff}} - A_{0\b 0\a}^{\text{eff}} A_{0\a 0\b}^{\text{eff}}}.
\label{eq:N0betauntrunc}
\end{align}

Here, the $\pm$ indicates that solutions may suggest either positive or negative abundances, depending on whether both strains survive. This corresponds to the untruncated joint distribution. We can evaluate the denominator of \ref{eq:N0alphauntrunc} and \ref{eq:N0betauntrunc} as they are the same (henceforth called $\Delta$) as it essentially is the determinant of an interaction matrix with effective parameterS of a 2-strain GLV model. To simplify the calculation, we introduce some shorthand notation





We henceforth define $g = K - \mu(\langle N_\a \rangle + \langle N_\b \rangle)$, the deterministic part of the effective growth rate and $\s_g^2 = (\s^2\left \langle M^2 \right \rangle + \l^2 \left \langle D^2 \right \rangle)$, the fluctuating part. We also define $C = (1-\tilde{\mu})(1+ \tilde{\mu} - 4\s^2\rho \nu)$ and $D = \l^2$. Lastly, this assumption allows us to simply write down the elements of the effective $A$ matrix as follows,
\begin{subequations}
    \begin{align}
        A_{0\a0\a}^{\text{eff}} &= A_{0\b0\b}^{\text{eff}} = 1-2\s^2\rho \nu = \xi\\
        A_{0\a0\b}^{\text{eff}} &= (\tilde{\mu} - \l y - 2\s^2\rho \nu) = {\tmu^{\mathrm{eff}}} -\l y \\
        A_{0\b0\a}^{\text{eff}} &= (\tilde{\mu} + \l y - 2\s^2\rho \nu) = {\tmu^{\mathrm{eff}}}+\l y
    \end{align}
\end{subequations}
 Further we will sometimes interchangeably call the determinant $C+Dy^2$ or $\Delta$ to improve legibility of the expression. Using this new notation, it becomes clear that $C =(\xi^2-({\tmu^{\mathrm{eff}}})^2)$.

Using this notation, we can first substitute (see \ref{eq:eff0aGLV}) and (\ref{eq:eff0bGLV}) and calculate the first moment of $N_{0 \a}^\pm$. 

\begin{equation}
    N_{0\a}^\pm = \frac{\xi (g+\s_g z^{(\mathrm{int})}_{0\a}+\s_K z_{0\a})-({\tmu^{\mathrm{eff}}} - \l y)\big( g+\s_g(\rho_{\mathrm{int}} z^{(\mathrm{int})}_{0\a} + \sqrt{1-\rho_{\mathrm{int}}^2} z^{(\mathrm{int})}_{0\b})+\s_K z_{0\b} \big)}{(\xi^2-({\tmu^{\mathrm{eff}}})^2) + \l^2y^2}
\end{equation}

Taking the expectation value of this expressions gives us
\begin{equation}
    \langle N_{0\a}^\pm \rangle = \left\langle \frac{1}{\Delta}(g (\xi - ({\tmu^{\mathrm{eff}}} - \l y) \right\rangle
\end{equation}

This leads to 

\be 
\langle N_{0\a}^\pm \rangle = g \left(\xi - {\tmu^{\mathrm{eff}}}\right) \left\langle \frac{1}{\Delta} \right\rangle + {\cancelto{}{\l\left\langle \frac{y}{\Delta}\right\rangle}} 
\ee

which following evaluation of the expectation values for $y \sim \mathcal {N}(0,1)$ results in the expression

\begin{equation}
    \langle N_{0\a}^\pm \rangle = g(1-\tilde{\mu})\sqrt{\frac{\pi}{2(\xi^2-({\tmu^{\mathrm{eff}}})^2)\l^2}}e^{\frac{(\xi^2-({\tmu^{\mathrm{eff}}})^2)}{2\l^2}}\, \mathrm{erfc}\biggr(\sqrt{\frac{(\xi^2-({\tmu^{\mathrm{eff}}})^2)}{2\l^2}}\biggr).
    \label{eq:first_mom_Nalpha}
\end{equation}

where $C$ is defined as earlier with $C > 0$ and ``erfc'' is the complementary error function. We find that $ \langle N_\b \rangle = \langle N_\a \rangle$. Next, we require to calculate the second moments as well as the cross-moment of $N_\a$ and $N_\b$. Calculating these quantities will bring us closer to determining the sign and magnitude of abundance correlations between closely-related strains. 

We calculate the untruncated second moment as well as the cross-moment by squaring the expressions in (\ref{eq:N0alphauntrunc}) and (\ref{eq:N0betauntrunc}). First, to calculate the second moment of $\langle N_\a^2 \rangle$ we can write 
\be
\langle {N_{0\a}^\pm}^2 \rangle = \left\langle \biggr( \f{A_{0\b 0\b}^{\text{eff}} K_{0\a}^{\text{eff}} - A_{0\a 0\b}^{\text{eff}} K_{0\b}^{\text{eff}}}{A_{0\a 0\a}^{\text{eff}} A_{0\b 0\b}^{\text{eff}} - A_{0\b 0\a}^{\text{eff}} A_{0\a 0\b}^{\text{eff}}} \biggr)^2 \right\rangle. 
\ee 

This can be split into 

\begin{align}
\langle {N_{0\a}^\pm}^2\rangle &=\left\langle\frac{\xi^2(g+\s_g z^{(\mathrm{int})}_{0\a} + \s_K z_{0\a})^2} {\Delta^2}\right\rangle \nn \\
&- 2 \left\langle \frac{\xi({\tmu^{\mathrm{eff}}}-\l y)(g+\s_g z^{(\mathrm{int})}_{0\a} + \s_K z_{0\a})(g+\s_g (\rho_{\mathrm{int}} z^{(\mathrm{int})}_{0\a} + \sqrt{1-\rho_{\mathrm{int}}^2} z^{(\mathrm{int})}_{0\b}) + \s_K z_{0\b})}{\Delta^2} \right\rangle \nonumber\\
&+ \left\langle \frac{({\tmu^{\mathrm{eff}}}-\l y)^2(g+\s_g (\rho_{\mathrm{int}} z^{(\mathrm{int})}_{0\a} + \sqrt{1-\rho_{\mathrm{int}}^2} z^{(\mathrm{int})}_{0\b}) + \s_K z_{0\b})^2}{\Delta^2} \right\rangle.    
\end{align}


Expanding this, we get 

\begin{align}
\xi^2 \left\langle \frac{1}{\Delta^2}\right\rangle \left(g^2+\s_g^2 + \s_K^2\right) - 2\xi{\tmu^{\mathrm{eff}}} (g^2+ \s_g^2 \rho_{\mathrm{int}})\left\langle \frac{1}{\Delta^2}\right\rangle + ({\tmu^{\mathrm{eff}}})^2(g^2+\s_g^2\rho_{\mathrm{int}}^2 + \s_g^2(1-\rho_{\mathrm{int}}^2) + \s_K^2) \left\langle \frac{1}{\Delta^2}\right\rangle \nonumber\\+ \l^2 (g^2+\s_g^2\rho_{\mathrm{int}}^2 + \s_g^2(1-\rho_{\mathrm{int}}^2) + \s_K^2)\left\langle \frac{y^2}{\Delta^2}\right\rangle.
\end{align}.

We can note that the coefficients of the third and fourth term reduce to $(g^2+\s_g^2+\s_K^2)$ i.e. the coefficients of the first term. Grouping all the terms with $g^2 \langle\frac{1}{\Delta^2}\rangle$, this reduces to
\be
g^2\left\langle\frac{1}{\Delta^2}\right\rangle (\xi-{\tmu^{\mathrm{eff}}})^2 
- 2\s_g^2 \xi {\tmu^{\mathrm{eff}}} \rho_{\mathrm{int}} \left\langle\frac{1}{\Delta^2}\right\rangle 
+ (\s_K^2 + \s_g^2)(\xi^2 +({\tmu^{\mathrm{eff}}})^2) \left\langle\frac{1}{\Delta^2}\right\rangle
+ \l^2 (g^2+\s_g^2 + \s_K^2)\left\langle\frac{y^2}{\Delta^2}\right\rangle.
\ee

From the definition, we can see that $(\xi-{\tmu^{\mathrm{eff}}})$ is equal to $(1-\tilde{\mu})$. Also note that $\rho_{\mathrm{int}} \s_g^2 = \s^2\langle M^2 \rangle$ and $(1-\rho_{\mathrm{int}})\s_g^2 = \l^2 \langle D^2\rangle$. Rearranging and substituting the coefficients of $\frac{1}{\Delta^2}$, we get 
\begin{align}
\left\langle {{N_{0\a}^\pm}^2} \right\rangle &= \left(g^2 +\rho_{\mathrm{int}} \s_g^2 \right) \left(1 - \tilde{\mu}\right)^2 \left\langle \frac{1}{\Delta^2} \right\rangle 
+ \left( \s_g^2 (1-\rho_{\mathrm{int}}) + \s_K^2 \right) \left\langle \frac{1}{\Delta^2} \right\rangle 
+ \lambda^2 \left(g^2 + \s_g^2 +\s_K^2 \right) \left\langle \frac{y^2}{\Delta^2} \right\rangle 
\label{eq:sec_mom_Nalpha}\\
\left\langle {N_{0\b}^\pm}^2 \right\rangle &= \left(g^2 +\rho_{\mathrm{int}} \s_g^2 \right) \left(1 - \tilde{\mu}\right)^2 \left\langle \frac{1}{\Delta^2} \right\rangle 
+ \left( \s_g^2 (1-\rho_{\mathrm{int}}) + \s_K^2 \right) \left\langle \frac{1}{\Delta^2} \right\rangle 
+ \lambda^2 \left(g^2 + \s_g^2 +\s_K^2 \right) \left\langle \frac{y^2}{\Delta^2} \right\rangle 
\label{eq:sec_mom_Nbeta}
\end{align}

Note that these calculations of first and second moments are of the untruncated abundances of the $\alpha$ and $\b$ strains. Next,we calculate the untruncated cross-moment of the strains $\langle N_{0\a}N_{0\b} \rangle$. This is given by  

\be
\left\langle N_{0\a}N_{0\b} \right\rangle = \left\langle \frac{(A_{0\b 0\b}^{\text{eff}} K_{0\a}^{\text{eff}} - A_{0\a 0\b}^{\text{eff}} K_{0\b}^{\text{eff}}) (A_{0\a 0\a}^{\text{eff}} K_{0\b}^{\text{eff}} - A_{0\b 0\a}^{\text{eff}} K_{0\a}^{\text{eff}})}{(A_{0\a 0\a}^{\text{eff}} A_{0\b 0\b}^{\text{eff}} - A_{0\b 0\a}^{\text{eff}} A_{0\a 0\b}^{\text{eff}})^2}\right\rangle.
\ee 

Following substitution and decomposing into four components we ge,

\begin{multline}
    \left\langle \frac{\xi^2(g+\s_g z^{(\mathrm{int})}_{0\a} +\s_Kz_{0\a})(g+\s_g (\rho_{\mathrm{int}} z^{(\mathrm{int})}_{0\a} + \sqrt{(1-\rho_{\mathrm{int}}^2)}z^{(\mathrm{int})}_{0\b}) +\s_K z_{0\b})}{\Delta^2} \right\rangle 
    - \left\langle \frac{\xi({\tmu^{\mathrm{eff}}}+\l y)(g+\s_g z^{(\mathrm{int})}_{0\a} +\s_Kz_{0\a})^2}{\Delta^2} \right\rangle \\
    - \left\langle \frac{\xi({\tmu^{\mathrm{eff}}}-\l y)(g+\s_g (\rho_{\mathrm{int}} z^{(\mathrm{int})}_{0\a} + \sqrt{(1-\rho_{\mathrm{int}}^2)}z^{(\mathrm{int})}_{0\b}) +\s_K z_{0\b})^2}{\Delta^2} \right\rangle \\  
    + \left\langle \frac{({\tmu^{\mathrm{eff}}}+\l y)({\tmu^{\mathrm{eff}}}-\l y)(g+\s_g z^{(\mathrm{int})}_{0\a} +\s_Kz_{0\a})(g+\s_g (\rho_{\mathrm{int}} z^{(\mathrm{int})}_{0\a} + \sqrt{(1-\rho_{\mathrm{int}}^2)}z^{(\mathrm{int})}_{0\b}) +\s_K z_{0\b})}{\Delta^2} \right\rangle
\end{multline}

Making the appropriate substitutions and taking the expectation values yields

\[ \xi^2(g^2 + \rho_{\mathrm{int}} \s_g^2)\left\langle\f{1}{\Delta^2}\right\rangle
- 2\xi{\tmu^{\mathrm{eff}}}(g^2+\s_g^2 +\s_K^2) \left\langle \f{1}{\Delta^2} \right\rangle + ({\tmu^{\mathrm{eff}}})^2 (g^2 + \rho_{\mathrm{int}} \s_g^2) \left\langle\f{1}{\Delta^2}\right\rangle - \l^2(g^2 + \rho_{\mathrm{int}} \s_g^2) \left\langle\f{y^2}{\Delta^2}\right\rangle\]. 

Making a similar substitution as earlier where $(\xi-{\tmu^{\mathrm{eff}}}) = (1- \tilde{\mu})$ and rearranging terms gives us
\begin{multline}
    \left\langle N_{0\alpha} N_{0\beta} \right\rangle = \left(g^2+\s_g^2 \rho_{\mathrm{int}} \right)\left(1 - \tilde{\mu}\right)^2 \left\langle \frac{1}{\Delta^2}\right\rangle
- {2 {\tmu^{\mathrm{eff}}} \xi \left( (1-\rho_{\mathrm{int}})\, \s_g^2 + \s_K^2 \right)} \left\langle \frac{1}{\Delta^2} \right\rangle
- \lambda^2 (g^2 + \rho_{\mathrm{int}} \s_g^2) \left\langle \frac{y^2}{\Delta^2} \right\rangle.
\label{eq:cross_mom_Nalphabeta}
\end{multline}  

To get complete expressions for (\ref{eq:sec_mom_Nalpha}),(\ref{eq:sec_mom_Nbeta}), and (\ref{eq:cross_mom_Nalphabeta}), we still require values for the expectation of $\frac{1}{\Delta^2}$ and $\frac{y^2}{\Delta^2}$. Recall $\Delta = C+Dy^2$, i.e., the difference between the product of the self-regulation terms of each strain and the product of the cross-strain interaction terms. Also recall $C = (\xi^2 - ({\tmu^{\mathrm{eff}}})^2), D=\l^2$.  Further, $y \sim $ Normal(0,1). We find these expectation values are
\begin{align}
    \left\langle\frac{1}{\Delta^2}\right\rangle &= \frac{1 - (C-D)\sqrt\frac{\pi}{2CD}e^{\frac{C}{2D}}\,\mathrm{erfc}\sqrt{\frac{C}{2D}}} {2CD} 
    \label{eq:exp_1_del2}\\
    \left\langle\frac{y^2}{\Delta^2}\right\rangle &= \frac{-1 - (C+D)\sqrt\frac{\pi}{2CD}e^{\frac{C}{2D}}\,\mathrm{erfc}\sqrt{\frac{C}{2D}}} {2D^2}.
    \label{eq:exp_y2_del2}
\end{align}
By substituting the expressions from (\ref{eq:exp_1_del2}) and (\ref{eq:exp_y2_del2}) into (\ref{eq:sec_mom_Nalpha}),(\ref{eq:sec_mom_Nbeta}), and (\ref{eq:cross_mom_Nalphabeta}), we can get complete expressions for the untruncated second and cross-moments of the variables. The cross moment offers a clear picture and partitioning of the multiple effects driving the correlation patterns of the competing strains in a community. Term 1 is the ``filtering'' contribution. When the community creates a favourable environment (high mean effective growth rate `g') or when species-level interaction dissimilarity is high i.e. high $\s^2 \langle M^2 \rangle$ and it fluctuates in ways that affect both strains equally, their abundances fall and rise together leading to positive correlations. Term 2 captures the fact that within the species, strains can experience slightly different environments (even if they are mildly dissimilar ($\l$) or due to intrinsic variation in growth rates ($\s_K$)). This leads to a decorrelation of strain abundances resulting in a negative contribution to the cross-moment. Term 3 is the "sorting" contribution to variance. When $\l$ is large, environmental fluctuations push strains in opposite directions i.e. one strain benefits while the other suffers. This is also a mechanism that flips correlations from positive to negative as $\l$ increases.

The covariance of abundance between the two conspecific strains of the focal species is obtained by subtracting the product of the means from the calculated cross--moment. Using
\begin{equation}
\big\langle N_{0\alpha}\big\rangle \big\langle N_{0\beta}\big\rangle
=
g^2(1-\tilde{\mu})^2\, \left\langle \frac{1}{\Delta} \right\rangle^2,
\end{equation}
we obtain
\begin{align}
\mathrm{Cov}\big(N_{0\alpha},N_{0\beta}\big)
&=
\big\langle N_{0\alpha}N_{0\beta}\big\rangle
-
\big\langle N_{0\alpha}\big\rangle
\big\langle N_{0\beta}\big\rangle
\\[3pt]
&=
g^2(1-\tilde{\mu})^2 \left( \left\langle\frac{1}{\Delta^2}\right\rangle - \left\langle \frac{1}{\Delta} \right\rangle^2\right)
+\s_g^2 \rho_{\mathrm{int}}(1-\tilde{\mu})^2  \left\langle\frac{1}{\Delta^2}\right\rangle
\nonumber \\
&\quad -2 {\tmu^{\mathrm{eff}}} \xi\left( \left( 1-\rho_{\mathrm{int}}\right)\, \s_g^2 + \s_K^2 \right) \left\langle \frac{1}{\Delta^2} \right\rangle 
-
\lambda^2\big(g^2 + \rho_{\mathrm{int}} \s_g^2\big)\,\left\langle \frac{y^2}{\Delta^2} \right\rangle.
\label{eq:cov-full}
\end{align}

 Conveniently for us, $\s_g^2 \rho_{\mathrm{int}} = \s^2\left \langle (N_{0\a} +N_{0\b})^2 \right  \rangle$ and $(1-\rho_{\mathrm{int}})\s_g^2 = \l^2\left \langle (N_{0\a} - N_{0\b})^2 \right \rangle$. Also, the first term of the covariance is exactly 0 as $\l \rightarrow 0$ and remains close to 0 for a large range of $\l$ (upto O(1)). It contributes negligibly to the covariance relative to the other three terms. Thus, we can approximate the covariance of conspecific strain abundances as,

 \begin{align}
     \mathrm{Cov}\big(N_{0\alpha},N_{0\beta}\big) &= \left[\s^2(1-\tilde{\mu})^2\left \langle (N_{0\a} +N_{0\b})^2 \right  \rangle  -2\l^2 (1- 2 \s^2 \rho \nu) (\tilde{\mu}- 2 \s^2 \rho \nu) \langle (N_{0\a} - N_{0\b})^2 \rangle  \right]\left\langle\frac{1}{\Delta^2}\right\rangle \nn \\
     &+ 2\s_K^2 (1- 2 \s^2 \rho \nu) (\tilde{\mu}- 2 \s^2 \rho \nu)\left\langle\frac{1}{\Delta^2}\right\rangle - \l^2 \left( g^2 + \s^2\left \langle (N_{0\a} +N_{0\b})^2 \right  \rangle \right)\left\langle\frac{y^2}{\Delta^2}\right\rangle
 \end{align}

Following appropriate substitution of quantities, we can rewrite it as,
\begin{align}
    \mathrm{Cov}\big(N_{0\alpha},N_{0\beta}\big) &= \underbrace{(\s^2\langle (N_{0\a} +N_{0\b})^2 \rangle) (1-\tilde{\mu})^2 \left\langle\frac{1}{\Delta^2}\right\rangle}_{\text{Shared species dissimilarity effects on conspecific strains (Positive)}} - \underbrace{ 2 \s_K^2(1- 2 \s^2 \rho \nu) (\tilde{\mu} ~ - 2 \s^2 \rho \nu)  \left\langle\frac{1}{\Delta^2}\right\rangle}_{\text{Intrinsic growth rate differences (Negative)}} \nn \\
    & - \l^2 \underbrace{\biggr[2 (1- 2 \s^2 \rho \nu) (\tilde{\mu}- 2 \s^2 \rho \nu) \langle (N_{0\a} - N_{0\b})^2 \rangle \left\langle\frac{1}{\Delta^2}\right\rangle
    + \big((K - \mu \left \langle N_{0\a} +N_{0\b} \right  \rangle)^2 + \s^2\langle (N_{0\a} +N_{0\b})^2 \rangle\big) \left\langle\frac{y^2}{\Delta^2}\right\rangle \biggr]}_{\text{Strain interaction dissimilarity driven effect (only when $\l > 0$; Negative)}}  
    \label{eq:cov-full-explicit}
\end{align}


The corresponding correlation between the two strains is then
\begin{equation}
\label{eq:rhoab-def}
\rho_{\alpha\beta}
 = 
\frac{\mathrm{Cov}\big(N_{0\alpha},N_{0\beta}\big)}
{\sqrt{\mathrm{Var}(N_{0\alpha})\;\mathrm{Var}(N_{0\beta})}},
\end{equation}
where the numerator is given by Eq.~\eqref{eq:cov-full} and the denominator can be obtained from the single--strain moments Eq. (\ref{eq:sec_mom_Nalpha}, \ref{eq:sec_mom_Nbeta}), or directly measured from simulations conditioned on two--strain coexistence.

However, biologically we are only interested in the non-negative abundances of the strains. Thus, given the untruncated expressions along with the truncation of the variables $N_\a$ and $N_\b$ at (0,0), we need to calculate the truncated first, second and cross-moments of the variables to get their accurate distributions. Previous work has calculated these truncated moments from the untruncated variables for bivariate, standard normal random variables \cite{begier1971correlation, rosenbaum1961moments}. We make use of these expressions and back-transform these variables to our required variables $N_\a$ and $N_\b$ with untruncated means and variances to obtain the necessary expressions. Thus the cavity framework yields a closed system of 12 self-consistency equations relating the following unknowns i.e., 1) Truncated first moments: $\langle N_\alpha \rangle$, $\langle N_\beta \rangle$ , 2) truncated second moments: $\langle N_\alpha^2 \rangle$, $\langle N_\beta^2 \rangle$, 3) survival probabilities: $\phi_\alpha$, $\phi_\beta$, $\phi_{\alpha\beta}$, 4) susceptibilities: $\nu_{\alpha\alpha}$, $\nu_{\beta\beta}$, $\nu_{\alpha\beta}$, $\nu_{\beta\alpha}$, and 5) the truncated cross-moment: $\langle N_\alpha N_\beta \rangle$. The self consistency equations which were solved together are as follows
\begin{align}
    \langle N_\alpha \rangle &= \langle N_{0\a}^{\pm} \rangle + \sqrt{\mathrm{Var}(N_{0\alpha}^{\pm})} \, \langle X \mid X > h, Y > k \rangle \\
    \langle N_\alpha^2 \rangle &= \langle {N_{0\a}^{\pm}}^2 \rangle + 2 N_{0\a}^{\pm} \sqrt{\mathrm{Var}(N_{0\alpha}^{\pm})} \, \langle X \mid X > h, Y > k \rangle + \mathrm{Var}(N_{0\alpha}^{\pm}) \, \langle X^2 \mid X > h, Y > k \rangle \\
    \langle N_\b \rangle &= \langle N_{0\b}^{\pm} \rangle + \sqrt{\mathrm{Var}(N_{0\beta}^{\pm})} \, \langle X \mid X > h, Y > k \rangle \\
    \langle N_\b^2 \rangle &= \langle {N_{0\b}^{\pm}}^2 \rangle + 2 N_{0\b}^{\pm} \sqrt{\mathrm{Var}(N_{0\beta}^{\pm})} \, \langle X \mid X > h, Y > k \rangle + \mathrm{Var}(N_{0\beta}^{\pm}) \, \langle X^2 \mid X > h, Y > k \rangle \\
    \langle N_\a N_\b\rangle &= \langle N_{0 \alpha} \rangle \langle N_{0 \beta} \rangle + \sqrt{\mathrm{Var}(N_{0\alpha}^{\pm})} \sqrt{\mathrm{Var}(N_{0\beta}^{\pm})}\langle XY \mid X > h, Y > k \rangle \nn \\&=+\sqrt{\mathrm{Var}(N_{0\beta}^{\pm})} \langle Y \mid X > h, Y > k \rangle \langle N_{0 \alpha} \rangle + \sqrt{\mathrm{Var}(N_{0\alpha}^{\pm})} \langle X \mid X > h, Y > k \rangle \langle N_{0 \beta} \rangle \\
        \phi_\alpha &= \Phi(-h) = P(N_\alpha > 0) \\
    \phi_\beta &= \Phi(-k) = P(N_\beta > 0) \\
    \phi_{\alpha\beta} &= L = P(N_\alpha > 0, N_\beta > 0) \\
    \langle \nu_{\alpha\alpha} \rangle &= \phi_\alpha \xi \, \left\langle \frac{1}{\Delta} \right\rangle \\
    \langle \nu_{\beta\beta} \rangle &= \phi_\beta \xi \, \left\langle \frac{1}{\Delta} \right\rangle \\
    \langle \nu_{\alpha\beta} \rangle &= -\phi_{\alpha\beta} {\tmu^{\mathrm{eff}}} \, \left\langle \frac{1}{\Delta} \right\rangle \\
    \langle \nu_{\beta\alpha} \rangle &= -\phi_{\alpha\beta} {\tmu^{\mathrm{eff}}} \, \left\langle \frac{1}{\Delta} \right\rangle  
\end{align}

where $\langle X \rangle$ and $\langle Y \rangle$ are the first moments of two standard normal random variables 'X' and 'Y' (that correspond to the $\a$ and $\b$ strains respectively), $\langle X^2 \rangle, \langle Y^2 \rangle$ correspond to their second moments, and $\langle XY \rangle$ corresponds to their cross-moments. We standardized the untruncated distribution by defining the limits $h$ and $k$ as follows:
\begin{equation}
    h = -\frac{\langle N_{0\a}^{\pm}\rangle}{\sqrt{\mathrm{Var}(N_{0\alpha}^{\pm})}}, \qquad k = -\frac{\langle N_{0\b}^{\pm} \rangle}{\sqrt{\mathrm{Var}(N_{0\beta}^{\pm})}}
\end{equation}
so that the truncation region becomes $\{X > h, Y > k\}$. The marginal survival probabilities for each strain type are then given by $\phi_\alpha$ and $\phi_\beta$ in the above equations where $\Phi(\cdot)$ is the standard normal cumulative distribution function. The joint survival probability, representing the fraction of species in which both strains coexist, is given by $\phi_{\alpha\beta}$. Lastly, the terms $\mathrm{Var}(N_{0\alpha}^{\pm})$ and $\mathrm{Var}(N_{0\beta}^{\pm})$ are the variances of the untruncated distributions of the conspecific strain abundances.


\begin{figure*} [] 
    \centering
    \includegraphics[width=0.65\textwidth]{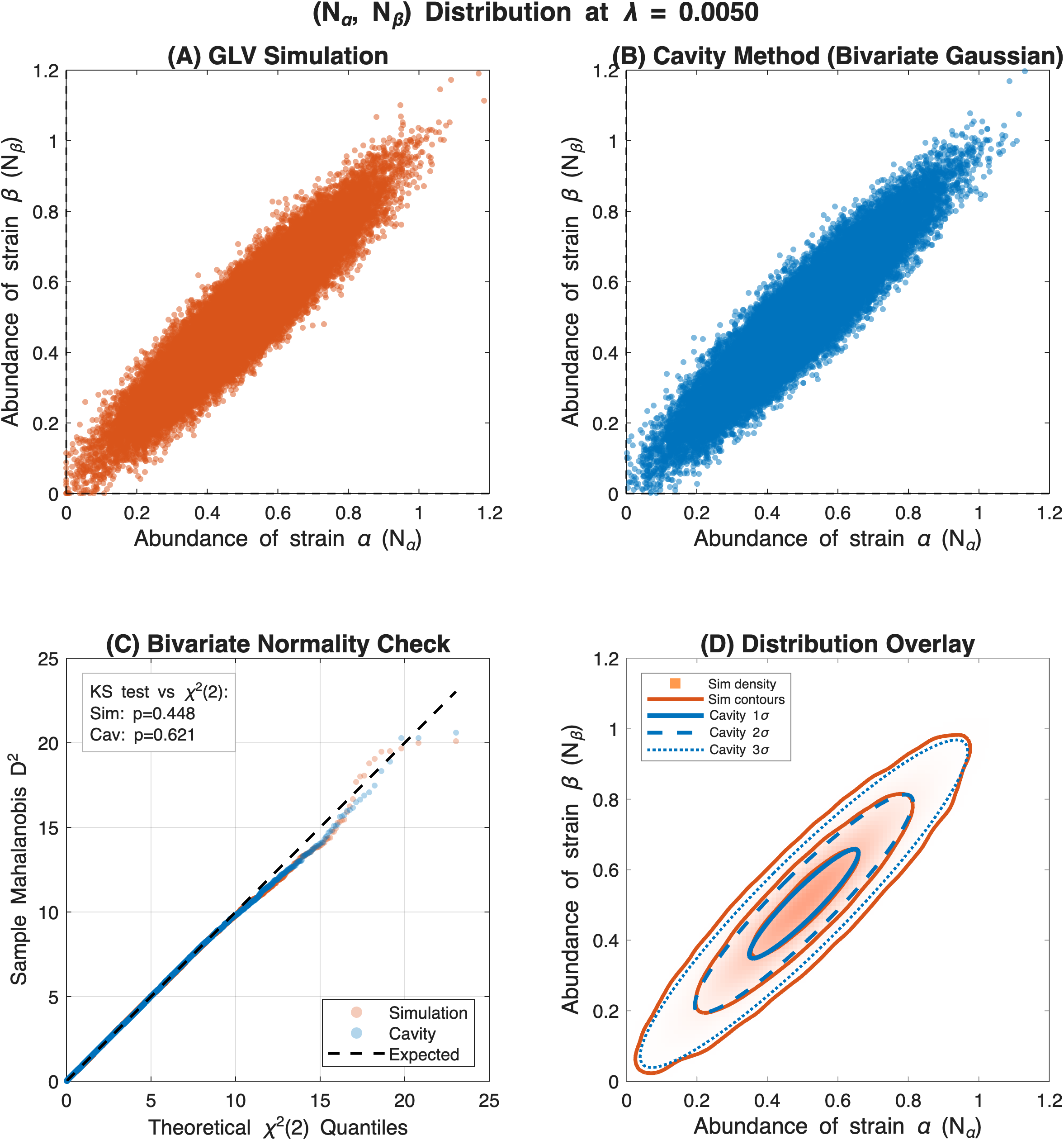}
    \caption{\justifying\textsf{\textbf{Joint distribution of strain abundances is a truncated  bivariate Gaussian for closely-related strains (low $\l$).} Panels in the top row show scatter plots for the abundances of strains $\a$ and $\b$ of the same species at steady-state within a diverse community with $S=100$ species. On the left, we show results from simulations of the full community with $2S$ strains; on the right, theoretical predictions from the cavity method for the same parameter set. The points in the right panel were generated using Monte Carlo sampling based on a truncated bivariate Gaussian distribution with moments as calculated from the cavity method. Bottom left panel shows a QQ plot to confirm that the distributions (both analytic and simulations) were bivariate normal. To check for the normality of the distributions we plotted the square of the Mahalanobis distance of the sample distributions against a theoretical $\chi^2$ distribution following which we performed a KS test to check for deviations from the theoretical distributions. Inset box shows $p$-values; $p > 0.1$ suggests that both joint distributions --- from simulations and cavity theory --- are consistent with a truncated bivariate Gaussian. In the bottom right panel we show contours for the joint distributions obtained from simulations and cavity theory at $1\s$, $2\s$, and $3\s$. Wiggles indicate low sampling. Parameters: $\mu = 0.3, \tilde{\mu} = 0.5, \s = 0.2, \s_K = 0.02, S=100, \rho = 0.99, \mathrm{Realizations} = 500$}}
    \label{fig:supp_scatter_lo_lam}
\end{figure*}

\begin{figure*} [t] 
    \centering
    \includegraphics[width=0.95\textwidth]{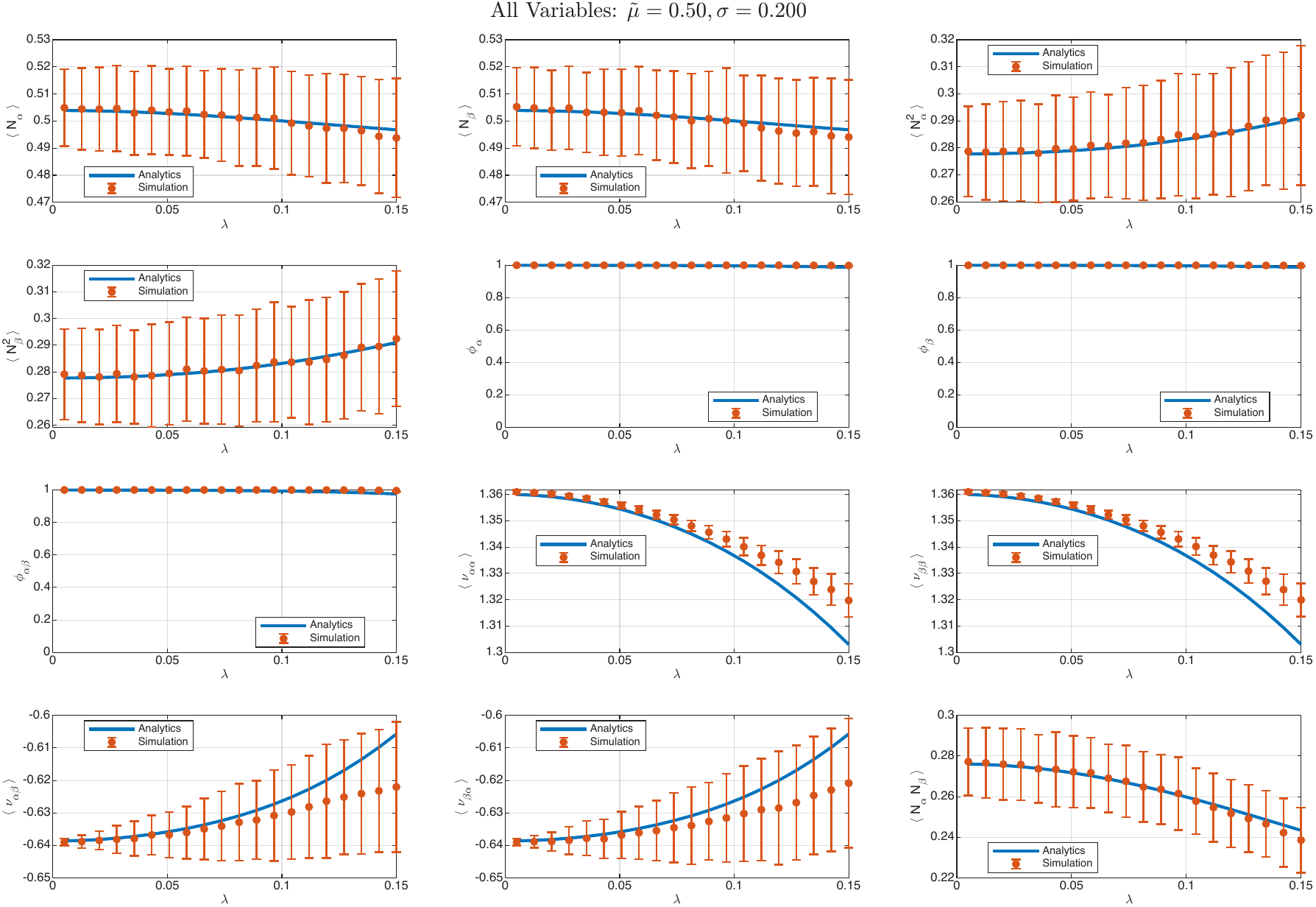}
    \caption{\justifying\textsf{\textbf{Numerical simulations match analytical calculations for self-consistency variables.} Orange dots show mean value of variables with error bars showing standard error of mean across all realizations. Blue line shows the analytical values derived from cavity calculations (see below in Appendix F; eq. S164-S179). We find that this analytical approach matches well with numerical simulations (see Figure \ref{fig:cavity_sims_match}). At large $\l$, simulations deviate from analytics due to breakdown of 2D Gaussian ansatz (see Figure \ref{fig:supp_scatter_hi_lam}) . Parameters used in simulations (and analytics): $\mu = 0.3, \tilde{\mu} = 0.5, \s = 0.2, \s_K = 0.02, S=100, \rho = 0.99, \mathrm{Realizations} = 500$.}}
    \label{fig:cavity_sims_match}
\end{figure*}

\begin{figure*} [t] 
    \centering
    \includegraphics[width=0.95\textwidth]{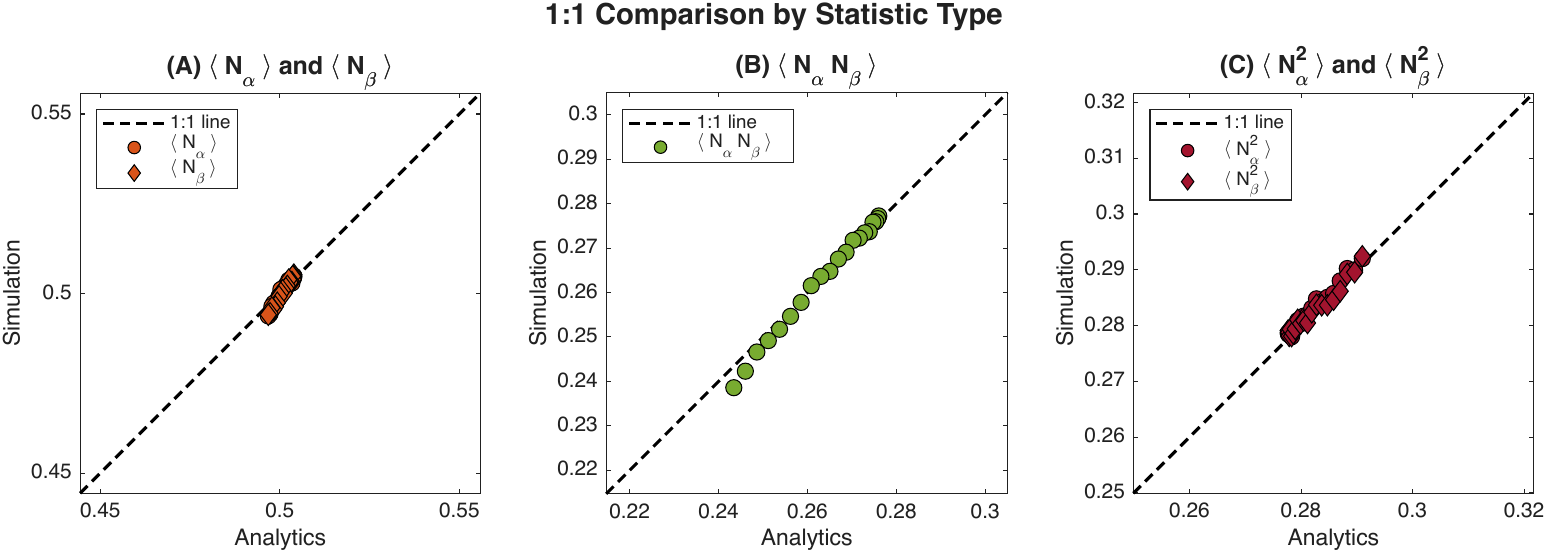}
    \caption{\justifying\textsf{\textbf{Strain abundance correlation between numerics and analytics match closely for first, second, and cross-moments.} From left to right: panels show the comparison of first, cross, and second moments of the two strains generated from simulations (y-axis) and cavity calculations (analytics; x-axis). Each data point refers to the average value over all realizations (S=500) of the quantity for a given $\l$ value. All the moments lie on the 1:1 line barring few minute deviations of two cross-moment terms. Parameters: $\mu = 0.3, \tilde{\mu} = 0.5, \s = 0.2, \s_K = 0.02, S=100, \rho = 0.99, \mathrm{Realizations} = 500, \l \in [0.005,0.15]$.}}
    \label{fig:one_to_one_type}
\end{figure*}

\begin{figure*} [t] 
    \centering
    \includegraphics[width=0.99\textwidth]{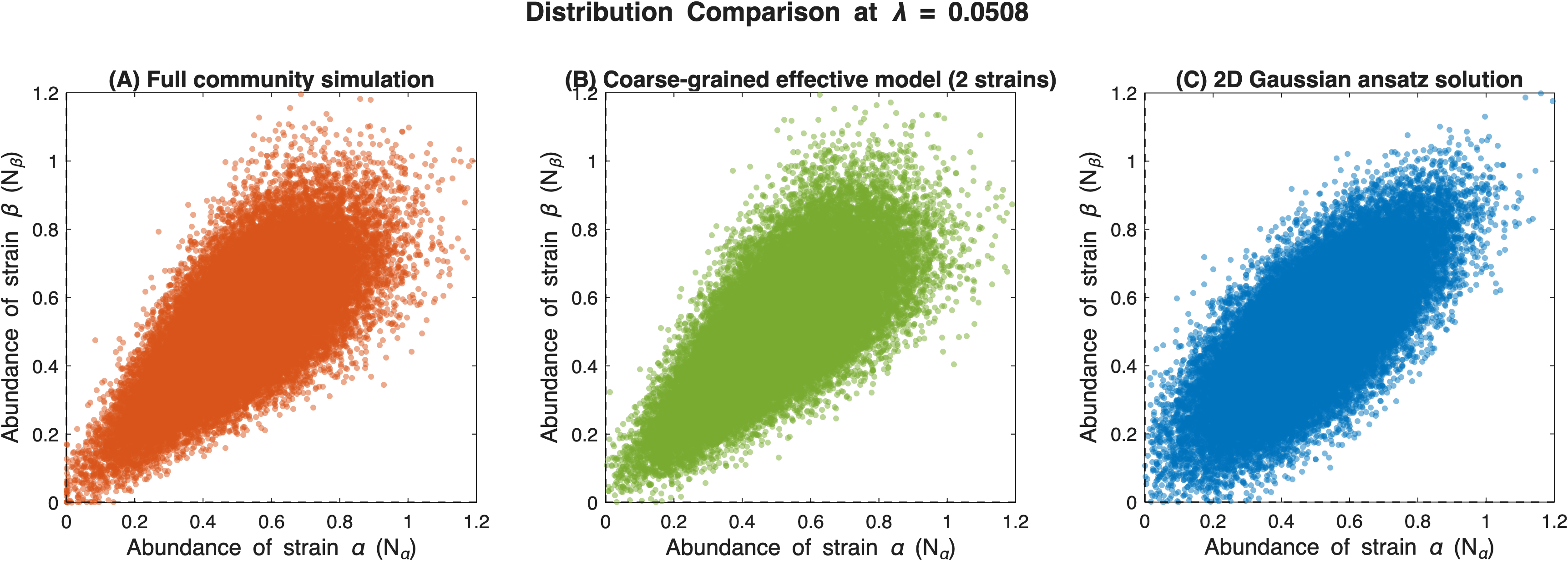}
    \caption{\justifying\textsf{\textbf{Truncated bivariate Gaussian description of joint distribution of strain abundances breaks down at larger$\l$.} Left and center panels show joint abundances scatter plots  for the same set of disorder parameters via numerical simulations of the entire community (2S species) and the effective 2-strain model respectively. The observed distribution and deviation from truncated bivariate Gaussianity is exactly similar. In the right panel, we see the truncated bivariate Gaussian ansatz constraining the analytical solution and leading to qualitatively different abundance scatterplots. These points in the right panel were generated using Monte Carlo sampling based on a truncated bivariate Gaussian distribution with moments as calculated from the cavity method. Parameters: $\mu = 0.3, \tilde{\mu} = 0.5, \s = 0.2, \s_K = 0.02, S=100, \rho = 0.99, \mathrm{Realizations} = 500$.}}
    \label{fig:supp_scatter_hi_lam}
\end{figure*}

\subsection*{Solution of self-consistency equations using a truncated bivariate Gaussian ansatz}

The effective two-strain model derived via the cavity method (Eqs.~(\ref{eq: 2-spp-glv-sigG_alpha}) and (\ref{eq: 2-spp-glv-sigG_beta})) suggests that the joint distribution of $(N_\alpha, N_\beta)$ is a bivariate Gaussian truncated below zero in both variables.  We first verify this ansatz directly. Fig.~\ref{fig:supp_scatter_lo_lam}A--B shows scatter plots of $(N_\alpha, N_\beta)$ at steady state from simulations of the full $2S$-strain community (A) and from Monte Carlo sampling of the cavity-predicted truncated bivariate Gaussian (B). The two distributions are visually indistinguishable. To test Gaussianity quantitatively, we plot the squared Mahalanobis distance of the sample distributions against a theoretical $\chi^2$ distribution (Fig.~\ref{fig:supp_scatter_lo_lam}C). KS test $p$-values exceed $0.1$ for both the simulation and cavity distributions, confirming consistency with a truncated bivariate Gaussian. The $1\sigma$, $2\sigma$, and $3\sigma$ contours from theory and simulation also agree closely (Fig.~\ref{fig:supp_scatter_lo_lam}D).


In Fig.~\ref{fig:cavity_sims_match}, we compare the analytical solutions of the self-consistency equations with numerical simulations across the full range of $\lambda$. All 12 quantities---first moments, second moments, cross-moment, survival probabilities, and susceptibilities---largely agree with simulations. Fig.~\ref{fig:one_to_one_type} further confirms this agreement via one-to-one comparisons of the first, second, and cross-moments, which fall on the identity line across the full range of $\lambda$ explored.

However, for some quantities there is a slight discrepancy at larger strain interaction dissimilarity $\lambda$. This can be understoof by noting that the denominator of the abundance expressions in Eqs.~(\ref{eq:N0alphauntrunc}) and (\ref{eq:N0betauntrunc}) is $\Delta = C + \lambda^2 y^2$, where $C = \xi^2 - (\tilde{\mu}^{\mathrm{eff}})^2$ and $y \sim \mathcal{N}(0,1)$. At low $\lambda$, $\lambda^2 y^2 \ll C$, so $\Delta$ is effectively constant across realizations and the ratio of a Gaussian numerator by a constant denominator inherits Gaussianity. As $\lambda$ increases, $\lambda^2 y^2$ becomes comparable to $C$, introducing a $\chi^2$ random variable in the denominator. This produces heavier tails and mild right skew in the marginal distributions, visible as a characteristic ``fanning out'' of the joint abundance scatter plots relative to the elliptical Gaussian contours (Fig.~\ref{fig:supp_scatter_hi_lam}).

Two observations are important here. First, this deviation is a property of the true two-strain dynamics, not an artifact of the cavity approximation: the same non-Gaussian shape appears in simulations of both the full $2S$-strain community and the effective two-strain model (Fig.~\ref{fig:supp_scatter_hi_lam}, left and center panels). The cavity method correctly identifies the effective model; it is the subsequent Gaussian ansatz that becomes approximate. Second, despite this deviation in distributional shape, the low-order moments that determine observable quantities of interest such as strain abundance correlations remain accurately predicted throughout the parameter range. Small deviations appear only in the cross-moment at the largest values of $\lambda$ (Fig.~\ref{fig:one_to_one_type}). As a result, the cavity theory accurately captures the strain abundance correlation, including the sign transition at $\lambda_c$ and its dependence on $\sigma$ and $\tilde{\mu}$ (Figs.~3 and 4 in the main text). Developing an improved distributional ansatz that captures the non-Gaussian tails at large $\lambda$---for instance, using scale mixtures of normals to account for the stochastic denominator---is a natural extension but is not necessary to support the conclusions of this work.

\section*{Appendix D: Mapping to Modern Coexistence Theory} \label{app:MCT setup}

Modern coexistence theory (MCT) decomposes coexistence requirements into stabilizing mechanisms (niche differentiation) and equalizing mechanisms (fitness ratios). We first establish the general framework and derive the coexistence criterion, then connect our effective two-strain models to these canonical definitions.

\subsection*{Standard Two-Species LV Form and Mutual Invasibility}

We consider our effective two strain model which is akin to a generic two-species Lotka-Volterra competition model:
\begin{align}
\frac{dN_{\alpha}}{dt} &= N_{\alpha} \left( K_{\alpha}^{\mathrm{eff}} - A_{\alpha\alpha}^{\mathrm{eff}} N_{\alpha} - A_{\alpha\beta}^{\mathrm{eff}} N_{\beta} \right), \label{eq:supp_effModel1}\\
\frac{dN_{\beta}}{dt} &= N_{\beta} \left( K_{\beta}^{\mathrm{eff}} - A_{\beta\beta}^{\mathrm{eff}} N_{\beta} - A_{\beta\alpha}^{\mathrm{eff}} N_{\alpha} \right).
\label{eq:supp_effModel2}
\end{align}
where $K_{\alpha}^{\mathrm{eff}}, K_{\beta}^{\mathrm{eff}}$ are the effective growth rates of the two focal strains following coarse-graining of the community and $A_{\a\b}^{\mathrm{eff}}, A_{\b\a}^{\mathrm{eff}}$ are inter-strain interaction coefficient and  $A_{\a\a}^{\mathrm{eff}}, A_{\b\b}^{\mathrm{eff}}$ are self-regulation coefficients. For simplicity, we will drop the superscript for the rest of this section. 


Strain $\a$ can invade when rare in a resident $\b$ population if its per-capita growth rate is positive when $N_{\a} \to 0$ and $N_{\b}$ is at its single-strain equilibrium $N_{\b}^* = K_{\b}/A_{\b\b}$:
\begin{equation}
K_{\a} - A_{\a\b}\frac{K_{\b}}{A_{\b\b}} > 0 \quad \Longleftrightarrow \quad \frac{K_{\a}}{K_{\b}} > \frac{A_{\a\b}}{A_{\b\b}} \equiv \gamma_{\a\b}
\label{eq:invasion_alpha}
\end{equation}
where we define the `relative limitation ratio' $\gamma_{\a\b} = A_{\a\b}/A_{\b\b}$, measuring how sensitive strain $\b$'s equilibrium is to competition from strain $\a$.

Similarly, strain $\b$ invades strain $\a$ when rare if:
\begin{equation}
\frac{K_{\b}}{K_{\a}} > \frac{A_{\b\a}}{A_{\a\a}} = \gamma_{\b\a} \quad \Longleftrightarrow \quad \frac{K_{\a}}{K_{\b}} < \frac{1}{\gamma_{\b\a}}
\label{eq:invasion_beta}
\end{equation}

Coexistence via mutual invasibility therefore requires:
\begin{equation}
\gamma_{\a\b} < \frac{K_{\a}}{K_{\b}} < \frac{1}{\gamma_{\b\a}}
\label{eq:mutual_inv}
\end{equation}

\subsection*{Niche Overlap and Fitness Ratio}

Following standard operating definitions in MCT, the niche overlap is defined as the geometric mean of the relative limitation ratios \cite{chesson2018updates}:
\begin{equation}
\tilde{\mathcal{N}} = \sqrt{\gamma_{\a\b}\gamma_{\b\a}} = \sqrt{\frac{A_{\a\b} A_{\b\a}}{A_{\a\a} A_{\b\b}}}
\label{eq:niche_overlap_def}
\end{equation}
This quantity measures how similar the interspecific effects are to the intraspecific effects; $\tilde{\mathcal{N}} = 1$ corresponds to identical niches (neutral competition), while $\tilde{\mathcal{N}} < 1$ indicates niche differentiation.

The coexistence condition \eqref{eq:mutual_inv} can be rewritten in terms of $\tilde{\mathcal{N}}$ by noting that:
\begin{equation}
\gamma_{\a\b} = \tilde{\mathcal{N}}\sqrt{\frac{\gamma_{\a\b}}{\gamma_{\b\a}}}, \quad \frac{1}{\gamma_{\b\a}} = \frac{1}{\tilde{\mathcal{N}}}\sqrt{\frac{\gamma_{\a\b}}{\gamma_{\b\a}}}
\end{equation}

Substituting into \eqref{eq:mutual_inv}:
\begin{equation}
\tilde{\mathcal{N}}\sqrt{\frac{\gamma_{\a\b}}{\gamma_{\b\a}}} < \frac{K_{\a}}{K_{\b}} < \frac{1}{\tilde{\mathcal{N}}}\sqrt{\frac{\gamma_{\a\b}}{\gamma_{\b\a}}}
\end{equation}

Dividing through by $\sqrt{\gamma_{\a\b}/\gamma_{\b\a}}$, we obtain the standard MCT form:
\begin{equation}
\tilde{\mathcal{N}} < \frac{K_{\a}}{K_{\b}}\sqrt{\frac{\gamma_{\b\a}}{\gamma_{\a\b}}} < \frac{1}{\tilde{\mathcal{N}}}
\label{eq:MCT_coexistence}
\end{equation}

The quantity 
\begin{equation}
\mathcal{F} \equiv \frac{K_{\a}}{K_{\b}}\sqrt{\frac{\gamma_{\b\a}}{\gamma_{\a\b}}} = \frac{K_{\a}}{K_{\b}}\sqrt{\frac{A_{\b\a}/A_{\a\a}}{A_{\a\b}/A_{\b\b}}} = \frac{K_{\a}}{K_{\b}}\sqrt{\frac{A_{\b\a}A_{\b\b}}{A_{\a\b}A_{\a\a}}}
\label{eq:fitness_ratio_def}
\end{equation}
is the fitness ratio in MCT. It combines the intrinsic growth rate ratio $K_{\a}/K_{\b}$ with a factor $\sqrt{\gamma_{\b\a}/\gamma_{\a\b}}$ that accounts for asymmetries in competitive relative limitation ratios.

The coexistence condition $\tilde{\mathcal{N}} < \mathcal{F} < 1/\tilde{\mathcal{N}}$ states that coexistence requires the fitness ratio to lie within a window determined by niche overlap. When $\tilde{\mathcal{N}}$ is small (strong niche differentiation), the window is wide and coexistence is robust to fitness ratios. When $\tilde{\mathcal{N}} \to 1$ (identical niches), the window collapses and only $\mathcal{F} = 1$ (perfect fitness equality) permits coexistence.

For equal self-regulation $A_{\a\a} = A_{\b\b} \equiv A$, the relative limitation ratios simplify to:
\begin{equation}
\gamma_{\a\b} = \frac{A_{\a\b}}{A}, \quad \gamma_{\b\a} = \frac{A_{\b\a}}{A}
\end{equation}
and the fitness ratio becomes:
\begin{equation}
\mathcal{F} = \frac{K_{\a}}{K_{\b}}\sqrt{\frac{A_{\b\a}}{A_{\a\b}}}
\label{eq:fitness_ratio_symmetric}
\end{equation}

When additionally the cross-interactions are equal, $A_{\a\b} = A_{\b\a} \equiv B$, we have $\gamma_{\a\b} = \gamma_{\b\a} = B/A = \tilde{\mathcal{N}}$, and the fitness ratio reduces to:
\begin{equation}
\mathcal{F} = \frac{K_{\a}}{K_{\b}}
\end{equation}
In this fully symmetric case, the coexistence condition becomes simply:
\begin{equation}
\tilde{\mathcal{N}} < \frac{K_{\a}}{K_{\b}} < \frac{1}{\tilde{\mathcal{N}}}
\label{eq:coex_fully_symmetric}
\end{equation}

For analytical tractability with Gaussian-distributed growth rates, we convert the ratio condition to a difference condition. This conversion is made to avoid dividing two Gaussians leading to a Cauchy-like distribution (ratio distribution). Such distributions have heavy tails and makes moment calculations difficult thereby precluding analytical expressions for coexistence probabilities (which Gaussians allow for especially when $\s_K << K$).

In the symmetric case \eqref{eq:coex_fully_symmetric}, define $\bar{K} = (K_{\a} + K_{\b})/2$ and $\Delta K = K_{\a} - K_{\b}$, so that $K_{\a} = \bar{K} + \Delta K/2$ and $K_{\b} = \bar{K} - \Delta K/2$. The ratio becomes:
\begin{equation}
\frac{K_{\a}}{K_{\b}} = \frac{\bar{K} + \Delta K/2}{\bar{K} - \Delta K/2}
\end{equation}

The lower bound $\tilde{\mathcal{N}} < K_{\a}/K_{\b}$ rearranges to:
\begin{align}
\tilde{\mathcal{N}}(\bar{K} - \Delta K/2) &< \bar{K} + \Delta K/2 \nonumber\\
\Delta K \cdot \frac{1+\tilde{\mathcal{N}}}{2} &> \bar{K}(\tilde{\mathcal{N}} - 1) \nonumber\\
\Delta K &> -2\bar{K}\frac{1-\tilde{\mathcal{N}}}{1+\tilde{\mathcal{N}}}
\end{align}
Similarly, the upper bound $K_{\a}/K_{\b} < 1/\tilde{\mathcal{N}}$ yields $\Delta K < 2\bar{K}\frac{1-\tilde{\mathcal{N}}}{1+\tilde{\mathcal{N}}}$. Combining:
\begin{equation}
|\Delta K| < 2\bar{K}\frac{1-\tilde{\mathcal{N}}}{1+\tilde{\mathcal{N}}}
\label{eq:coex_difference}
\end{equation}

For $K_{\a}, K_{\b} \sim \mathcal{N}(K, \sigma_K^2)$ independently, we have $\bar{K} \sim \mathcal{N}(K, \sigma_K^2/2)$, so $\bar{K} \approx K$ when $\sigma_K \ll K$. The coexistence threshold is thus approximately $2K(1-\tilde{\mathcal{N}})/(1+\tilde{\mathcal{N}})$. This means that when the difference in strain growth rates is within the calculated threshold, we can expect conspecific strain coexistence. More generally, $|\Delta K|$ tells us how different the two strains are in their growth rates upon which strain competition and exclusion can act on. The RHS of eqn \ref{eq:coex_difference} tells us that when niches overlap completely $\tilde{\mathcal{N}} \rightarrow 1$, even minimal growth rate asymmetry can lead to exclusion of a strain. However, when niche overlap is minimal, $\tilde{\mathcal{N}} \ll 1$, the window of coexistence becomes larger and strains can coexist despite relatively larger growth rate differences.

Thus, the question that follows from this analysis is how community-mediated feedbacks can reshape the conditions for coexistence. In other words, does the presence of a community narrow or widen the coexistence window relative to the isolation case? In Appendix E, we attempt to answer this question.
\section*{Appendix E: Expressing the effective two-strain model in the MCT framework}

\subsection*{Two-strain model in the absence of a community (Baseline Case)}

We consider two closely-related strains $\alpha$ and $\beta$ with GLV dynamics:
\begin{align}
\frac{dN_\alpha}{dt} &= N_\alpha[K_\alpha - N_\alpha - (\tilde{\mu} - \lambda y)N_\beta] \label{eq:baseline_alpha}\\
\frac{dN_\beta}{dt} &= N_\beta[K_\beta - N_\beta - (\tilde{\mu} + \lambda y)N_\alpha] \label{eq:baseline_beta}
\end{align}
where $K_\alpha, K_\beta \sim \mathcal{N}(K, \sigma_K^2)$ are intrinsic growth rates, $\tilde{\mu}$ is mean inter-strain competition, $\lambda$ quantifies strain interaction dissimilarity, and $y \sim \mathcal{N}(0, 1)$ is a random interaction modifier.

Identifying with the general form \eqref{eq:supp_effModel1} and \ref{eq:supp_effModel2}: $A_{\alpha\alpha} = A_{\beta\beta} = 1$, $A_{\alpha\beta} = \tilde{\mu} + \lambda y$, $A_{\beta\alpha} = \tilde{\mu} - \lambda y$.

The relative limitation ratios are:
\begin{equation}
\gamma_{\alpha\beta} = \frac{A_{\alpha\beta}}{A_{\beta\beta}} = \tilde{\mu} + \lambda y, \quad \gamma_{\beta\alpha} = \frac{A_{\beta\alpha}}{A_{\alpha\alpha}} = \tilde{\mu} - \lambda y
\end{equation}

The niche overlap is:
\begin{equation}
\tilde{\mathcal{N}}_{\text{baseline}} = \sqrt{\gamma_{\alpha\beta}\gamma_{\beta\alpha}} = \sqrt{(\tilde{\mu} + \lambda y)(\tilde{\mu} - \lambda y)} = \sqrt{\tilde{\mu}^2 - \lambda^2 y^2}
\label{eq:rho_baseline}
\end{equation}

The fitness ratio \eqref{eq:fitness_ratio_symmetric} is:
\begin{equation}
\mathcal{F}_{\text{baseline}} = \frac{K_\alpha}{K_\beta}\sqrt{\frac{\gamma_{\beta\alpha}}{\gamma_{\alpha\beta}}} = \frac{K_\alpha}{K_\beta}\sqrt{\frac{\tilde{\mu} - \lambda y}{\tilde{\mu} + \lambda y}}
\label{eq:fitness_baseline}
\end{equation}

The coexistence condition $\tilde{\mathcal{N}} < \mathcal{F} < 1/\tilde{\mathcal{N}}$ becomes:
\begin{equation}
\sqrt{\tilde{\mu}^2 - \lambda^2 y^2} < \frac{K_\alpha}{K_\beta}\sqrt{\frac{\tilde{\mu} - \lambda y}{\tilde{\mu} + \lambda y}} < \frac{1}{\sqrt{\tilde{\mu}^2 - \lambda^2 y^2}}
\label{eq:baseline_coexistence_full}
\end{equation}

For small $\lambda$ (similar strains), $\gamma_{\alpha\beta} \approx \gamma_{\beta\alpha} \approx \tilde{\mu}$, so $\sqrt{\gamma_{\beta\alpha}/\gamma_{\alpha\beta}} \approx 1$ and the fitness ratio simplifies to $\mathcal{F} \approx K_\alpha/K_\beta$. The niche overlap becomes $\tilde{\mathcal{N}} \approx \tilde{\mu}$, and the coexistence condition reduces to:
\begin{equation}
\tilde{\mu} < \frac{K_\alpha}{K_\beta} < \frac{1}{\tilde{\mu}}
\label{eq:baseline_coexistence_symmetric}
\end{equation}

Converting to difference form:
\begin{equation}
|K_\alpha - K_\beta| < 2K\frac{1-\tilde{\mu}}{1+\tilde{\mu}}
\label{eq:baseline_coexistence_diff}
\end{equation}

\subsection*{Two-strain model in the presence of a community (Community Feedback Model)}

Using the cavity method, the effective dynamics for strains embedded in a diverse community ($S \to \infty$ species) become:
\begin{align}
0 &= N_{\a}[K^{\text{eff}}_{\alpha} - \xi N_{\a} - ({\tmu^{\mathrm{eff}}} - \lambda y)N_{\beta}] \label{eq:cfm_alpha}\\
0 &= N_{\beta}[K^{\text{eff}}_{\beta} - \xi N_{\beta} - ({\tmu^{\mathrm{eff}}} + \lambda y)N_{\a}] \label{eq:cfm_beta}
\end{align}
with effective parameters:
\begin{align}
K^{\text{eff}}_{\alpha} &= g + \sqrt{\sigma_g^2}\,z^{(\mathrm{int})}_{0\a} + \sqrt{\sigma_K^2}\,z_{0\a} \label{eq:Keff_alpha}\\
K^{\text{eff}}_{\beta} &= g + \sqrt{\sigma_g^2}\left(\rho_{\mathrm{int}} z^{(\mathrm{int})}_{0\a} + \sqrt{1-\rho_{\mathrm{int}}^2}\,z^{(\mathrm{int})}_{0\b}\right) + \sqrt{\sigma_K^2}\,z_{0\b} \label{eq:Keff_beta}\\
\xi &= 1 - 2\sigma^2\rho\, \nu, \quad {\tmu^{\mathrm{eff}}} = \tilde{\mu} - 2\sigma^2\rho\, \nu \label{eq:xi_eta}\\
g &= K - \mu\langle N_\alpha + N_\beta\rangle \label{eq:g}
\end{align}
where $z^{(\mathrm{int})}_{0\a}, z^{(\mathrm{int})}_{0\b}, z_{0\a}, z_{0\b} \sim \mathcal{N}(0,1)$ are independent, $\sigma^2$ is inter-species interaction variance, $\mu$ is mean inter-species competition, $\rho$ is inter-species interaction correlation, and $\nu$ is the community susceptibility.

Identifying with the general form: $A_{\alpha\alpha} = A_{\beta\beta} = \xi$, $A_{\alpha\beta} = {\tmu^{\mathrm{eff}}} - \lambda y$, $A_{\beta\alpha} = {\tmu^{\mathrm{eff}}} + \lambda y$.

The relative limitation ratios are:
\begin{equation}
\gamma_{\alpha\beta} = \frac{{\tmu^{\mathrm{eff}}} - \lambda y}{\xi}, \quad \gamma_{\beta\alpha} = \frac{{\tmu^{\mathrm{eff}}} + \lambda y}{\xi}
\end{equation}

The niche overlap is:
\begin{equation}
\tilde{\mathcal{N}}_{\text{CFM}} = \sqrt{\gamma_{\alpha\beta}\gamma_{\beta\alpha}} = \frac{\sqrt{({\tmu^{\mathrm{eff}}} - \lambda y)({\tmu^{\mathrm{eff}}} + \lambda y)}}{\xi} = \frac{\sqrt{({\tmu^{\mathrm{eff}}})^2 - \lambda^2 y^2}}{\xi}
\label{eq:rho_CFM}
\end{equation}

The fitness ratio is:
\begin{equation}
\mathcal{F}_{\text{CFM}} = \frac{K^{\text{eff}}_{0\alpha}}{K^{\text{eff}}_{0\beta}}\sqrt{\frac{\gamma_{\beta\alpha}}{\gamma_{\alpha\beta}}} = \frac{K^{\text{eff}}_{0\alpha}}{K^{\text{eff}}_{0\beta}}\sqrt{\frac{{\tmu^{\mathrm{eff}}} + \lambda y}{{\tmu^{\mathrm{eff}}} - \lambda y}}
\label{eq:fitness_CFM}
\end{equation}

For small $\lambda$, $\gamma_{\alpha\beta} \approx \gamma_{\beta\alpha} \approx {\tmu^{\mathrm{eff}}}/\xi$, so the fitness ratio simplifies to $\mathcal{F} \approx K^{\text{eff}}_{0\alpha}/K^{\text{eff}}_{0\beta}$ and the niche overlap becomes $\tilde{\mathcal{N}}_{\text{CFM}} \approx {\tmu^{\mathrm{eff}}}/\xi$.

The ratio ${\tmu^{\mathrm{eff}}}/\xi$ is:
\begin{equation}
\frac{{\tmu^{\mathrm{eff}}}}{\xi} = \frac{\tilde{\mu} - 2\sigma^2\rho\, \nu}{1 - 2\sigma^2\rho\, \nu}
\end{equation}

The coexistence condition in difference form becomes:
\begin{equation}
|K^{\text{eff}}_{0\alpha} - K^{\text{eff}}_{0\beta}| < 2g\,\frac{1 - {\tmu^{\mathrm{eff}}}/\xi}{1 + {\tmu^{\mathrm{eff}}}/\xi} = 2g\,\frac{\xi - {\tmu^{\mathrm{eff}}}}{\xi + {\tmu^{\mathrm{eff}}}} = 2g\,\frac{1 - \tilde{\mu}}{1 + \tilde{\mu} - 4\sigma^2\rho\, \nu}
\label{eq:cfm_coexistence}
\end{equation}
where we used $\xi - {\tmu^{\mathrm{eff}}} = 1 - \tilde{\mu}$ and $\xi + {\tmu^{\mathrm{eff}}} = 1 + \tilde{\mu} - 4\sigma^2\rho \nu$. We can see the effects of the community on shaping coexistence by the $4\s^2\r\nu$ term in the denominator of \ref{eq:cfm_coexistence}. As the variance of species interactions increases, the denominator reduces leading to a larger coexistence window.

We can now compute the coexistence probabilities of the strains competing in isolation which we call `Baseline Model' and the strains competing in the presence of the community and associated feedbacks `Community Feedback Model (CFM)'). By calculating their probabilities and taking their ratio ($R_{\mathrm{coexist}}$) we can identify whether the community enhances ($R_{\mathrm{coexist}} > 1$) or hinders strain coexistence ($R_{\mathrm{coexist}}<1$). Note that this form of easy decomposition of fitness and niche differences arises when $\l$ is small. However, with increasing $\l$, it is no longer easy to partition the effects of filtering and sorting into these constituents as they are interconnected in a complicated fashion \cite{song2019consequences}.

For the baseline model in the symmetric limit, the coexistence probability is determined by how often $|K_\alpha - K_\beta|$ falls within the threshold $2K(1-\tilde{\mu})/(1+\tilde{\mu})$.

Since $K_\alpha, K_\beta \sim \mathcal{N}(K, \sigma_K^2)$ independently, $\Delta K = K_\alpha - K_\beta \sim \mathcal{N}(0, 2\sigma_K^2)$.

For $X \sim \mathcal{N}(0, \sigma^2)$, we have $P(|X| < T) = \text{erf}(T/(\sigma\sqrt{2}))$. With $\sigma^2 = 2\sigma_K^2$ and $T = 2K(1-\tilde{\mu})/(1+\tilde{\mu})$:
\begin{equation}
P_{\text{baseline}} = \text{erf}\left(\frac{2K(1-\tilde{\mu})}{(1+\tilde{\mu})\sqrt{2}\cdot\sqrt{2}\sigma_K}\right) = \text{erf}\left(\frac{K(1-\tilde{\mu})}{\sigma_K(1+\tilde{\mu})}\right)
\label{eq:P_baseline}
\end{equation}

For the CFM, from Eqs.~\eqref{eq:Keff_alpha}--\eqref{eq:Keff_beta}:
\begin{equation}
K^{\text{eff}}_{0\alpha} - K^{\text{eff}}_{0\beta} = \sqrt{\sigma_g^2}\left[(1-\rho_{\mathrm{int}})z^{(\mathrm{int})}_{0\a} - \sqrt{1-\rho_{\mathrm{int}}^2}\,z^{(\mathrm{int})}_{0\b}\right] + \sqrt{\sigma_K^2}(z_{0\a} - z_{0\b})
\end{equation}

Since $z^{(\mathrm{int})}_{0\a}, z^{(\mathrm{int})}_{0\b}, z_{0\a}, z_{0\b}$ are independent standard normals:
\begin{align}
\text{Var}(\Delta K^{\text{eff}}) &= \sigma_g^2\left[(1-\rho_{\mathrm{int}})^2 + (1-\rho_{\mathrm{int}}^2)\right] + 2\sigma_K^2 \nonumber\\
&= \sigma_g^2\left[2 - 2\rho_{\mathrm{int}}\right] + 2\sigma_K^2 = 2\sigma_g^2(1-\rho_{\mathrm{int}}) + 2\sigma_K^2
\end{align}

With the coexistence conditions in the community feedback model being $2g(1-\tilde{\mu})/(1+\tilde{\mu}-4\sigma^2\rho \nu)$ and variance $2[\sigma_g^2(1-\rho_{\mathrm{int}}) + \sigma_K^2]$:
\begin{equation}
P_{\text{CFM}} = \text{erf}\left(\frac{g(1-\tilde{\mu})}{(1+\tilde{\mu}-4\sigma^2\rho \nu)\sqrt{\sigma_g^2(1-\rho_{\mathrm{int}}) + \sigma_K^2}}\right)
\label{eq:P_CFM}
\end{equation}

Next, we define the arguments of the error functions:
\begin{equation}
\Theta_{\text{baseline}} = \frac{K(1-\tilde{\mu})}{\sigma_K(1+\tilde{\mu})}, \quad \Theta_{\text{CFM}} = \frac{g(1-\tilde{\mu})}{(1+\tilde{\mu}-4\sigma^2\rho \nu)\sqrt{\sigma_g^2(1-\rho_{\mathrm{int}}) + \sigma_K^2}}
\end{equation}

Since $\text{erf}(x)$ is monotonically increasing, $P_{\text{CFM}} > P_{\text{baseline}}$ iff $\Theta_{\text{CFM}} > \Theta_{\text{baseline}}$. The ratio is:
\begin{align}
\frac{\Theta_{\text{CFM}}}{\Theta_{\text{baseline}}} &= \frac{g}{K} \cdot \frac{1+\tilde{\mu}}{1+\tilde{\mu}-4\sigma^2\rho \nu} \cdot \frac{\sigma_K}{\sqrt{\sigma_g^2(1-\rho_{\mathrm{int}}) + \sigma_K^2}} \nonumber\\[6pt]
&= \underbrace{\frac{g}{K}}_{\substack{\text{mean fitness} \\ \text{reduction}}} \times \underbrace{\frac{1+\tilde{\mu}}{1+\tilde{\mu}-4\sigma^2\rho \nu}}_{\substack{\text{niche enhancement} \\ \text{}}} \times \underbrace{\sqrt{\frac{\sigma_K^2}{\sigma_g^2(1-\rho_{\mathrm{int}}) + \sigma_K^2}}}_{\substack{\text{variance ratio} \\ \text{(measure of dispersion of fitness ratios)}}}
\label{eq:theta_ratio}
\end{align}

As shown by the underbraces in \ref{eq:theta_ratio}, there are three different biological mechanisms that contribute to the coexistence probability ratio that are different for the baseline model and CFM respectively. Note that these three mechanisms are not independent but partitioned in a manner to highlight the different drivers of coexistence. 

The first factor, $\frac{g}{K}$, captures mean fitness reduction. Because strains embedded in a community competes with strains from other species, their effective mean growth rate g is lower than the intrinsic carrying capacity K. This shrinks the coexistence window for strain. This effect is always present in competitive communities $\mu >0 $. 

The second factor shows that community feedbacks reduce the effective inter-strain competition and self-regulation of the strain i.e. $\tmu \rightarrow \tmueff$ and $1 \rightarrow\xi$. The ultimate effect is a decrease in the effective niche overlap which increases the probability of coexistence and increases with increasing species interaction dissimilarity ($\s$).

The third factor captures relative strengths of equalization of strain growth rates and asymmetry in competitive ability between strains. When strains are vary similar ($\l \rightarrow 0$), $\r_{\mathrm{int}} \approx 1$ and this ratio approaches 1. Here, community feedbacks equalize strain growth rates. However, with increasing $\l$, strains get more dissimilar making their competitive ability asymmetric. This leads either strain excluding the other thereby reducing the probability of coexistence.

\begin{figure*} [ht] 
    \RaggedRight
    \includegraphics[width=1.0\textwidth]{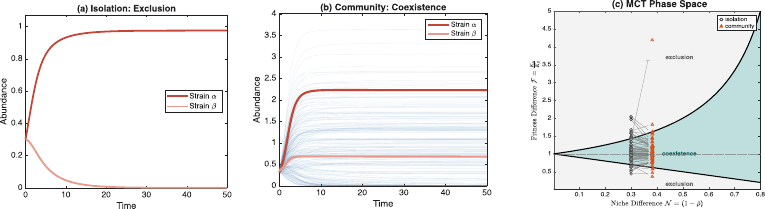}
    \caption{\justifying\textsf{\textbf{Community feedback promotes strain coexistence.} Panel (A) shows two strains ($\alpha$ and $\beta$) of a focal species competing in isolation where one goes extinct. Panel (B) shows that when embedded in a diverse community of $S=100$ species (light blue trajectories), the same strain pair can undergo coexistence. Panel (C) Modern Coexistence Theory axes (i.e. Fitness-Niche Difference space) showing niche difference ($\mathcal{N} = 1-\tilde\rho$) versus fitness ratio ($\mathcal{F}$) for all strain pairs. Open circles indicate isolation; filled triangles indicate community context. Arrows connect corresponding pairs, showing how community feedback shifts their positions. The teal region permits coexistence ($\rho < \mathcal{F} < 1/\rho$); gray regions lead to exclusion. Community interactions, via equalization of growth rates can reduce the variance in fitness ratios while also modifying reducing niche differences. Parameters: $\sigma = 0.38$, $\sigma_K = 0.2$, $\lambda = 0.10$, $\tilde{\mu} = 0.70$, $\rho = 0.99$.}}
    \label{fig:supp_coexistence_gain}
\end{figure*}

\begin{figure*} [t] 
    \centering
    \includegraphics[width=0.95\textwidth]{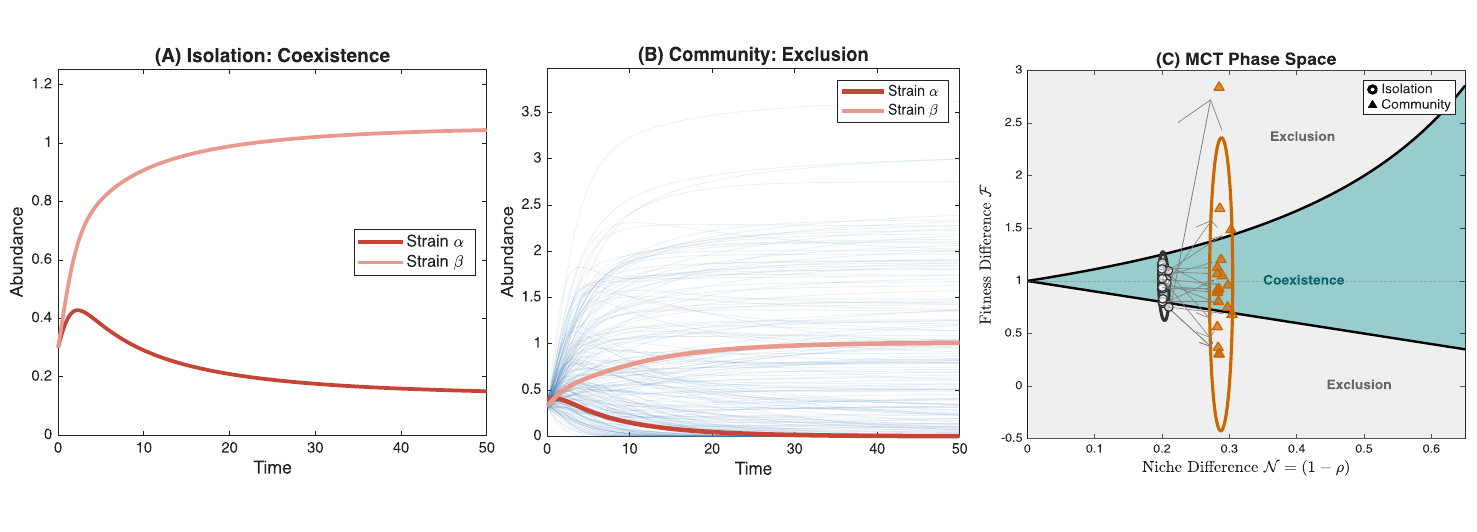}
    \caption{\justifying\textsf{\textbf{Community feedbacks can also suppress strain coexistence of dissimilar strains.} Panel (A) shows two strains ($\alpha$ and $\beta$) of a focal species coexist stably when competing in isolation. Panel (B) shows that when embedded in a diverse community of $S=100$ species (light blue trajectories), the same strain pair undergoes competitive exclusion, with one strain driven to extinction. Panel (C) Modern Coexistence Theory axes (i.e. Fitness-Niche Difference space) showing niche difference ($\mathcal{N} = 1-\rho$) versus fitness ratio ($\mathcal{F}$) for all strain pairs. Open circles indicate isolation; filled triangles indicate community context. Arrows connect corresponding pairs, showing how community feedback shifts their positions. The teal region permits coexistence ($\rho < \mathcal{F} < 1/\rho$); gray regions lead to exclusion. Ellipses show two standard deviation confidence regions. Community interactions can push strain pairs outside the coexistence boundary by amplifying fitness ratios while also modifying reducing niche differences. Parameters: $\sigma = 0.38$, $\sigma_K = 0.15$, $\lambda = 0.10$, $\tilde{\mu} = 0.80$, $\rho = 0.99$.}}
    \label{fig:supp_coexistence_loss}
\end{figure*}

Rearranging the terms, we can obtain an inequality where CFM promotes coexistence by requiring:
\begin{equation}
\left(\frac{g}{K}\right)^2 \left(\frac{1+\tilde{\mu}}{1+\tilde{\mu}-4\sigma^2\rho \nu}\right)^2 > 1 + \frac{\sigma_g^2(1-\rho_{\mathrm{int}})}{\sigma_K^2}
\label{eq:CFM_promotes}
\end{equation}

This expression however, contains a number of emergent parameters from the model which obfuscate its biological meaning.   Using the expressions from equations \ref{eq:rho_K} and from the definition of g and $\s_g$, we can rewrite \ref{eq:CFM_promotes} as

\begin{equation}
    \left(1-\frac{\mu(\langle N_\a +N_\b \rangle)}{K}\right)^2\left(\frac{1+\tilde{\mu}}{1+\tilde{\mu} - 4\s^2\r\nu}\right)^2 > 1+ \frac{\l^2 \left \langle (N_\a - N_\b)^2 \right \rangle}{\s_K^2}
\end{equation}

In the limit of low $\l \rightarrow 0$ i.e. very similar strains, we can simplify the above equation leading to

\begin{equation}
    \left(1 - \frac{2\mu\langle N \rangle)}{K}\right)^2 \left(\frac{1+\tilde{\mu}}{1+\tilde{\mu} - 4\s^2\r\nu}\right)^2 > 1
\end{equation}

In competitive communities, $\mu > 0$ and the abundances of the strains are greater than 0. This implies that the first term on the LHS is less than 1. Further, the second term is greater than 1 in the presence of a community (i.e. when $\s > 0$). Their product requires to be greater than the RHS. At low $\l$, we can approximate the bounds of the RHS leading to being equal to 1. The lower values of the RHS increase the likelihood that CFM promotes coexistence relative to the baseline model. In the other limit, as $\l \rightarrow \s$, we see that the RHS increases from 1 monotonically to an upper bound $1+\frac{\s^2}{\s_K^2}\left \langle (N_\a - N_\b)^2 \right \rangle$ implying that the requirement of LHS is now greater to ensure that the community enhances coexistence. In this limit, biologically, we see that the ratio of the variances of species interaction dissimilarity ($\s^2$)to independent growth rate fluctuations ($\s_K^2)$ is a key factor in determining coexistence enhancement or hindrance by a community (relative to isolated strain pairs).

In the main text in Figure \ref{fig:coexistence}d, we show an instance of how community enhances the probability of coexistence at low $\l$ of the two strains. In Figure \ref{fig:supp_coexistence_loss}, we show the opposite outcome at a larger value of $\l$. We also show how, in this case, the community reduces coexistence probability by increasing fitness ratios to a greater degree than increasing niche differences.

We also explore the relative effects and interplay of species interaction dissimilarity ($\s$) and independent variation in growth rates $\s_K^2$ in determining increased or lowered probability of coexistence ($R_{\mathrm{coexist}} >\mathrm{or} < 1$). From the inequality in eqn \ref{eq:CFM_promotes}, we can identify that $\s_K$ sets the scale of intrinsic growth rate variation between strains. When $\s_K$ is small (Figure ~\ref{fig:variance effects}a,c), growth rates are nearly identical across strains, so even a small community-mediated perturbation to the growth rate difference can tip the competitive balance from coexistence to exclusion. The community's sorting effect $\l^2\langle D^2 \rangle$, becomes the dominant source of fitness ratios, and the stronger the species interaction dissimilarity ($\s$), the stronger these effects due to increased differences in strain abundances i.e. higher $\langle D^2 \rangle$. By amplifying the variance of the strains, $\s$ while strongly correlating their abundances, also increases the probability of both strains going extinct. This results in the community maintaining (or reducing; as $\s$ increases) the coexistence probability of two interacting strains relative to the isolation case. When $\s_K$ is large (Figure ~\ref{fig:variance effects}b,d), the intrinsic growth rate variation already dominates the fitness ratio, and the additional community effects through $\s$ are a relatively small perturbation. In this regime, the community's niche-enhancing effect via $\s$, which reduces effective niche overlap independently of the fitness ratio, can operate without being undermined by the sorting effects from interactions.


\begin{figure*} [t] 
    \centering
    \includegraphics[width=0.95\textwidth]{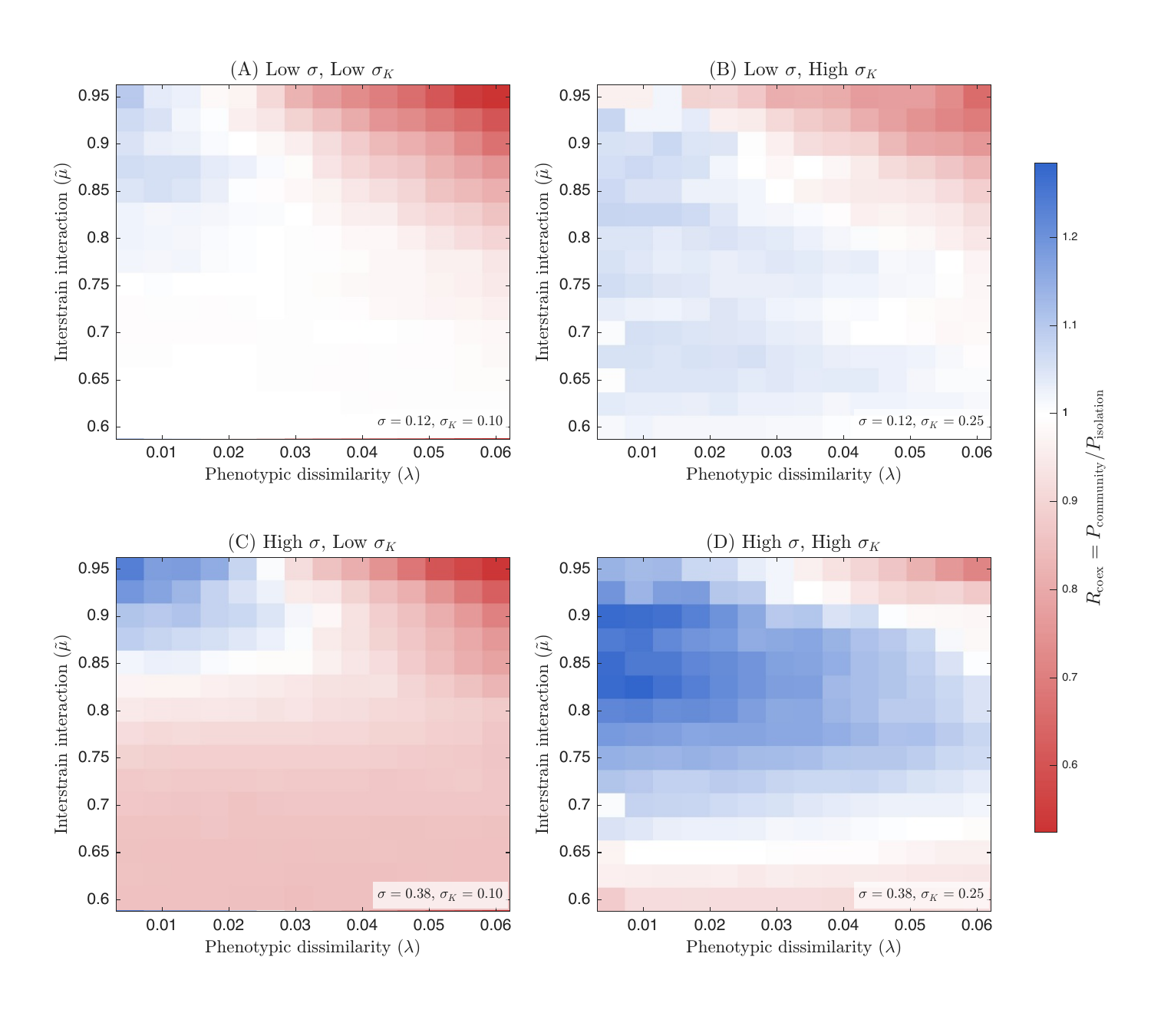}
    \caption{\justifying\textsf{\textbf{Effect of interaction variance $\s$ and growth rate heterogeneity $\s_K$ on community-mediated coexistence.} Phase diagrams showing the coexistence ratio $R_{\mathrm{coex}} = P_{\mathrm{community}}/P_{\mathrm{isolation}}$ across strain interaction dissimilarity ($\lambda$) and inter-strain interaction strength ($\tilde{\mu}$) for four parameter combinations. In Panel(A), we show $R_{\mathrm{coexist}}$ heatmap for low inter-species interaction variance ($\sigma = 0.12$) and low growth rate heterogeneity ($\sigma_K = 0.10$). In Panel (B), we look at the case of low inter-species interaction variance ($\sigma = 0.12$) and high growth rate heterogeneity ($\sigma_K = 0.25$). In Panel (C), we look at the case of high inter-species interaction variance ($\sigma = 0.38$) and low growth rate heterogeneity ($\sigma_K = 0.01$).  In Panel (D), we look at the case of high inter-species interaction variance ($\sigma = 0.38$) and high growth rate heterogeneity ($\sigma_K = 0.25$). Blue regions ($R_{\mathrm{coex}} > 1$) indicate community-enhanced coexistence; red regions ($R_{\mathrm{coex}} < 1$) indicate community-suppressed coexistence; white corresponds to $R_{\mathrm{coex}} = 1$ (no net effect). Higher inter-species interaction variance ($\sigma_g$) generally strengthens community feedback effects, while growth rate heterogeneity ($\sigma_K$) modulates the balance between enhancement and suppression. All panels share a common colour scale. Parameters: $S = 100$ species, $\rho = 0.99$.}}
    \label{fig:variance effects}
\end{figure*}

The comparison between baseline and CFM coexistence probabilities reveals competing mechanisms that can either promote or hinder strain coexistence. Firstly, we know that the mean effective growth rate $g$ satisfies a self-consistency condition. At equilibrium:
\begin{equation}
\langle N_\alpha + N_\beta \rangle = \frac{2g(\xi - {\tmu^{\mathrm{eff}}})}{\xi^2 - ({\tmu^{\mathrm{eff}}})^2 + \lambda^2 y^2}
\end{equation}
Substituting into $g = K - \mu\langle N_\alpha + N_\beta \rangle$ and working in the low $\l$ regime, we can solve and obtain:
\begin{equation}
g = \frac{K}{1 + \frac{2\mu(\xi - {\tmu^{\mathrm{eff}}})}{\xi^2 - ({\tmu^{\mathrm{eff}}})^2}}
\label{eq:g_self_consistent}
\end{equation}

Therefore, we can understand the 3 mechanisms shown in \ref{eq:theta_ratio} at the low $\l$ regime. Firstly, for small $\mu$: $g/K \approx 1 - 2\mu/(\xi + {\tmu^{\mathrm{eff}}})$, showing that mean fitness reduction scales linearly with $\mu$. Thus, the net community feedback ultimately reduces the mean effective growth rate relative to the growth rates of the carrying capacities of strains in isolation.

Secondly, community feedback reduces the effective niche overlap from $\tilde{\mu}$ to ${\tmu^{\mathrm{eff}}}/\xi < \tilde{\mu}$ (when $\nu > 0$). This can be verified: ${\tmu^{\mathrm{eff}}}/\xi = (\tilde{\mu} - 2\sigma^2\rho \nu)/(1 - 2\sigma^2\rho \nu) < \tilde{\mu}$ requires $\tilde{\mu} - 2\sigma^2\rho \nu < \tilde{\mu}(1 - 2\sigma^2\rho \nu)$, which simplifies to $1 > \tilde{\mu}$, always true. This stabilizing effect scales as $\sigma^2\rho \nu$.

Thirdly, when $\lambda \approx 0$, the correlation $\rho_{\mathrm{int}} \to 1$, synchronizing fitness fluctuations. The variance of fitness ratios $2\sigma_g^2(1-\rho_{\mathrm{int}}) + 2\sigma_K^2$ approaches the baseline $2\sigma_K^2$. When $\lambda > 0$ and is increased, $\rho_{\mathrm{int}}$ decreases, increasing the uncorrelated variance contribution. Further, increasing $\l$ also increases the variance in growth rates ($\s_g$) in addition decorrelating fluctuations.




Thus, community feedback stabilizes through emergent niche differentiation and correlated fluctuations, but destabilizes through mean fitness reduction and variance amplification. Thus, we predict that high-variance environments in weakly competitive communities and similar strains favor CFM whereas low-variance environments with strongly competing  communities and divergent strains disfavor it.

\section*{Appendix F: Strain abundance correlations in a community versus in isolation} \label{app:isolation}

In this section, we first calculate the how the strain abundance correlation changes with increasing strain interaction dissimilarity between strains embedded in a community. Next, we identify an expression for the value of $\l_c$, a critical value of lambda beyond which positive correlations between strains turn negative. We present a model of two competing strains in isolation (for comparison) and show that it is almost impossible to obtain positive correlations between these strains. Lastly, we show that even if we relax our assumption that the growth rates of strains in isolation are uncorrelated, the presence of a community only further enhances the positive correlation between them. 

\subsection*{Crossover strain interaction dissimilarity value for transition from positive to negative values of strain abundance correlation for competing strains in a community}

The correlation $\rho_{\alpha\beta}(\lambda)$ is entirely controlled by the sign of the covariance~\eqref{eq:cov-full}. We therefore define the crossover value $\lambda_c$ as the first value of $\lambda$ at which the covariance changes sign:
\begin{equation}
\mathrm{Cov}\big(N_{0\alpha},N_{0\beta}\big)\big|_{\lambda=\lambda_c}
=
0.
\end{equation}
Using Eq.~\eqref{eq:cov-full}, this condition reads
\begin{equation}
\label{eq:lambda_c_equation}
g^2(1-\tilde{\mu})^2 \left( \left\langle\frac{1}{\Delta^2}\right\rangle - \left\langle \frac{1}{\Delta} \right\rangle^2\right)
+\s_g^2 \rho_{\mathrm{int}}(1-\tilde{\mu})^2  \left\langle\frac{1}{\Delta^2}\right\rangle
-
2 {\tmu^{\mathrm{eff}}} \xi\left( \left( 1-\rho_{\mathrm{int}}\right)\, \s_g^2 + \s_K^2 \right) \left\langle \frac{1}{\Delta^2} \right\rangle
-
\lambda^2\big(g^2 + \rho_{\mathrm{int}} \s_g^2\big)\,\left\langle \frac{y^2}{\Delta^2} \right\rangle
=
0,
\end{equation}
where $\left\langle \frac{1}{\Delta} \right\rangle$, $\left\langle \frac{1}{\Delta^2} \right\rangle$ and $\left\langle \frac{y^2}{\Delta^2} \right\rangle$ are the averages defined above, evaluated at
\begin{equation}
C = \xi(\lambda_c)^2 - {\tmu^{\mathrm{eff}}}(\lambda_c)^2,
\qquad
B = \lambda_c^2,
\end{equation}
and the dependence of $\xi$, ${\tmu^{\mathrm{eff}}}$ and $\rho_{\mathrm{int}}$ on $\lambda$ enters via the self--consistency of the cavity solution for $\nu$, $g^2$ and $\s_g^2$.

In the full cavity theory, all of the quantities $g^2$, $\s_g^2$, $\nu$, $\rho_{\mathrm{int}}$ and hence $(\xi,{\tmu^{\mathrm{eff}}},C,\left\langle \frac{1}{\Delta} \right\rangle,\left\langle \frac{1}{\Delta^2} \right\rangle,\left\langle \frac{y^2}{\Delta^2} \right\rangle)$ are determined self--consistently as functions of $(\tilde{\mu},\s,\lambda)$ and the underlying random--matrix parameters. In Figure \ref{fig:dynamics}a, we plot our theoretical line (in black) by solving the equation \ref{eq:lambda_c_equation}, for the necessary parameter values. For intuition, we show below a simplified description of $\l_c$, by identifying that the first term of \ref{eq:cov-full} is, in the parameter regimes we work in, 0 (simulations show it is orders of magnitude smaller than the other terms of the expression). This provides a simplified expression for Eq.~\eqref{eq:lambda_c_equation} which we can write as
\begin{equation}
\label{eq:lambda_c_explicit_formal}
\lambda_c^2
=
\frac{
\left[\s_g^2 \rho_{\mathrm{int}}(1-\tilde{\mu})^2 
-
2 {\tmu^{\mathrm{eff}}} \xi\left( \left( 1-\rho_{\mathrm{int}}\right)\, \s_g^2 + \s_K^2 \right)\right] \left\langle \frac{1}{\Delta^2} \right\rangle
}{
\big(g^2 + \rho_{\mathrm{int}}\,\s_g^2\big)\,\left\langle \frac{y^2}{\Delta^2} \right\rangle
},
\end{equation}
which makes explicit how the crossover emerges. In particular, as $\s_K \rightarrow 0 $, the expression further reduces to 

\begin{equation}
\label{eq:lambda_c_explicit_simple}
\lambda_c^2
=
\frac{
\left[\s_g^2 \left(\rho_{\mathrm{int}}(1-\tilde{\mu})^2 
-
\left( 1-\rho_{\mathrm{int}}\right)2 {\tmu^{\mathrm{eff}}} \xi \right)\right] \left\langle \frac{1}{\Delta^2} \right\rangle
}{
\big(g^2 + \rho_{\mathrm{int}}\,\s_g^2\big)\,\left\langle \frac{y^2}{\Delta^2} \right\rangle
},
\end{equation}

shows that the numerator (and its sign) arises from (i) a balance between two quantities weighted by $\rho_{\mathrm{int}}$, which correlates the fluctuations of the strains and  (ii) the purely strain--level sorting contribution in the denominator. A higher value of $\rho_{\mathrm{int}}$ guarantees a larger $\l_c$ required for a crossover. Therefore, greater correlation of strains (generated by larger variation in species interaction dissimilarity $\s^2$ will require greater strain interaction dissimilarity ($\l$) to crossover into negatively correlated abundances.

We next derive the condition that separates the region where correlations can transition from positive to negative (as $\lambda$ increases) from the region where correlations are always negative. For Eq. ~\ref{eq:cov-full} as $\lambda \to 0$, we have $D = \lambda^2 \to 0$, so $\Delta = C + Dy^2 \to C$ becomes deterministic, and $\rho_{\mathrm{int}} \to 1$ from equation since $\lambda^2\langle D^2\rangle \to 0$. The expectations over $y$ therefore simplify as $\langle 1/\Delta \rangle$ becomes $ 1/C$, $\langle 1/\Delta^2 \rangle$ becomes $1/C^2$, and consequently $\langle 1/\Delta^2 \rangle - \langle 1/\Delta \rangle^2 $ goes to 0. At $\rho_{\mathrm{int}} = 1$,the sorting term $\lambda^2\langle y^2/\Delta^2 \rangle = 0$ since $\lambda^2$ goes to $0$ faster than any potential divergence in the expectation which implies $\s_g^2 = \s^2\left \langle \left(N_\alpha + N_\beta \right)^2 \right \rangle$.

Combining these results, the limiting covariance of the strain abundances is
\begin{equation}
\lim_{\lambda \to 0} \text{Cov}(N_\alpha, N_\beta) = \frac{1}{C^2}\left[ \s^2(1-\tilde{\mu})^2\left \langle \left(N_\alpha + N_\beta \right)^2 \right \rangle - 2\xi{\tmu^{\mathrm{eff}}}\sigma_K^2\right].
\end{equation}
Since $C^2 > 0$ assuming the system is stable, the sign of the covariance is determined by $\text{sign}(\text{Cov}) = \text{sign}[\s^2 \langle M^2 \rangle(1-\tilde{\mu})^2 - 2\xi{\tmu^{\mathrm{eff}}}\sigma_K^2]$.

The phase boundary between regimes occurs when $\s^2(1-\tilde{\mu})^2\left \langle \left(N_\alpha + N_\beta \right)^2 \right \rangle = 2\xi{\tmu^{\mathrm{eff}}}\sigma_K^2$, where $\xi = 1 - 2\sigma^2\rho \nu$ and ${\tmu^{\mathrm{eff}}} = \tilde{\mu} - 2\sigma^2\rho \nu$. When $\sigma^2(1-\tilde{\mu})^2\left \langle \left(N_\alpha + N_\beta \right)^2 \right \rangle > 2\xi{\tmu^{\mathrm{eff}}}\sigma_K^2$, correlations are positive at small $\lambda$ and can transition to negative as $\lambda$ increases. When $\sigma^2(1-\tilde{\mu})^2\left \langle \left(N_\alpha + N_\beta \right)^2 \right \rangle < 2\xi{\tmu^{\mathrm{eff}}}\sigma_K^2$, correlations are negative even at small $\lambda$ and remain negative throughout. When this inequality becomes an equality, the limiting covariance of the conspecific strain abundances expression serves as the basis for the theory line calculated in the phase diagram in Figure 4 (panel c) which tells us when correlations transition from positive to negative.

\subsection*{Comparison to the two-strain model in isolation}

To show that the presence of the community serves to correlate the strain growth rates, we can study the dynamics of the two strains in isolation. The equivalent two strain model can be written as follows:

\begin{align}
    \frac{dN_\alpha}{dt} &= N_\alpha\bigl[K_\alpha - N_\alpha - (\tmu - \lambda y)\,N_\beta\bigr], \label{eq:dyn_alpha}\\[6pt]
    \frac{dN_\beta}{dt} &= N_\beta\bigl[K_\beta - N_\beta - (\tmu + \lambda y)\,N_\alpha\bigr]. \label{eq:dyn_beta}
\end{align}

At steady state, ($dN_\alpha/dt = dN_\beta/dt = 0$), and assuming both strains coexist ($N_\alpha, N_\beta > 0$), we set the per-capita growth rates to zero:
\begin{align}
    K_\alpha &= N_\alpha + (\tmu - \lambda y)\,N_\beta, \label{eq:ss1}\\
    K_\beta  &= N_\beta + (\tmu + \lambda y)\,N_\alpha. \label{eq:ss2}
\end{align}
 where $K_\alpha, K_\beta \sim \mathcal{N}(K, \sigma_K^2)$ are independently drawn intrinsic growth rates, $\tmu$ is the mean inter-strain competition strength, $\lambda$ quantifies strain interaction dissimilarity between strains, and $y \sim \mathcal{N}(0,1)$ is a random variable that modifies interactions from realisation to realisation.
The determinant of the 2x2 interaction matrix is:
\begin{equation}
    \bar{\Delta} = 1 - (\tmu - \lambda y)(\tmu + \lambda y) = 1 - \tmu^2 + \lambda^2 y^2.
\end{equation}
which can be solved at equilibrium to give 
\begin{align}
    N_\alpha^* &= \frac{K_\alpha - (\tmu - \lambda y)\,K_\beta}{\bar{\Delta}}, \label{eq:Nalpha}\\[6pt]
    N_\beta^*  &= \frac{K_\beta - (\tmu + \lambda y)\,K_\alpha}{\bar{\Delta}}. \label{eq:Nbeta}
\end{align}

We can take the expectation values of these quantities, their squares as well as their product i.e. $\langle N_\a \rangle$, $\langle N_\a^2 \rangle$ (and similarly for strain $\b$) and $\langle N_\a N_\b\rangle$. We obtain,

$$\langle N_\alpha^*    \rangle = \frac{K(1-\tilde{\mu}+\lambda y)}{\bar{\Delta}}, \qquad \langle N_\beta^*    \rangle = \frac{K(1-\tilde{\mu}-\lambda y)}{\bar{\Delta}}$$

with the cross-moment being

$$\langle N_\alpha^* N_\beta^* \rangle = \frac{K^2(1-\tilde{\mu}+\lambda y)(1-\tilde{\mu}-\lambda y) - 2\tilde{\mu}\,\sigma_K^2}{\bar{\Delta}^2}$$

and the second moments being

$$\langle N_\alpha^{*2}    \rangle = \frac{K^2(1-\tilde{\mu}+\lambda y)^2 + \sigma_K^2[1 + (\tilde{\mu}-\lambda y)^2]}{\bar{\Delta}^2}$$

$$\langle N_\beta^{*2}    \rangle = \frac{K^2(1-\tilde{\mu}-\lambda y)^2 + \sigma_K^2[1 + (\tilde{\mu}+\lambda y)^2]}{\bar{\Delta}^2}$$

Averaging over $y$, we get 1st moments

\begin{equation}
    \langle N_\alpha^* \rangle = \langle N_\beta^* \rangle = K(1-\tilde{\mu})\left\langle \frac{1}{\bar{\Delta}} \right\rangle
\end{equation}

cross-moment

\begin{equation}
    \langle N_\alpha^* N_\beta^* \rangle = K^2\left\langle\frac{(1-\tilde{\mu})^2}{\bar{\Delta}^2} \right\rangle {-\; 2\tilde{\mu}\,\sigma_K^2\left\langle\frac{1}{ \bar{\Delta}^2}\right\rangle} - \lambda^2 \left \langle \frac{ y^2}{\bar{\Delta}^2} \right \rangle
\end{equation}

and 2nd moments

\begin{equation}
    \langle N_\alpha^{*2} \rangle = \langle N_\b^{*2} \rangle =K^2\left\langle\frac{(1-\tilde{\mu})^2}{\bar{\Delta}^2} \right\rangle+ \lambda^2 \left \langle \frac{ y^2}{\bar{\Delta}^2} \right \rangle+ \sigma_K^2\left\langle\frac{1+\tilde{\mu}^2+\lambda^2 y^2}{  \bar{\Delta}^2}\right\rangle.
\end{equation}

Thus, the covariance of the abundances of the strains at equilibrium is simply

\begin{equation}
    \mathrm{Cov}(N_\a^*,N_\b^*) = K^2(1-\tilde{\mu})^2 \left(\left\langle\frac{1}{\bar{\Delta}^2}  \right\rangle - \left\langle\frac{1}{\bar{\Delta}}\right\rangle^2 \right) {-\; 2\tilde{\mu}\,\sigma_K^2\left\langle\frac{1}{  \bar{\Delta}^2}\right\rangle} - \lambda^2 \left \langle \frac{ y^2}{\bar{\Delta}^2} \right \rangle.
\end{equation}

In the limit of low $\l$, this can be approximately simplified to 

\begin{figure*} [t] 
    \centering
    \includegraphics[width=0.95\textwidth]{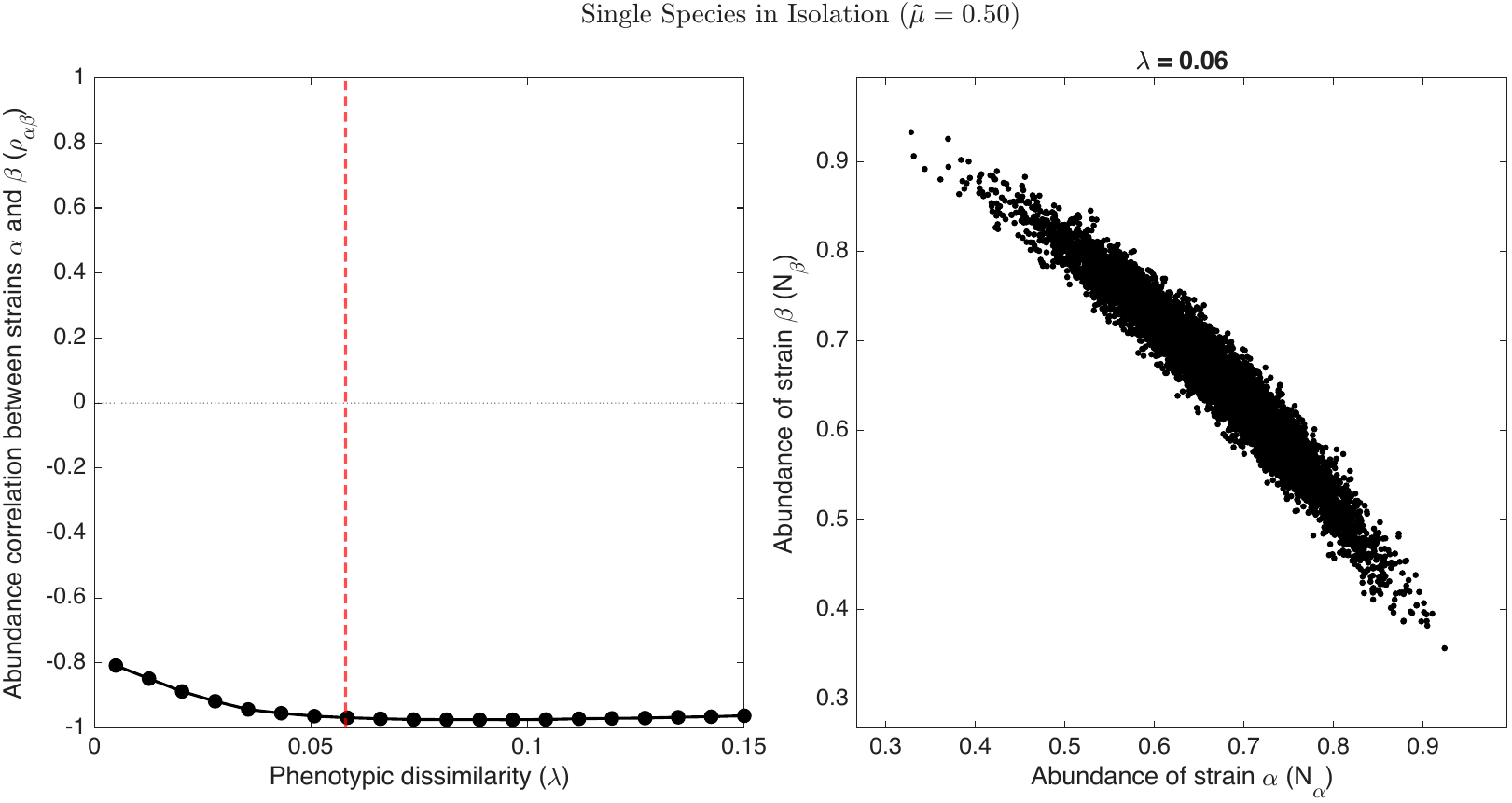}
    \caption{\justifying\textsf{\textbf{Negative abundance correlation between 2 strains ($\a$ and $\b$) within a single species in isolation, i.e., in the absence of a community.} Left panel shows the average steady-state  abundance correlation between the two strains $\rho_{\a\b} = \mathrm{corr}(N_\a,N_\b)$ as a function of strain interaction dissimilarity $\l$. Strain abundance correlations are negative even at low $\l$ and become increasingly negative with increasing $\l$. Right panel shows a scatter plot of strain abundances of the same species $(N_\a,N_\b)$ at steady state for 1,000 separate realizations (scatter shown for $\l=0.06$). Parameters: $\tilde{\mu} = 0.5, \l \in [0,0.15], \s_K = 0.02$.}}
    \label{fig: correlation_isolation}
    \end{figure*}

\begin{figure*} [t] 
    \centering
    \includegraphics[width=1.0\textwidth]{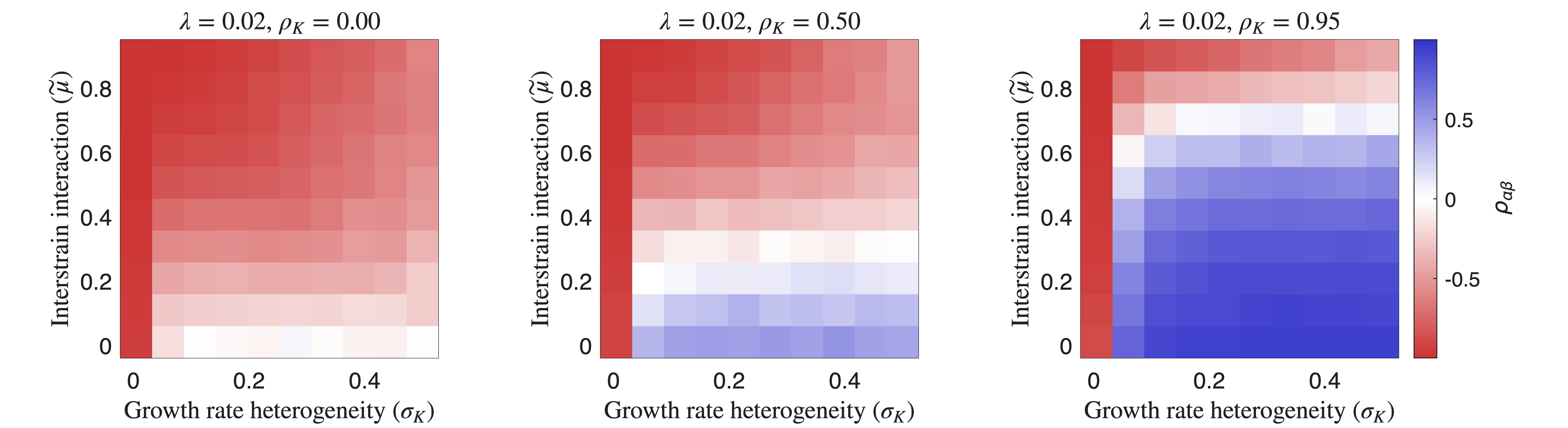}
    \caption{\justifying\textsf{\textbf{Strain abundance correlation $\rho_{\alpha\beta}$ in pairwise isolation across inter-strain interaction strength ($\tilde{\mu}$) and growth rate heterogeneity ($\sigma_K$)}, for increasing carrying-capacity correlations $\rho_K$ (0.00, 0.50, 0.95) at fixed strain interaction dissimilarity $\lambda = 0.02$. Each heatmap is computed from simulations integrated to steady-state, with mean carrying capacity $\bar{K} = 1$. As $\rho_K$ increases, negative correlations at several values of (until $\tilde{\mu} \sim 0.7$) are progressively replaced by positive correlations, particularly at low $\tilde{\mu}$ and high $\sigma_K$. Parameters: $\tilde{\mu} \in [0.01, 0.90]$, $\sigma_K \in [0.005, 0.50]$ with 10 equally spaced values.}}
    \label{fig: mu_tilde_vs_sigK}
\end{figure*}

\begin{equation}
    \mathrm{Cov}(N_\a^*,N_\b^*) = \frac{-2\tmu\s_K^2 - \l^2}{1-\tmu^2}
\end{equation}
The difference between the first and second moments of $\frac{1}{  \bar{\Delta}}$ is almost exactly zero (observed from calculations for all values of $\l$); see Appendix C for expressions) and hence we are left with the second and third terms of the covariance which are both less than 0. This implies that barring values of $\tmu \approx 0$, the correlations between strains in isolation are negative (or at best 0). Since we work in the regime of $\tmu \sim O(1)$  noting that strains of a species compete stronger with one another than with different species, we need not worry about the case of $\tmu \approx 0$. The negative abundance correlation of conspecific strains could also be lost due to another extreme case where $\s_K > K$ which leads to large probabilities of extinction of one or both strains. Here, the truncation of abundances at 0 leads to an inflation of the strain abundance correlation leading to it being close to 0. We do not work in this unrealistic region of parameter space either. 

In Fig.  \ref{fig: correlation_isolation}, we show the decrease in correlation with increasing strain interaction dissimilarity ($\l$) as well as the scatter of abundances for both strains across different realizations for a given $\l$. This serves as a baseline to compare the transition in correlation sign and strength in the presence of a community as shown in Fig. \ref{fig:dynamics}.

\subsection*{Correlations between strains generically increases in a community relative to them in isolation}

Here, we relax the assumption that the growth rates of strains in isolation are uncorrelated i.e. Corr$(K_\a, K_b) = 0$ and assume they have some underlying correlation $\r_{\text{iso}}$.  Let $K_\a = K + \sigma_K z_\a$ and $K_\b = K + \sigma_K z_\b$, where $z_\a$ and $z_\b$ are correlated random variables with $\text{Corr}(z_\a, z_\b) = \rho_{\text{iso}}$. This intrinsic correlation persists whether the strains are in isolation or embedded in a community, giving $\text{Cov}(K_\a, K_\b) = \rho_{\text{iso}} \sigma_K^2$.

In isolation, the effective growth rate correlation is simply $\rho_{\text{iso}}$. In a community, the effective growth rates acquire additional fluctuations mediated by interactions with other species (see Appendix B, eqns. S78 and S79):

\begin{align}
K_\a^{\text{eff}} &= g + \sigma_g z_\a^{(\text{int})} + \sigma_K z_\a \\
K_\b^{\text{eff}} &= g + \sigma_g \left( \rho_{\text{int}} z_\a^{(\text{int})} + \sqrt{1 - \rho_{\text{int}}^2} \, z_\b^{(\text{int})} \right) + \sigma_K z_\b
\end{align}
where $z_\a^{(\text{int})}$ and $z_\b^{(\text{int})}$ are independent standard normals representing community-mediated fluctuations, while $z_\a$ and $z_\b$ retain their intrinsic correlation $\rho_{\text{iso}}$. The community-induced correlation is $\rho_{\text{int}} = \sigma^2 M^2 / (\sigma^2 M^2 + \lambda^2 D^2)$.

The variance of each effective growth rate is $\text{Var}(K_\a^{\text{eff}}) = \sigma_g^2 + \sigma_K^2$, and the covariance between strain effective growth rates is $\text{Cov}(K_\a^{\text{eff}}, K_\b^{\text{eff}}) = \rho_{\text{int}} \sigma_g^2 + \rho_{\text{iso}} \sigma_K^2$. The total effective growth rate correlation in the community is therefore

\begin{equation}
\rho_K = \frac{\rho_{\text{int}} \sigma_g^2 + \rho_{\text{iso}} \sigma_K^2}{\sigma_g^2 + \sigma_K^2}.
\end{equation}

The community enhances correlation relative to isolation when $\rho_K > \rho_{\text{iso}}$:
\begin{equation}
\frac{\rho_{\text{int}} \sigma_g^2 + \rho_{\text{iso}} \sigma_K^2}{\sigma_g^2 + \sigma_K^2} > \rho_{\text{iso}} \implies \rho_{\text{int}} \sigma_g^2 + \rho_{\text{iso}} \sigma_K^2 > \rho_{\text{iso}} \sigma_g^2 + \rho_{\text{iso}} \sigma_K^2 \implies \rho_{\text{int}} > \rho_{\text{iso}}.
\end{equation}

Thus the community increases effective growth rate correlation if and only if the community-mediated correlation exceeds the intrinsic correlation. We can substitute the expression for $\rho_{\mathrm{int}}$ giving us,

\begin{equation}
    \frac{\s^2 \langle (N_\a +N_\b)^2 \rangle}{\s^2 \langle (N_\a +N_\b)^2 \rangle + \l^2 \langle (N_\a -N_\b)^2 \rangle} > \rho_{\mathrm{iso}}.
    \label{eq: rho_int_rho_iso}
\end{equation}

When conspecific strains are very similar i.e. where $\l \ll \s$, we have $\rho_{\text{int}} \approx 1$ (as the second term in the denominator of \ref{eq: rho_int_rho_iso} goes to 0 and $\l \rightarrow 0$). This exceeds any underlying strain correlation (unless $\rho_{\text{iso}} = 1$), implying that community embedding generically enhances equalization especially between closely related strains.

\section*{Appendix G: Materials and methods} \label{app:numerical_methods}

\subsection*{Numerical simulation of complete GLV system and model overview}

We performed numerical simulations of a generalized Lotka-Volterra model to investigate the population dynamics, coexistence, and correlation patterns of species and strains in microbial communities consisting of $S$ species, each harboring two distinct strains (denoted $\a$ and $\b$). The full system therefore comprises $2S$ dynamical variables representing strain-level abundances. All simulations were implemented in MATLAB v. R2025b and executed in parallel across multiple realizations for each parameter condition. All relevant code to reproduce figures and simulations can be accessed at \url{https://github.com/eltanin4/strains-filtering-sorting-cavity}.

The temporal evolution of strain abundances follows the generalized Lotka-Volterra equations as decribed in \ref{eq:main_GLV_dynamics}:
\begin{equation}
\frac{dN_{i\alpha}}{dt} = N_{i\alpha} \left( K_{i\alpha} - \sum_{\substack{j=1}}^S A_{i\alpha j\alpha} N_{j\alpha} - \sum_{\substack{j=1}}^S A_{i\alpha j\beta} N_{j\beta} \right)
\end{equation}
where $N_{i\a}$ denotes the abundance of strain $\a$ of specie `i', $K_{i\a}$ represents the growth rate, and $A_{i\a j\b}$ quantifies the interaction strength between strains $i\a$ and $j\b$. To prevent numerical extinction artifacts during integration, a small constant ($10^{-20}$) was added to the right-hand side. The system was integrated using MATLAB's \texttt{ode45} adaptive Runge-Kutta solver over a time span of $[0, 2000]$ time units, which was sufficient for the system to reach equilibrium across all parameter regimes explored.

The interaction matrix $A$ encodes both within-species (intra-specific) and between-species (inter-specific) interactions, with a correlation structure that captures the relationship between strains of the same species. For interaction term details, refer to Appendix A. Simulations were conducted with $S = 100$ species (200 total strains) across 100 independent realizations per parameter condition. The asymmetry parameter $\lambda$ was varied from 0.005 (as close to 0 yet not zero to prevent degeneracy of strains) to $\sigma$ (the unscaled standard deviation of interaction coefficients). We make the biologically realistic assumption that strains cannot e more dissimilar than species. growth rates were set to a constant value ($K = 1$ for all strains, with $\sigma_K = 0.02$). Initial conditions were drawn uniformly at random from $[0, 1]$ for each strain.

Following numerical integration, equilibrium abundances were determined from the final time point. Strains with abundances below a threshold of $10^{-14}$ were classified as extinct and removed from subsequent analyses. For the surviving community, the theoretical equilibrium abundances were computed by solving the reduced linear system:
\begin{equation}
    {N}^* = {A}_{\text{red}}^{-1} {K}_{\text{red}}
\end{equation}
where ${A}_{\text{red}}$ and ${K}_{\text{red}}$ denote the interaction matrix and growth rate vector restricted to surviving strains, respectively. The absolute value of the difference between the numerical and theoretical equilibria was recorded to verify convergence.

Realizations producing any strain abundance exceeding 100 at equilibrium were rejected as biologically implausible and regenerated with new random interaction matrices, ensuring that all analyzed results correspond to stable, bounded ecological equilibria.

For each realization, we computed the susceptibility matrix as the inverse of the reduced interaction matrix:
\begin{equation}
    {\nu} = {A}_{\text{red}}^{-1}
\end{equation}
The susceptibility $\nu_{ij}$ quantifies the linear response of strain $i$'s equilibrium abundance to a perturbation in the growth rate of strain $j$. We extracted four categories of susceptibilities for each species where both strains survived namely the self-susceptibilities: $\nu_{\alpha\alpha}$ and $\nu_{\beta\beta}$ and the cross-susceptibilities: $\nu_{\alpha\beta}$ and $\nu_{\beta\alpha}$. These were averaged across species within each realization to obtain community-level susceptibility statistics. For each value of $\lambda$ , we computed the mean abundances ($\langle N_\alpha \rangle$, $\langle N_\beta \rangle$), second moments ($\langle N_\alpha^2 \rangle$, $\langle N_\beta^2 \rangle$), cross-moment ($\langle N_\alpha N_\beta \rangle$) (all conditioned on both strains surviving) as well as the fraction of $\a$ and $\b$ strains surviving ($\phi_\alpha$ and $\phi_\beta$ respectively) and the fraction of species where both strains coexist ($\phi_{\alpha\beta}$).



Correlation coefficients:
The correlation between $\alpha$ and $\beta$ strain abundances was computed both across all strain pairs and restricted to coexisting pairs, using the standard formula:
\begin{equation}
    \rho_{\alpha\beta} = \frac{\langle N_\alpha N_\beta \rangle - \langle N_\alpha \rangle \langle N_\beta \rangle}{\sqrt{\left(\langle N_\alpha^2 \rangle - \langle N_\alpha \rangle^2\right)\left(\langle N_\beta^2 \rangle - \langle N_\beta \rangle^2\right)}}
\end{equation}

Lastly, simulations were parallelized across asymmetry parameter values (i.e., $\l$) using MATLAB's Parallel Computing Toolbox (\texttt{parfor}), with each worker independently generating interaction matrices, integrating the dynamical system, and computing summary statistics for a given $\lambda$ value.

\subsection*{Numerically solving the cavity self-consistency equations}

Concomitantly, we analytically calculate and solve the 12 essential quantities (first, second, and cross-moments as well as survival probabilities and susceptibilities) of the interacting strains obtained using the cavity method (see equations in Appendix C). Due to the system having 12 self-consistency quations and 12 unknowns, we the use nonlinear least squares method. We define the objective function to be the sum of the squares of the differences between the left and right-hand sides of the self-consistency equations.

Prior to imposing the constraint that abundances must be positive, the cavity framework yields Gaussian-distributed abundances for each strain. The untruncated means, variances, and cross-moments of untruncated quantities have been shown in extensive detail from equations \ref{eq:first_mom_Nalpha},  \ref{eq:sec_mom_Nalpha}, \ref{eq:sec_mom_Nbeta}, and \ref{eq:cross_mom_Nalphabeta}.
From these moments, the untruncated covariance and correlation of the abundances of the conspecific strains follow as:
\begin{align}
    \mathrm{Cov}_{\alpha\beta}^{\text{untrunc}} &= \langle N_\alpha N_\beta \rangle^{\text{untrunc}} - \langle N_{0\a}^{\pm} \rangle \langle N_{0\b}^{\pm} \rangle \\
    \rho_{\alpha\beta}^{\text{untrunc}} &= \frac{\mathrm{Cov}_{\alpha\beta}^{\text{untrunc}}}{\sqrt{\mathrm{Var}(N_{0\alpha}^{\pm}) \, \mathrm{Var}(N_{0\beta}^{\pm})}}
\end{align}

Since only strains with positive equilibrium abundances survive, the observed statistics should correspond to moments of a truncated bivariate normal distribution over the region $\{N_\alpha > 0, N_\beta > 0\}$. We standardized the untruncated distribution by defining:
\begin{equation}
    h = -\frac{\langle N_{0\a}^{\pm} \rangle}{\sqrt{\mathrm{Var}(N_{0\alpha}^{\pm})}}, \qquad k = -\frac{\langle N_{0\b}^{\pm} \rangle}{\sqrt{\mathrm{Var}(N_{0\beta}^{\pm})}}
\end{equation}
so that the truncation region becomes $\{X > h, Y > k\}$ for standard bivariate normal variables $(X, Y)$ with correlation $\rho_{\alpha\beta}^{\text{untrunc}}$.

The truncated moments were computed using the formulas of Rosenbaum \cite{rosenbaum1961moments} and Begier and Hamdan \cite{begier1971correlation} for bivariate normal distributions truncated to upper rectangles. Specifically, $\langle X \rangle$ and $\langle Y \rangle$ are the first moments of two standard normal random variables 'X' and 'Y' (that correspond to the $\a$ and $\b$ strains respectively), $\langle X^2 \rangle, \langle Y^2 \rangle$ correspond to their second moments, and $\langle XY \rangle$ corresponds to their cross-moments., the joint survival probability is:
\begin{equation}
    L = P(X > h, Y > k)
\end{equation}
computed via the bivariate normal cumulative distribution function. The conditional first moments are:
\begin{align}
    \langle X \mid X > h, Y > k \rangle &= \frac{\phi(h) Q(a) + \rho \phi(k) Q(b)}{L} \\
    \langle Y \mid X > h, Y > k \rangle &= \frac{\rho \phi(h) Q(a) + \phi(k) Q(b)}{L}
\end{align}
where $\phi(\cdot)$ is the standard normal density, $Q(\cdot) = 1 - \Phi(\cdot)$ is the upper tail probability, and
\begin{equation}
    a = \frac{k - \rho h}{\sqrt{1-\rho^2}}, \qquad b = \frac{h - \rho k}{\sqrt{1-\rho^2}}
\end{equation}

The conditional second moments are:
\begin{align}
    \langle X^2 \mid X > h, Y > k \rangle &= \frac{L + h \phi(h) Q(a) + \rho^2 k \phi(k) Q(b) + e}{L} \\
    \langle Y^2 \mid X > h, Y > k \rangle &= \frac{L + k \phi(k) Q(b) + \rho^2 h \phi(h) Q(a) + e}{L}
\end{align}
where the edge term is:
\begin{equation}
    e = \frac{\rho \sqrt{1-\rho^2}}{\sqrt{2\pi}} \phi\left( \sqrt{\frac{h^2 - 2\rho h k + k^2}{1-\rho^2}} \right)
\end{equation}

The conditional cross-moment is:
\begin{equation}
    \langle XY \mid X > h, Y > k \rangle = \frac{\rho L + \rho h \phi(h) Q(a) + \rho k \phi(k) Q(b) + e'}{L}
\end{equation}
where $e' = \frac{\sqrt{1-\rho^2}}{\sqrt{2\pi}} \phi\left( \sqrt{\frac{h^2 - 2\rho h k + k^2}{1-\rho^2}} \right)$.

These standardized truncated moments were then transformed back to obtain the truncated abundance statistics (see Appendix C).

Thus, the cavity framework yields a closed system of 12 self-consistency equations relating the following unknowns i.e., 1) Truncated first moments: $\langle N_\alpha \rangle$, $\langle N_\beta \rangle$ , 2) truncated second moments: $\langle N_\alpha^2 \rangle$, $\langle N_\beta^2 \rangle$, 3) survival probabilities: $\phi_\alpha$, $\phi_\beta$, $\phi_{\alpha\beta}$, 4) susceptibilities: $\nu_{\alpha\alpha}$, $\nu_{\beta\beta}$, $\nu_{\alpha\beta}$, $\nu_{\beta\alpha}$, and 5) the truncated cross-moment: $\langle N_\alpha N_\beta \rangle$

Each equation takes the form of a residual $r_i( {x}) = x_i - f_i( {x})$, where $x_i$ is the $i$-th unknown and $f_i( {x})$ is its predicted value from the cavity framework given the current estimates of all unknowns. The self-consistent solution satisfies $ {r}( {x}^*) =  {0}$.

The system was solved by minimizing the sum of squared residuals:
\begin{equation}
    \min_{ {x}} \frac{1}{2} \sum_{i=1}^{12} r_i( {x})^2
\end{equation}
subject to bound constraints ensuring physical feasibility (e.g., non-negative moments, survival probabilities in $[0,1]$). Optimization was performed using MATLAB's \texttt{fmincon} with the sequential quadratic programming (SQP) algorithm. For parameter sweeps over the asymmetry parameter $\lambda$, solutions were computed sequentially with a warm-start: the solution at each $\lambda$ served as the initial guess for the subsequent value, improving convergence and ensuring continuity of the solutions.

The correlation $\rho_{\mathrm{int}}$ between cavity fields was computed directly from the current estimates of the abundance moments at each iteration, ensuring internal consistency of the solution.

\section*{Appendix H: Parameters used in main text figures}

In the main text and SI, we run all our simulations in the regime where $\rho = 1$ i.e., the interactions between species are symmetric. 

\paragraph*{Figure 1 (schematic) details.}\

The illustrations in this figure were made by AG and NV using the Inkscape software. Equations in the figure were converted from Latex text to .svg format via \url{https://viereck.ch/latex-to-svg/}.

\paragraph*{Figure 2 (community mediated coexistence) details}\

Simulations of community-mediated coexistence were performed with $S = 100$ species (200 total strains). The mean inter-species interaction strength was $\mu = 0.05$, with interaction correlation $\rho = 0.99$. growth rates were drawn from a normal distribution with mean $K = 1$ and standard deviation $\sigma_K = 0.2$. The focal strain pair in Panels A and B used within-species cross-strain interaction $\tilde{\mu} = 0.7$ and asymmetry parameter $\lambda = 0.02$. For the structured community condition, the inter-species interaction standard deviation was $\sigma = 0.38$. The system was integrated over the time interval $[0, 2000]$ with extinction threshold $10^{-14}$. The phase diagram in Panel D swept over $\tilde{\mu} \in [0.6, 0.95]$ in increments of 0.05 and $\lambda \in [0.005, 0.06]$ in increments of 0.005, with 100 independent realizations per parameter point. 

\paragraph*{Figure 3. Strain abundance correlation patterns}\

The correlation structure across parameter space was characterized using $S = 100$ species (200 total strains). The mean inter-species interaction strength was $\mu = 0.05$, with interaction correlation $\rho = 0.99$. growth rates were drawn from a normal distribution with mean $K = 1$ and standard deviation $\sigma_K = 0.022$. The within-species cross-strain interaction was fixed at $\tilde{\mu} = 0.5$.  The system was integrated until steady state for 100 different realizations. Panels (a) and (b) show scatter plots of the abundances of pair of strains in a species in isolation and in a community respectively (with $\s = 0.30$). The ellipse drawn around the points represent 2 standard deviations of the distribution generated from the analytically predicted truncated bivariate Gaussian distribution (derived from the cavity method). The ellipse is constructed by eigendecomposition of the covariance matrix where eigenvalues and eigenvectors describe the size (along each principal axis) and orientation of the theoretical ellipse. In panel (c), we show how the same pair of strains in isolation (white circle) having negative correlation has positive correlation in a community (orange triangle) through stabilization (reduced $\tmueff$) and equalization (higher $\rho_K$) due to the community. These points are overlaid atop a heatmap of strain abundance correlations obtained by numerically solving the two-strain GLV for pairs of values of ($\tmueff, \rho_K$) which are averaged across 500 independent realizations. In panel (d), we show that strain abundance correlations ($\r_{\a\b}$) increase with increasing species interaction dissimilarity ($\s$). The prediction from the analytic cavity solution (line) for the correlation is overlaid on top of correlation derived from simulations (dots). In panel (d) inset, we use $\rho_K$ and $\left \langle \frac{1}{\xi^2 - (\tmueff)^2 + \l^2y^2} \right \rangle$ as measures of equalization and stabilization and show that they both increase with increasing $\s$.


\paragraph*{Figure 4. Mechanisms driving correlation transitions}\

For panel (a), simulations were run for communities using $S = 100$ species (200 total strains). The mean inter-species interaction strength was $\mu = 0.30$, with interaction correlation $\rho = 0.99$. growth rates were drawn from a normal distribution with mean $K = 1$ and standard deviation $\sigma_K = 0.02$. The system was integrated over the time interval $[0, 1000]$ with extinction threshold $10^{-14}$. Further, we fixed $\s = 0.06$ to show positive to negative transition in correlations. Panel (b) uses $\rho_K$ and $\left \langle \frac{1}{\xi^2 - (\tmueff)^2 + \l^2y^2} \right \rangle$ as measures of equalization and stabilization and show that they both increase with increasing $\l \in [0.005, 0.11]$. Panel (c) shows a phase diagram in ($\s,\l$) space with parameter sweep covering $\sigma \in [0.005, 0.20]$ and $\lambda \in [0.005, 0.20]$, both in increments of 0.005, restricted to $\lambda \leq \sigma$. The phase diagram in panel (d) swept over $\tilde{\mu} \in [0.1, 0.95]$ in increments of 0.03 and $\sigma \in [0.01, 0.31]$ in increments of 0.01. Regimes were classified by evaluating correlations at $\lambda_{\min} = 0.01$ and $\lambda_{\max} = \sigma$: positive-to-negative (blue) if $\rho_{\alpha\beta}(\lambda_{\min}) > 0$ and $\rho_{\alpha\beta}(\lambda_{\max}) < 0$; always negative (red) if $\rho_{\alpha\beta}(\lambda_{\max}) < 0$ without sign change; and positive-to-zero/positive (green) otherwise (after checking for monotonicity). The theoretical line was calculated by solving the cavity self-consistency equations for each value of $\s$ and finding a concomitant $\tilde{\mu}$ where the covariance goes to 0 using a binary search algorithm (faster than brute force search). 


\end{widetext}

\end{widetext}

\end{document}